\documentclass[iop, numberedappendix]{emulateapj}
\shorttitle{Isothermal dusty gas dynamics}
\shortauthors{M-K.\ Lin, A.N. Youdin}

\usepackage{amsmath}
\usepackage{paralist}
\usepackage{url}
\usepackage{bm}
\usepackage{tikz}
\usetikzlibrary{shapes.geometric, arrows}
\usepackage{multirow}

\newcommand{\p}{\partial}
\newcommand{\zmax}{z_\mathrm{max}}

\newcommand{\ii}{\mathrm{i}}

\newcommand{\dd}{\delta}
\newcommand{\real}{\operatorname{Re}}
\newcommand{\imag}{\operatorname{Im}}
\newcommand{\sgn}{\operatorname{sgn}}

\newcommand{\rich}{\mathrm{Ri}}
\newcommand{\Htilde}{H_\epsilon}

\newcommand{\Hgas}{H_\mathrm{g}}
\newcommand{\hgas}{h_\mathrm{g}}
\newcommand{\tepsilon}{f_\mathrm{d}}
\newcommand{\rhod}{\rho_\mathrm{d}}
\newcommand{\rhog}{\rho_\mathrm{g}}
\newcommand{\tstop}{t_\mathrm{s}}
\newcommand{\taus}{\tau_\mathrm{s}}
\newcommand{\st}{\mathrm{St}}

\newcommand{\fdust}{f_\mathrm{d}}
\newcommand{\fg}{f_\mathrm{g}}

\newcommand{\Hg}{H_\mathrm{g}}
\newcommand{\Hd}{H_\mathrm{d}}
\newcommand{\calP}{\mathcal{P}}
\newcommand{\calQ}{\mathcal{Q}}
\newcommand{\calA}{\mathcal{A}}
\newcommand{\calB}{\mathcal{B}}
\newcommand{\seff}{S_\mathrm{eff}}
\newcommand{\tcool}{t_\mathrm{cool}} 

\defcitealias{lin12}{L12} 
\defcitealias{lin15}{LY15}
\defcitealias{nelson13}{N13}
\defcitealias{barker15}{BL15}
\defcitealias{mcnally14}{MP14}
\defcitealias{goldreich67}{GS67}

\def \OmK {\Omega_{\rm K}}

\begin{document}

%\title{On the entropy of 
%  isothermal dusty gas: application to protoplanetary disks}  
\title{A thermodynamic view of dusty protoplanetary disks}
 % \title{Dust-free modeling of dusty protoplanetary disks I:
%axisymmetric stability}
\author{Min-Kai Lin\altaffilmark{1,2,3}}
\author{Andrew N. Youdin\altaffilmark{2}}
%\affil{Department of Astronomy and Steward Observatory, 
%University of
 % Arizona, Tucson, AZ 85721, USA\\
%Institute of Astronomy and Astrophysics, Academia Sinica, 
%Taipei 10617, Taiwan}
\altaffiltext{1}{Institute of Astronomy and Astrophysics, Academia Sinica, Taipei 10617, Taiwan}
\altaffiltext{2}{Department of Astronomy and Steward Observatory, University of Arizona, Tucson, AZ 85721, USA}
\altaffiltext{3}{Steward Theory Fellow}

\email{mklin@asiaa.sinica.edu.tw}

\begin{abstract}
  Small solids embedded in gaseous protoplanetary disks are subject
  to strong dust-gas friction. Consequently, tightly-coupled dust
  particles almost follow the gas flow.
  %AY: Edited b/c "This is" was too vague
  This near conservation of dust-to-gas ratio along streamlines
  is analogous to the near conservation of entropy along flows of (dust-free) gas with weak heating and cooling.   
  We develop this thermodynamic analogy into a
  framework to study dusty gas dynamics in protoplanetary disks. We  
  show that an isothermal dusty gas behaves like an  
  adiabatic pure gas; and that finite dust-gas coupling may be
  regarded as an effective 
  heating/cooling. We exploit this correspondence 
  to deduce that   
\begin{inparaenum}[1)] 
\item 
 perfectly coupled, thin dust layers cannot cause axisymmetric 
 instabilities; 
\item 
 radial dust edges are unstable if the dust is vertically well-mixed;  
\item 
  the streaming instability necessarily involves a gas pressure
  response 
  that lags behind dust density; 
%  the streaming instability neccessarily involves a phase lag between
%  gas pressure and dust densities; 
\item dust-loading introduces buoyancy forces that generally 
  stabilizes the vertical shear 
  instability associated with global radial temperature gradients. 
\end{inparaenum}  
%growth rates, and 
%preferentially stabilize modes with small radial 
%wavelengths, as well as `surface modes' concentrated near the
%disk boundaries. 
%These results can be understood in terms of 
%buoyancy forces in an isothermal dusty disk. 
%We show that vertical shear arising from
%dust settling does not lead to axisymmetric instability. 
We also discuss %future applications of this thermodynamic framework in
%finding 
 dusty analogs of other hydrodynamic 
processes (e.g. Rossby wave instability, convective overstability, and 
zombie vortices), and how to simulate dusty protoplanetary disks
with minor tweaks to existing codes for pure gas dynamics.%, as well as possible generalizations
%of our analogy between dust-laden and nearly adiabatic flows. 

%We briefly describe extending this
%equivalence principle to polytropic and adiabatic gas. 

\end{abstract}

\section{Introduction}
%si, sgi as examples
Protoplanetary disks are comprised of a mixture of gas and dust
\citep{chiang10}. Although the overall dust content is small
($\sim 1\%$ by mass), their presence can have profound impacts on the 
gas dynamics and vice versa. Dust-gas friction introduces phenomena
that are absent in pure gas disks. Important examples include the
streaming instability 
\citep[SI,][]{youdin05a,youdin07b,johansen07}, secular gravitational
instabilities \citep[SGI,][]{ward00,youdin11,michikoshi12,takahashi14,taka16,latter17} and 
Kelvin-Helmholtz instabilities
\citep{goldreich1973, chiang08,barranco09,lee10}. 

These instabilities can trigger turbulence, and (in different regimes) promote or inhibit  
planetesimal formation. 
More recently, new 
dusty instabilities have appeared in numerical  
simulations \citep{loren15,loren16,lamb16} and analytical calculations \citep{squire17,hopkins17},   
but their physical underpinnings are not yet well 
understood.

%particles 
Current state-of-the-art models of dusty protoplanetary
disks directly simulate gas dynamics coupled to explicit
Lagrangian dust particles
\citep{johansen2006, nelson10,bai10,yang14,zhu14,gibbons15,simon16,baruteau16}. 
The equation of motion for each particle is solved 
including dust-gas drag; the strength of which is measured 
by a stopping time $\tstop$ --- the decay timescale for the 
relative velocity between gas and dust. Thus small $\tstop$
corresponds to tightly-coupled particles. 

%refs 

Another approach is to model the dust population as a continuous, pressureless
fluid 
\citep{barriere-fouchet05,paardekooper06b,ayliffe12,meheut12,laibe12,loren14,fu14b,surville16}. The 
hydrodynamic equations are evolved for two fluids: the gas and dust with density and
velocity $\rhog$, $\bm{v}_\mathrm{g}$ and $\rhod$, 
$\bm{v}_\mathrm{d}$, respectively. Source terms are introduced into
the momentum/energy equations to model dust-gas drag. This approach
can take advantage of existing numerical methods/codes for
simulating pure gas dynamics. A practical difficulty with this approach is the need to numerical stabilize the pressureless dust fluid with (e.g.) artificial diffusion.

%refs 

%AY: reworded b/c (a) I didn't think it was clear enough that the one fluid simplification wasn't exact and (b) that this tight coiupling limit needed to be emphasized more.

The two-fluid equations for dust and gas can be reformulated into 
equivalent dynamical equations for the center of mass velocity $\bm{v}$
and the relative dust-gas velocity  $\bm{v}_\mathrm{d}-\bm{v}_\mathrm{g}$ 
\citep{youdin05a}. In this reformulation, continuity equations for dust and gas can be replaced 
by the mass conservation of total density $\rho$ plus the evolution of the dust-to-gas ratio $\rhod/\rhog$ 
(or dust fraction, $\rhod/\rho$, \citealp{laibe14}).

This reformulation is particularly advantageous in the tight coupling limit of  $\tstop \ll 1/\Omega$, where the orbital frequency $\Omega$ sets the characteristic timescale in many disk dynamics problems.
Models with two fluids or with Lagrangian solids are numerically difficult to evaluate in this regime, due to stiff drag forces.  
In the reformulated equations, however, the center of mass motion does
not experience drag forces and thus is not stiff.  Moreover,
for $\tstop\Omega\ll 1$ the relative motion satisfies the terminal-velocity approximation and can be eliminated from the equations.  This single fluid framework for a well-coupled dust-gas mixture facilitates numerical calculations \citep{price15}; and also simplifies analytic calculations \citep{youdin05a,jacquet11}, which is useful for gaining insight.  

%It is possible to re-formulate the two-fluid equations into that for a
%single fluid with total density $\rho$ and center-of-mass
%velocity $\bm{v}$ of the dusty gas mixture
%\citep{laibe14}. The standard hydrodynamic equations for
%$\rho$ and $\bm{v}$ are supplemented by evolutionary equations for
%the dust-to-gas ratio $\rhod/\rhog$ and the relative dust-gas velocity
%$\Delta\bm{v}\equiv \bm{v}_\mathrm{d}-\bm{v}_\mathrm{g}$. The latter also contributes to 
%additional source terms in the momentum and energy equations.  
%
%The one-fluid approach is appropriate for 
%simulating gas tightly coupled to dust particles with small $\tstop$.  
%This can be numerically challenging for two-fluid models and 
%following Lagragian particles; but is straight forward with the 
%one-fluid formulation \citep{price15}.  The single fluid framework 
%also has the important advantage that it is 
%amenable to standard analytical methods \citep{youdin05a,jacquet11}.  
%This is the necessary approach to gain physical insight.  

%AY: Attempted to clarify a bit by separating more the analogy and it's applications.

In this work, we develop a parallel between this single-fluid description of dust-gas mixtures and standard hydrodynamics.
Specifically we recast the evolution of the dust-to-gas ratio as an energy equation for the thermal content of a fluid.
This parallel is most precise when the gas in the dust/gas mixture obeys a locally isothermal equation of state for the gas, $P = c_s^2\rhog$, where $c_s$ is a prescribed sound-speed profile.   This isothermal approximation holds when cooling times are short \citep{lin15}, e.g.\ in protoplanetary disks 
 whose temperature is set by external irradiation
\citep{chiang97,stam08}.

%are motivated to develop a general analytical
%framework to study the stability of dusty protoplanetary 
%disks by drawing on the physical and mathematical parallel between
%dusty gas and standard hydrodynamics. This is possible in certain limits of astrophysical
%interest, allowing us to interpret dusty phenomena through standard
%hydrodynamics.   

%AY: a bit detailed for an introduction, plus the idea of terminal velocity needed to be introduced earlier as the basis of the slngle fluid equations, I thought.
%We consider tightly coupled dust particles with $\tstop\Omega\ll
%1$, where $\Omega$ is the disk orbital frequency. We may then use the
%`terminal velocity approximation' to set the dust-gas relative drift
%to $\Delta \bm{v} = \tstop\nabla
%P/\rhog$, where $P$ is the gas pressure \citep{youdin05a, 
%  jacquet11,laibe14}. This approximation reflects the physical result
%that dust particles tend to drift towards pressure maximum. 
 
% We also consider irradiated protoplanetary disks 
% in which the temperature is set externally
%\citep{chiang97,stam08}. This allows us to adopt a locally isothermal
%equation of state, $P = c_s^2\rhog$, where $c_s$ is a prescribed
%sound-speed profile. This physically represents short cooling times
%\citep{lin15}.   

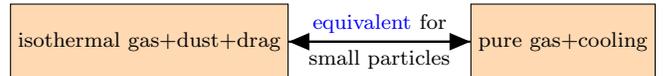
\begin{figure}
\tikzstyle{process} = [rectangle, minimum width=2cm, minimum height=1cm, text centered, draw=black, fill=orange!30]
  \tikzstyle{line} = [draw, -triangle 45,thick]  
  \begin{center}
    \begin{tikzpicture}[node distance=5cm]
      \node[process](ellip){isothermal gas+dust+drag};
      \node[process,right of=ellip, node distance = 5.5cm](gi){pure 
        gas+cooling};
      \path [line](ellip) -- node [yshift=.8em]
      {\textcolor{blue}{equivalent} for} (gi);
      \path [line](gi) -- node [yshift=-.8em] {small particles} (ellip);
    \end{tikzpicture}
 \end{center}
\caption{The central concept of this work. A mixture of small particles
  imperfectly coupled to isothermal gas behave similarly to an adiabatic, pure gas
  subject to cooling.  This
  permits one to use the equations for standard hydrodynamics to study
  dust-gas dynamics. \label{concept_chart}}
\end{figure}

Under these two approximations of strong drag and isothermal gas, we show that the equations of 
 dusty gas dynamics are equivalent to that of nearly adiabatic (i.e. slowly cooling) gas
dynamics without dust, see Fig.~\ref{concept_chart}.  The basis of the analogy is that the entrainment 
of tightly-coupled dust particles in gas flows is similar to the advection
of entropy in an adiabatic fluid. %We show that the   
%effective entropy of isothermal dusty gas scales with its gas
%fraction.  %AY: too technical for introduction.
Finite dust-gas drag act as effective heating/cooling: the relative
drift between dust and gas causes a fluid parcel to exchange dust
particles with its surroundings; an effect similar to heat exchange
between a gas parcel and its surroundings.   Needless to say the effective heating term must have a 
specific form, which we derive, for the equivalency to hold.

%suitable for 
%analytic studies 
%more importantly, can understood 

The purpose of our physical and mathematical parallel between
dusty gas and standard hydrodynamics is to gain a deeper understanding of the dynamics
and stability of dusty protoplanetary disks.  
We can apply established methods to generalize previous stability 
results for pure gas dynamics to dusty disks.  
We show that strictly isothermal disks with 
perfectly-coupled dust are generally 
stable to axisymmetric perturbations, unless the 
dust-to-gas ratio varies more rapidly in radius than in height.  
%, regardless of how  thin the dust
%layer is. 
%{\bf should we mention vertically well mixed case here?} 
We put forward a thermodynamical interpretation of 
dust-gas drag instabilities, and apply it to the streaming instability.   
We also study the effect of dust-loading on the vertical shear
instability previously considered in pure gas disks
\citep[VSI,][]{nelson13,lin15,barker15}.

This paper is organized as follows.  
In \S\ref{grow_osc} we describe a key insight of our study, that
pressure (i.e.\ ``PdV")  work {can} explain the basis of several
instabilities, including SI and VSI, in terms of 
pressure-density phase lags. { This motivates us to seek an 
  analogy between dust-gas mixtures and pure gas.}    
We develop our formalism in \S\ref{setup} by transforming the
two fluid equations of a tightly-coupled dusty gas into single fluid
hydrodynamic equations with a special cooling function.  Here we also
define the effective entropy and buoyancy of our model fluid. 
In \S\ref{limits} we discuss general stability properties of dusty
disks. We then analyze explicit disk models 
\S\ref{linear_problem}, were we show that radial dust edges are unstable,
and revisit the SGI and SI in the tight 
coupling regime. In \S\ref{results} we 
extend the VSI to dusty disks. We   
discuss future applications of the dusty/adiabatic gas equivalence in 
\S\ref{discussion} before summarizing in \S\ref{summary}.

\section{Growing oscillations by doing work}\label{grow_osc}
{
In this section we provide physical arguments to highlight the
similarity between single phase fluids and dust-gas mixtures. The mathematical
description of this section is deferred to \S\ref{dust_work}. 

It is
well-known that in the limit of perfect coupling the addition of dust
increases the gas inertia but not pressure. One can then regard 
dusty gas as a single fluid with a reduced sound-speed. Here, we
argue that when dust-gas coupling is imperfect, there exists another
similarity with pure gas; related the phase of pressure and density
evolution. Thus it is still possible to regard
partially-coupled dusty-gas as a single fluid.  
}
%dust_work

A general result in fluid dynamics is that work is done whenever
oscillations in the pressure and density of the fluid are not in
phase. If  the average work done is positive, then oscillation
amplitudes grow: the work done allows the system to `overshoot' beyond 
the amplitude of preceding oscillations.   

%To see this, we first compute the work done
%assuming periodic oscillations. We then show that if the work done is positive, then the
%oscillation amplitude in fact grows: the work done leads the
%system to `overshoot' past the amplitude of the preceeding oscillation
%cycle. 
%its equilibrium. {\bf should be overshoot the max displacement}

\begin{figure}
  \includegraphics[width=\linewidth]{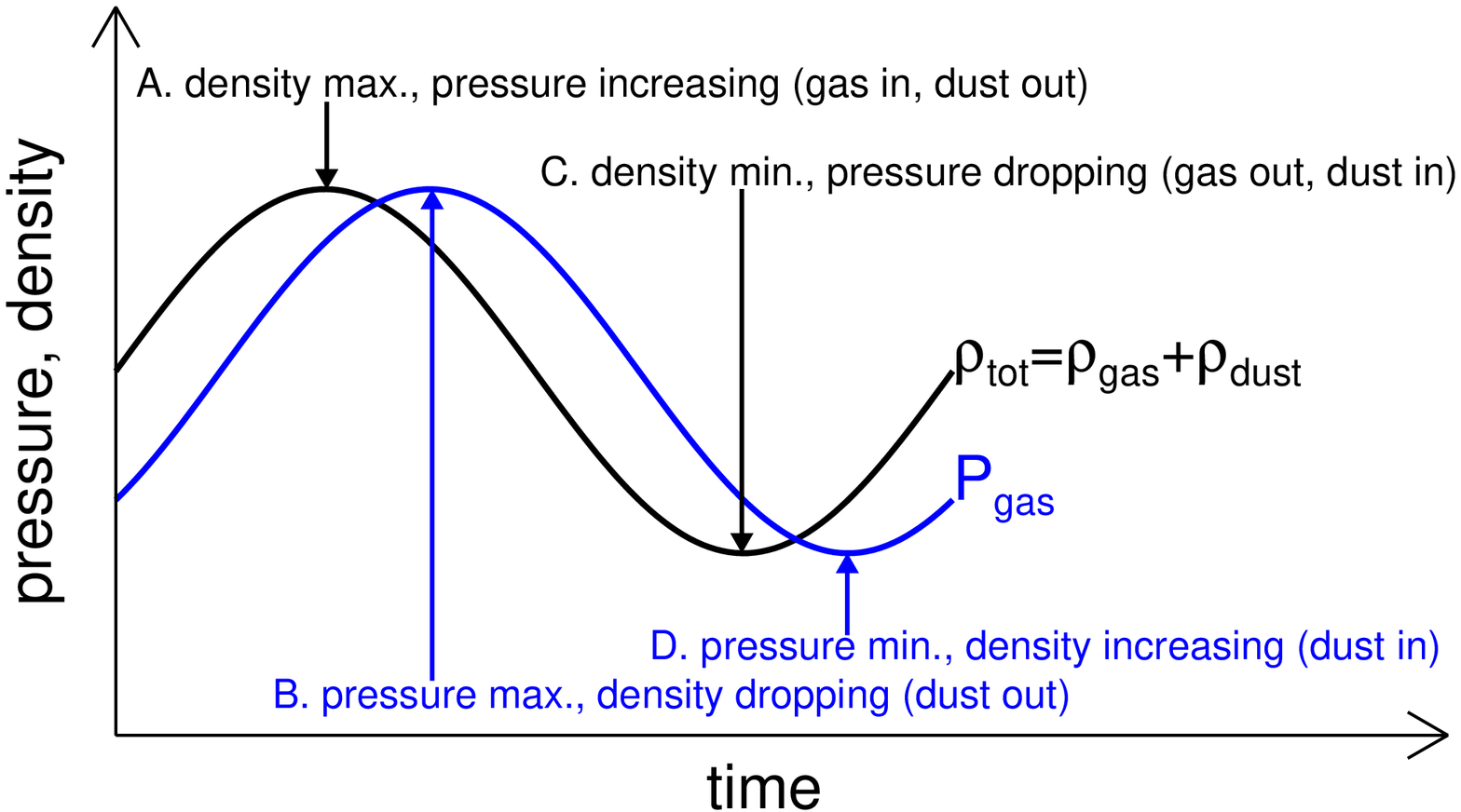}\\
\includegraphics[width=\linewidth]{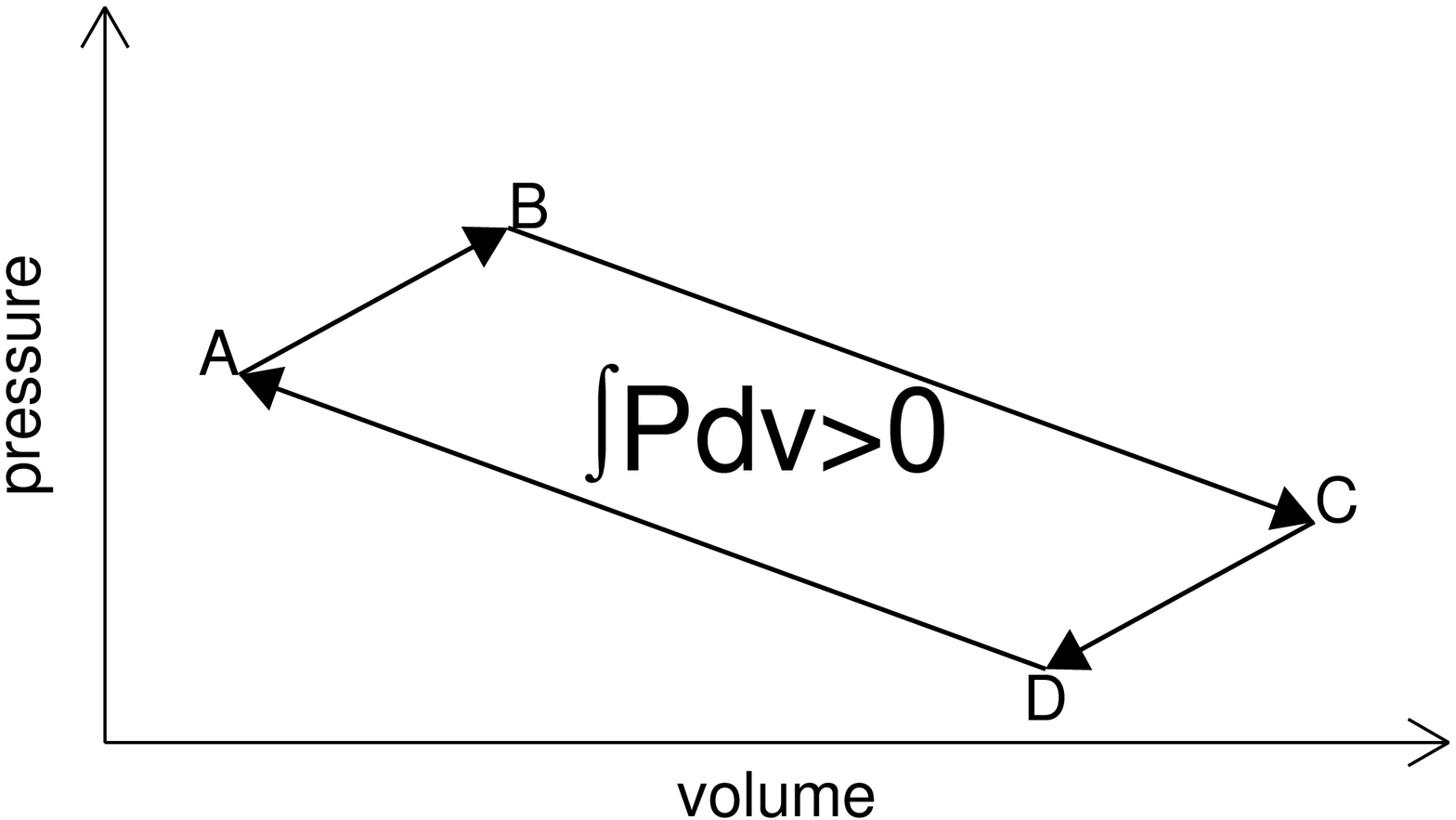}
  \caption{Thermodynamic interpretation of growing oscillations in a
    fluid. Top: pressure and density evolution in time. 
    Bottom: oscillation cycle in the P-V plane. In the case shown, 
    dust-drag causes  
    pressure (due to gas only) to lag behind total density. This results in a clockwise path in
    the P-V plane, implying positive work done by the fluid, which 
     would increase oscillation amplitudes.  
    \label{pdv_cartoon}
  }
\end{figure}

Fig. \ref{pdv_cartoon} gives a graphical demonstration that 
if pressure lags behind density, then positive is work done 
because it leads to a clockwise path in the 
`P-V' plane. The annotations consider the particular case where the
phase lag arises because the fluid has two components (gas and dust,
see below) but the following description is general. 

From $A$ to $B$, a fluid parcel is expanding to return to 
equilibrium, but pressure is still 
increasing. This over-compensation will cause the parcel to expand
beyond the maximum volume of the previous cycle. 
Similarly, from $C$ to $D$ the parcel is already contracting towards
equilibrium, but pressure 
is still dropping. This allows the  
parcel's contraction to over-shoot the maximum density attained in the
previous cycle. 
The overall positive work done leads to growth in the oscillation amplitude. 
This effect is similar to being pushed downwards on a swing when 
descending.

There are several situations where the pressure and density of a fluid
are not in phase. The obvious case is if the fluid is subject to
external heating/cooling. For example, in strongly irradiated
protoplanetary disks the
disk temperature  $T(r)$ is time-independent but varies with the cylindrical radius
$r$ from the star \citep{chiang97}. Since $P\propto \rhog T$ for an
ideal gas, %the pressure of a gas parcel varies with  
%position as well. 
there is no reason to expect $P$ and $\rhog$ to 
be in phase as a gas parcel oscillates between different radii and adopt the corresponding local temperatures. 
 In 
fact, we will show that this is a fundamental property of the vertical
shear instability \citep{lin15}.   

%temperature of a fluid parcel changes with position

Another possibility, as annotated in Fig. \ref{pdv_cartoon}, is a dusty gas. The
relevant density here is the total density $\rho = \rhod + \rhog$, 
but the fluid pressure $P$ is due to gas only. If dust were 
perfectly-coupled to the gas, then $P$ and $\rho$ would be in phase, 
and there is no work done. However, for finite dust-gas drag, $\rhod$ and
$\rhog$ are not necessarily in phase, because dust particles can
drift relative to the gas{, i.e. the dust-to-gas ratio evolves in time}. 
If this causes $P$ to lag 
$\rho$, then the positive work done would lead to growing
oscillations. Indeed, we show that this is true for the streaming
instability. 

In order to apply this thermodynamic interpretation of dust-gas drag
instabilities, we need to develop a formal analogy between dusty-gas
and pure hydrodynamics. We show this is possible in the limit of
strong drag and a fixed gas equation of state. 

\section{Single fluid description of dusty gas}\label{setup} 
We model an accretion disk as a mixture of gas with dust treated as a 
pressureless fluid. We denote their density and
velocity field as $(\rhog,\bm{v}_\mathrm{g})$ and
$(\rhod,\bm{v}_\mathrm{d})$, respectively. The 
mixture has total density \begin{align}
  \rho \equiv \rhog + \rhod,
\end{align}
and center of mass velocity, 
\begin{align}
  \bm{v} \equiv \frac{\rhog\bm{v}_\mathrm{g} + 
    \rhod\bm{v}_\mathrm{d}}{\rho}, 
\end{align}
a single temperature $T$, and its pressure $P$ arise solely from the  
gas component. Our goal is to obtain a set of equations
describing the dust-gas mixture that resembles standard, single-phase
hydrodynamics. 
%This is possible by considering
%strongly coupled dust particles (but not neccessarily perfectly
%coupled) and an isothermal gas equation of state.  

%\subsection{Terminal velocity approximation}

The dust and gas fluids interact via a drag force parameterized by the relative stopping 
time $\tstop$ such that 
\begin{align}  
  \rhod\left.\frac{\p \bm{v}_\mathrm{d}}{\p t}\right|_\mathrm{drag}= -
  \rhog\left.\frac{\p \bm{v}_\mathrm{g}}{\p t}\right|_\mathrm{drag}=
  - \frac{\rhog\rhod}{\rho}\frac{\left(\bm{v}_\mathrm{d} - \bm{v}_\mathrm{g}\right)}{\tstop}. 
\end{align}
is the dust-gas friction force per unit volume. %, where
%\begin{align}
%  \widetilde{\bm{v}} \equiv\bm{v}_\mathrm{d} - \bm{v}_\mathrm{g} 
%\end{align}
%is the dust-gas velocity difference. 
 Note that $\tstop$ 
differs slightly from the \emph{particle} stopping time $\tau_\mathrm{s} =
\tstop\rho/\rhog$ used in some studies
\citep[e.g.][]{youdin05a}. 

In general the relative velocity $\bm{v}_\mathrm{d} -
\bm{v}_\mathrm{g}$ obeys a complicated evolutionary equation 
\citep[see, e.g.][]{youdin05a}. % obtained from taking the difference
%between evolutionary equations for $\bm{v}_\mathrm{g}$ and
%$\bm{v}_\mathrm{d}$. 
 However, for tightly-coupled dust
particles with $\tstop\OmK\ll 1 $, where $\OmK$ is the Keplerian orbital
frequency (since we are interested in protoplanetary disks), we can use the 
`terminal velocity approximation' to set %The dust-gas velocity difference 
%Then so that the dust-gas velocity
%difference is 
\begin{align}
  \bm{v}_\mathrm{d} - \bm{v}_\mathrm{g} = \frac{\nabla
    P}{\rhog}\tstop\label{term_vel_approx} 
\end{align}
\citep{jacquet11}. This equation reflects the well-known effect of 
particle drift towards pressure maximum  \citep{weidenschilling77}. 

%\subsection{One-fluid equations for dusty gas}

Under the terminal velocity approximation the dust-gas mixture obeys   
the first-order (in $\tstop$) one-fluid equations: 
\begin{align} 
  &\frac{D\rho}{Dt} = -\rho\nabla\cdot\bm{v}, \label{masseq}\\ 
   &\frac{D\tepsilon}{Dt} = -\frac{1}{\rho} \nabla \cdot \left(\tepsilon 
     \tstop \nabla P \right),\label{dusteq}\\
  &\frac{D\bm{v}}{Dt} = - \nabla\Phi_\mathrm{tot} - \frac{1}{\rho}\nabla  P, \label{momeq}\\ 
  &\frac{D T}{D t} = - \left(\gamma-1\right)T\nabla\cdot\bm{v} +
  \mathcal{H} + \mathcal{H}_\mathrm{eff}  - \Lambda  \label{tempeq} 
\end{align} 
\citep[see ][for a detailed derivation from the two-fluid
equations]{laibe14,price15}, where $D/Dt \equiv \p_t +
\bm{v}\cdot\nabla$ is the Lagrangian 
derivative \emph{following the mixture}'s center-of-mass velocity $\bm{v}$. 
For an ideal gas the pressure is given 
by $P = \mathcal{R}\rhog T/\mu $, where $\mathcal{R}$ is
the gas constant and $\mu$ is the mean molecular weight.

Eq. \ref{dusteq} is obtained from the dust continuity equation,  
$\p_t\rhod + \nabla\cdot\left(\rhod\bm{v}_\mathrm{d}\right)= 0$, 
by writing 
$\bm{v}_\mathrm{d}=\bm{v} + \tstop\nabla P/\rho$, and eliminating  
$\rhod$ in favor of the dust fraction $\tepsilon$: 
\begin{align}
  \tepsilon \equiv \frac{\rhod}{\rho}  = \frac{\epsilon}{1+\epsilon},
\end{align}
where $\epsilon=\rhod/\rhog$ is the usual
dust-to-gas ratio. 
If $\tstop=0$ then the 
dust-to-gas ratio is conserved following the fluid. Otherwise, 
$\epsilon$ evolves due to particle drift in response to pressure
gradients. Note that Eq. \ref{dusteq} is equivalent to Eq. 46 of 
\cite{jacquet11}.

The total gravitational potential $\Phi_\mathrm{tot}=\Phi +
\psi$ includes that from a central star of mass $M_*$ and the disk's own
potential $\psi$. We adopt 
\begin{align}\label{thin_disk_potential}
  \Phi(r,z) =-\frac{GM_*}{\sqrt{r^2 + z^2}}\simeq
  -\frac{GM_*}{r}\left(1 - \frac{z^2}{2r^2}\right),  
\end{align}
where $(r,\phi, z)$ are cylindrical co-ordinates centered on the star, and 
$G$ is the gravitational constant. The second equality is the  
thin-disk approximation, appropriate for
$|z|\ll r$. We use this approximate form in order to
obtain explicit expressions for disk equilibria (\S\ref{steady_state}). 
The disk potential $\psi$ satisfies the Poisson equation
\begin{align}
  \nabla^2\psi = 4 \pi G \rho.\label{poisson}
\end{align}
We include $\psi$ for
completeness, but we will mostly neglect it unless stated
otherwise.

For the mixture's temperature evolution, Eq. \ref{tempeq}, 
$\gamma$ is 
the gas adiabatic index; $\mathcal{H}$ represents heating and $\mathcal{H}_\mathrm{eff}$ 
is an effective source term arising from transforming the gas energy equation (Eq. \ref{real_pressure}) 
from the 
two-fluid to one-fluid variables \citep[see][ for 
details]{laibe14}. We also include a simple model of radiative
cooling, 
\begin{align}
  \Lambda = \frac{T -
    T_\mathrm{ref}}{t_\mathrm{cool}}, \label{cooling} 
\end{align}
which relaxes the mixture back to a prescribed temperature profile   
$T_\mathrm{ref}$ on a timescale of $\tcool$. We will shortly simplify
the problem by considering rapid cooling, $\tcool\to0$. 

Eq. \ref{masseq}---\ref{tempeq} are not yet equivalent to standard 
hydrodynamics, which typically evolves two scalar fields: the density and
temperature (or pressure). By contrast, Eq. \ref{masseq}---\ref{tempeq}
involves three scalars: $\rho$, $\tepsilon$, and $T$. 
However, we can establish an  analogy by fixing the gas
equation of state, thus eliminating Eq. \ref{tempeq}; but then reform 
the dust-to-gas ratio evolution equation (\ref{dusteq}) into an
energy-like equation. 

\subsection{Locally isothermal equation of state}\label{loc_iso_eos}
We consider the limit of short cooling 
times, $\tcool\to 0$, appropriate for the outer parts of an irradiated
protoplanetary disk \citep{chiang97,lin15}. Then the disk temperature
$T = T_\mathrm{ref}$ at all times, and so we may
adopt a locally isothermal equation of state 
\begin{align}\label{eos}
  P = c_s^2(r,z)\rhog = c_s^{2}\left(1 - \tepsilon\right)\rho,   
\end{align}
where $c_s(r,z)= \sqrt{\mathcal{R}T_\mathrm{ref}/\mu}$ is a prescribed
sound-speed profile fixed in time. In most applications we consider vertically 
isothermal disks with \begin{align}\label{power_temp}
  c_s^2(r) \propto r^{q},
\end{align}
where $q$ is the power-law index for the disk temperature. For $q=0$
the disk is strictly isothermal.  

Notice Eq. \ref{eos} resembles an ideal gas equation of state but with  
a reduced temperature $\widetilde{T} = 
T_\mathrm{ref}\left(1-\tepsilon\right)$. %sound-speed $c_{s,\mathrm{eff}} = c_s\sqrt{\left(1 -
This reduced temperature decreases with dust-loading. 
%   \tepsilon\right)}$. 
%then the equation of state has the standard
%form $P=c_{s,\mathrm{eff}}^2\rho$. 
Since $\tepsilon$ typically decrease away from the midplane, we expect
vertically isothermal dusty disks to behave as if the temperature
\emph{increased} away from $z=0$.     

\subsection{Effective energy equation}\label{energy_analogy}
Although we have deleted the true energy
equation by fixing a locally isothermal
equation of state, we show that the mixture 
nevertheless obeys an effective energy evolution equation. This is
because advection of the dust-fraction, described by
Eq. \ref{dusteq}, can be transformed into an energy-like 
equation.  
%We show that for a prescribed temperature distribution, the mixture
%obeys an evolutionary energy equation, due to the advection of the
%dust-fraction. 
The equation of state, Eq. \ref{eos}, implies 
\begin{align*}
  \tepsilon = 1 - \frac{P}{c_s^2(r,z)\rho}.  
\end{align*}
Then Eq. \ref{dusteq} becomes
\begin{align}
%\frac{DP}{Dt} 
\frac{\p P}{\p t} + \bm{v}\cdot\nabla P  
&= - P \nabla\cdot\bm{v} + P\bm{v}\cdot\nabla\ln{c_s^2}
                + \mathcal{C},  \label{eff_energy} \\
\mathcal{C}&\equiv c_s^2 \nabla\cdot\left[\tstop\left(1 -
  \frac{P}{c_s^2\rho}\right)\nabla 
  P\right].
\end{align} 
For comparison with the standard energy equation in hydrodynamics,
re-writing Eq. \ref{tempeq} for a pure ideal gas with $P \propto
\rhog T$, and reverting back to gas velocities, gives  
%For later comparison it is convenient to re-write the temperature
%evolution, Eq. \ref{tempeq}, to that for the pressure: 
\begin{align}
  \frac{\p P}{\p t} + \bm{v}_\mathrm{g}\cdot\nabla P = - \gamma P
  \nabla\cdot \bm{v}_\mathrm{g}  +
  \frac{P}{T}\left(\mathcal{H} -
    \Lambda\right).\label{real_pressure} 
%\frac{\rho \mathcal{R}}{\mu}\frac{T
%  - T_\mathrm{ref}}{t_\mathrm{cool}}, \label{realenergy}
\end{align} 
We can thus interpret Eq. \ref{eff_energy} as the energy equation for
and ideal gas with adiabatic index $\gamma=1$, but now with the
imposed temperature gradient $\nabla c_s^2$ and dust-gas drag $\mathcal{C}$ 
acting as source terms. 

%playing the
%role of radiative cooling ($\Lambda$ in 
%Eq. \ref{tempeq}). 
%In this form, the function $\mathcal{C}$ can be interpreted
%as a cooling term. 
   
If we denote 
\begin{align}
  \bm{F} \equiv  - \frac{\nabla P}{\rho}
\end{align}
for the pressure forces, then
\begin{align*}
  \mathcal{C} = - c_s^2\nabla \left( \tepsilon \tstop \rho \bm{F}
  \right), 
\end{align*}
which is in the same form as cooling by radiative diffusion. In protoplanetary
disks the corresponding `heat flux', proportional to $\bm{F}$, is
directly radially outwards and vertically upwards. This is simply a
reflection of particle drift towards increasing pressure (inwards and 
downwards). Particle flux into a region contributes to `cooling' of
that region because the reduced temperature is lowered. 

% Eq. \ref{eff_energy} can be written in conservative form,
% \begin{align*}
%   \frac{\p P}{ \p t} + \nabla\cdot\left\{P\left[\bm{v} -
%       \tstop\left(c_s^2 - \frac{P}{\rho}\right)\nabla\ln{P}\right]
%     \right\}\\
%   = \left[P\bm{v} - \tstop\left(c_s^2 - \frac{P}{\rho}\right)\nabla
%     P\right]\cdot\nabla\ln{c_s^2}. 
% \end{align*} 
% We may thus re-interpret $P$ as the mixture's energy density, but the
% energy flux has an additional contribution from the pressure
% gradient and dust-gas friction. The term on the right-hand-side, owing
% to the imposed temperature profile, can be interpreted as an external
% heat source.  

%We comment that 

%An effective energy equation can also be derived for other 
%fixed gas equations of state, such as polytropic gas. We discuss this
%in \S\ref{gen_poly}.

\subsection{Entropy and buoyancy of isothermal dusty gas }\label{dusty_entropy}

The specific entropy of an ideal gas is 
$S~=~C_P\ln{\left(P^{1/\gamma}/\rhog\right)}$, where $C_P$ is the heat 
capacity at constant pressure. 

Since we have shown that a locally isothermal dusty gas effectively
has $\gamma=1$ we can define an effective 
entropy for the \emph{mixture} as 
\begin{align}
   \seff \equiv \ln \frac{P}{\rho} = \ln{\left[c_s^2(1-\tepsilon)\right]},  
\end{align} 
where the constant heat capacity has been absorbed into $\seff$. 
%This follows naturally by combining the effective 
%energy Eq. \ref{eff_energy} and Eq. \ref{masseq}. 
Then combining Eq. \ref{eff_energy} and \ref{masseq} gives  
\begin{align*}
  \frac{D \seff}{D t} = \bm{v}\cdot\nabla\ln{c_s^2} +
  \frac{c_s^2}{P}\nabla\cdot\left(\tstop\tepsilon\nabla P\right), 
\end{align*}
which is equivalent to entropy evolution in standard  
hydrodynamics, albeit with source terms. With an effective
entropy defined this way, many of the  
results concerning the stability of (locally) isothermal dusty gas
will have identical form and interpretations to that for pure gas.   

%If $c_s^2$ is constant and $\tstop=0$, then 
%$D\seff/ D t = D\tepsilon/D t=0$. In this case the effective 
%entropy is conserved because the dust-fraction is `frozen in' the
%flow. 

The physical reason for this analogy is that with strong drag
($\tstop\to0$), dust is almost perfectly entrained in the gas, but
there is some gain/loss of dust particles between different  
gas parcels. This is analogous to the entropy of an ideal 
pure gas subject to heating/cooling: entropy is conserved following
a gas parcel, except if there is heat exchange between a fluid parcel and the
surrounding. For strictly isothermal gas perfectly coupled to dust
(constant $c_s^2$ and $\tstop=0$) we have 
$D\seff/ D t = D\tepsilon/D t=0$. In this case the effective 
entropy is exactly conserved because the dust-fraction is `frozen in' the
flow.

%\subsection{Dusty buoyancy forces}
We can now define the vertical buoyancy frequency $N_z$ of the mixture
as    
%$N_z$ of the mixture is given by 
\begin{align}
  N_z^2 \equiv - \frac{1}{\rho}\frac{\p P}{\p z}\frac{\p \seff}{\p z}
  =c_s^2(r) \frac{\p \ln{\rhog}}{\p z}\frac{\p \tepsilon}{\p z},
%  \quad 
 % N_z^2 \equiv - \frac{1}{\rho}\frac{\p P}{\p z}\frac{\p \seff}{\p z}.
%&=
  %\frac{c_s^2(r)}{\left(1+\epsilon\right)^2}\frac{\p\ln\rhog}{\p 
%  z}\frac{\p\epsilon}{\p z} \\ &
  %                                =
 % \frac{\epsilon}{\left(1+\epsilon\right)^2}\left(\frac{z}{\Htilde}\right)^2\OmK^2\notag,  
\end{align}
%Typically in a thin, smooth disk we have $|N_r|\ll |N_z|$. 
where the second equality assumes a vertically isothermal disk. 
Protoplanetary disks have $\p_z\rhog<0$, thus stability against vertical 
convection ($N_z^2>0$) requires $\p_z\tepsilon <0 $, i.e. the dust density should drop faster 
than the gas density away from the midplane. 
This is equivalent to 
entropy increasing away from the midplane. A similar expression exists
for the radial buoyancy frequency $N_r$, but $|N_r|\ll |N_z|$ in thin,
smooth disks. 

A vertical buoyancy force exists even in vertically isothermal
dusty disks, because coupling gas to dust particles increases
the fluid's inertia, but pressure (i.e. restoring) forces are
unchanged. The increased weight of the fluid resists vertical
oscillations and hence there is an associated buoyancy force. However,
if $\tstop\neq0$ so the gas-dust coupling is imperfect, then gas is no
longer `weighed down' by the dust, and can slip past it. Thus finite
drag diminishes the dust-induced buoyancy. This is similar to reducing
gas buoyancy through cooling \citep{lin15}. 

%{\bf
\subsection{Physical disk conditions}
%physical disk conditions underwhich the above approx
%(small/well-coupled dust and isothermal gas) is applicable. smaller
%than mm at few tens of au. give dimensional numbers here
%}
The above correspondence between isothermal dusty gas and adiabatic
pure gas is derived under the terminal velocity approximation, which applies
to short stopping times, $\tstop\OmK\ll 1$; and the locally isothermal
approximation, which applies to short cooling times, $t_\mathrm{cool}\OmK\ll
1$. In typical protoplanetary disk
models such as the Minimum Mass Solar Nebula, $\tstop$ decreases for
smaller particles and/or towards smaller radii \citep{youdin11}. However,
$t_\mathrm{cool}$ is generally only small in the outer disk
\citep{lin15,malygin17}. Combining these results, we estimate the 
correspondence will be applicable to particles less than mm in size at a few
to 10s of AU in protoplanetary disks. However, note that the
isothermal approximation may be relaxed to other fixed equations of
state (see \S\ref{gen_poly}).

\section{General stability criteria for dusty gas}\label{limits}

In this section we discuss the stability properties of the 
dusty-gas mixture using a variational principle. This approach does not
require explicit solutions (i.e. a specific distribution of the density and dust-to-gas ratio)  
to the equilibrium equations. We consider axisymmetric systems and neglect self-gravity here.  

\subsection{Steady states}\label{eqm}
%We consider axisymmetric steady states as background equilibria to
%perturb. 

For a given distribution of the dust-fraction $\tepsilon$ (or
dust-to-gas ratio $\epsilon$), the 
mass and momentum Eqs. \ref{masseq}---\ref{momeq} admit     
solutions with $\rho(r,z)$ and 
$\bm{v}=r\Omega(r,z)\hat{\bm{\phi}}$ where $\Omega = v_\phi/r$, which satisfy 
\begin{align}
  r\Omega^2 &= \frac{\p \Phi}{\p r} + \frac{1}{\rho}\frac{\p P}{\p
    r},\label{steady_momr}\\
  0 & = \frac{\p\Phi}{\p z} + \frac{1}{\rho}\frac{\p P}{\p z},\label{steady_momz}
%  0 & = \nabla\cdot\left(\tepsilon\tstop\nabla P\right) \label{steady_dust}
\end{align}
with $P=P(\tepsilon,\rho)$ given by the equation of state
(Eq. \ref{eos}). An explicit solution is presented in 
\S\ref{steady_state}, when we analyze the stability of
protoplanetary disks. 

The mixture possess vertical shear. To see this, we eliminate $\Phi$
between Eq. \ref{steady_momr}---\ref{steady_momz} to 
obtain 
\begin{align}\label{vshear}
  r\frac{\p \Omega^2}{\p z} 
%&= \frac{\p\ln{\rho}}{\p r}\frac{\p}{\p
%    z}\left[c_s^2(1-\tepsilon)\right] - \frac{\p\ln{\rho}}{\p z}
%  \frac{\p}{\p r} \left[c_s^2(1-\tepsilon)\right]\\  
   &= \frac{1}{\rho}\left(\frac{\p P}{\p r}\frac{\p \seff}{\p z} -\frac{\p
    P}{\p z}\frac{\p \seff}{\p r} \right).% \\
%& = \frac{1}{\rhog}\left(\frac{\p P}{\p r}\frac{\p \seff}{\p z} -\frac{\p
%    P}{\p z}\frac{\p \seff}{\p r} \right)
\end{align}
and recall $\seff = \ln{\left[c_s^2\left(1-\tepsilon\right)\right]}$. It is well appreciated that 
vertical stratification of the dust layer contributes to vertical shear
\citep{chiang08}. Eq. \ref{vshear} shows that a radial dust
stratification ($\p_r\tepsilon$) also contributes to vertical shear.  

%Since $\seff$ depends on $\tepsilon$, we see that radial gradients in
%the dust-to-gas ratio also drives vertical shear \citep[cf. that 
%driven by $\p_z\epsilon$ usually discussed in dusty disks,  
%e.g.][]{chiang08}. 

Our equilibrium solutions satisfy the hydrostatic constraints of
Eq. \ref{steady_momr}---\ref{steady_momz} but are not in general
steady state solutions to the energy 
equation (\ref{eff_energy}), because the source term $\mathcal{C} \ne0$ for realistic disks.  
Limiting cases with $\mathcal{C} \equiv 0$ include \begin{inparaenum}[1)] 
\item 
  unstratified or 2D, razor-thin disk models with finite dust-gas drag; %(\S\ref{si});   
\item 
  perfectly-coupled dust with $\tstop=0$. %(\S\ref{results}). 
\end{inparaenum} 
The following analyses thus apply strictly to these limiting cases of
$\mathcal{C} = 0$ in equilibrium.  Our analysis is also a good approximation when 
$\tstop$ is sufficiently small such that the evolution of the background disk (e.g. dust-settling) 
occur on much longer timescales than the instability growth timescales.

%However, disk structures obtained from
%Eq. \ref{steady_momr}---\ref{steady_momz} generally do not satisfy  
%the steady state effective energy Eq. \ref{eff_energy} because typically 
%$\mathcal{C}\neq0$.  In order to have a well-defined stability problem, we formally
%consider equilibria with $\mathcal{C}\equiv 0$. Namely      
%\begin{inparaenum}[1)] 
%\item 
%  unstratified or 2D disks with finite dust-gas drag; %(\S\ref{si});   
%\item 
%  perfectly-coupled dust with $\tstop=0$. %(\S\ref{results}). 
%\end{inparaenum} 
%Alternatively, we are assuming that $\tstop$  is sufficiently small

 % N_z^2 
%\end{align}

%\section{Limiting behaviours}\label{limits}

\subsection{Integral relation}

%Our discussion is based on
%the corresponding axisymmetric linearized equations. 
We consider axisymmetric Eulerian perturbations to a variable $X$ of
the form  
\begin{align}
 \real\left[ \delta X(r,z)\exp{\left(-\ii\sigma t\right)}\right], 
\end{align}
where $\sigma$ is the complex mode frequency. We 
write 
\begin{align}
  \sigma = \ii s - \omega,
\end{align}
where $s$ and $\omega$ are the growth rate and real frequencies,
respectively. Then perturbations have time dependence $e^{st +
  \ii\omega t}$.  Thus for $\omega>0$, perturbations rotate
anti-clockwise in the complex plane. 

In Appendix \ref{var_prin} we linearize Eqs. \ref{masseq}, \ref{momeq}, and 
Eq. \ref{eff_energy} to derive the following integral relation: 
\begin{align}
%&  \sigma^2\int\rho\left(|\dd v_r|^2 + |\dd v_z|^2\right)dV \notag\\
  \sigma^2\mathcal{I}^2
&= \int\left[ \rho
  |\dd v_r|^2A + \rho  \left(\dd v_z \dd v_r^* + \dd v_z^*\dd v_r\right) B +
  \rho |\dd v_z|^2 D\phantom{\frac{1}{1}}\right. \notag\\
&\phantom{===}  \left. + \frac{1}{P}\Bigl\lvert \nabla\cdot\left(P\dd
  \bm{v}\right)\Bigr\rvert^2\right]dV  -\int \left(\nabla\cdot\dd\bm{v}^*\right)\dd\mathcal{C}dV \notag\\
&\phantom{===}
- \int P
  \left(\nabla\cdot\dd\bm{v}^*\right)\left(\dd\bm{v}\cdot\nabla\ln{c_s^2}\right)dV,\label{int_rel}
\end{align} 
%The real integral $\mathcal{I}^2>0$ 
where $^*$ denotes the complex conjugate, 
\begin{align}
  \mathcal{I}^2 \equiv \int\rho\left(|\dd v_r|^2 + |\dd v_z|^2\right)dV
\end{align}
is the meridional kinetic energy, 
and coefficients $A,B,D$ can be 
read off Eq. \ref{integral_ex}. %Note that $B=C$ from dynamical equilibrium
%(Eq. \ref{vshear}). 
We now consider various limits of Eq. \ref{int_rel}.

\subsection{Strictly isothermal gas perfectly coupled to dust}\label{iso_perfect}
When $c_s^2$ is a constant and $\tstop=0$ (so that $\mathcal{C} = 0$), the dusty-gas equations are
exactly equivalent to that for adiabatic hydrodynamics with unit
adiabatic 
index. Although the gas is strictly isothermal, the mixture behaves 
adiabatically because the dust fraction $\tepsilon$ is advected with 
the gas. This is similar to entropy conservation following an 
adiabatic gas without heating or cooling.  

In this case, the last two integrals in Eq. \ref{int_rel} vanish. 
Then the condition for axisymmtric \emph{stability}, $\sigma^2>0$, 
is met if the integrand coefficients satisfy $A+D>0$ and $AD-B^2>0$
\citep[][Section 11.6]{ogilvie16}, which becomes 
\begin{align}
  \kappa^2 + \frac{1}{\rhog}\nabla P \cdot\nabla\tepsilon &> 0,\label{dusty_solberg1}  \\
  -\frac{1}{\rhog}\frac{\p P}{\p z}\left(-\kappa^2\frac{\p\tepsilon}{\p
    z} + r\frac{\p\Omega^2}{\p z}\frac{\p \tepsilon}{\p r} \right) & > 0, \label{dusty_solberg2}
\end{align} 
%\begin{align}
%  \kappa^2 + c_s^2 \nabla\ln{\rhog}\cdot\nabla\tepsilon &> 0,\label{dusty_solberg1}  \\
%  -c_s^2\frac{\p\ln{\rhog}}{\p z}\left(-\kappa^2\frac{\p\tepsilon}{\p
%    z} + r\frac{\p\Omega^2}{\p z}\frac{\p \tepsilon}{\p r} \right) & > 0, \label{dusty_solberg2}
%\end{align} 
%\begin{align}
%  \kappa^2 - \frac{1}{\rho}\nabla P \cdot \nabla s &> 0, \label{dusty_solberg1}    \\
%  -\frac{1}{\rho}\frac{\p P}{\p z} \left(\kappa^2 \frac{\p s}{\p z} -
%  r\frac{\p\Omega^2}{\p z}\frac{\p s}{\p r}\right) &>0, \label{dusty_solberg2}
%\end{align} 
where %$s = \ln{P/\rho}$, 
$\kappa^2 \equiv r^{-3}\p_r\left(r^4\Omega^2\right)>0$ is the 
square of the epicylic frequency. Note that $\nabla P/\rhog =
c_s^2\nabla \ln{\rhog}$ for strictly isothermal gas considered here. 
 
Eq. \ref{dusty_solberg1}---\ref{dusty_solberg2} can in fact be obtained by
inserting $C_P\ln{\left(1 - 
    \tepsilon\right)}$ for the entropy into the standard
expression for the Solberg-Hoiland criteria 
\citep{tassoul78} for axisymmetric stability of adiabatic gas. 
This substitution is 
consistent with our definition of the effective entropy $\seff$ since
we are considering constant $c_s$. 

We expect Eq. \ref{dusty_solberg1} to be satisfied in typical
protoplanetary disks where the dust-to-gas ratio increases in the same
direction as the local pressure gradient, which is equivalent to the
gas density gradient for isothermal gas. 

On the other hand, Eq. \ref{dusty_solberg2} can be violated in
disks if the dust is vertically well-mixed but radially-stratified,  
such that $\p_z\tepsilon = 0$ but $\p_r\tepsilon\neq0$. In this case the
left-hand-side of Eq. \ref{dusty_solberg2} becomes
\begin{align}
-\left(\frac{1}{\rhog}\frac{\p P}{\p
    z}\right)^2\left(\frac{\p\tepsilon}{\p r}\right)^2<0.
\end{align}
%\begin{align}
%-c_s^4\left(\frac{\p\ln{\rhog}}{\p
%    z}\right)^2\left(\frac{\p\tepsilon}{\p r}\right)^2<0.
%\end{align}
Such an isothermal dusty disk is \emph{unstable} because there is no
effective vertical buoyancy to stabilize vertical motions ($N_z=0$),
which can tap into the free energy associated with  
vertical shear due to the radial gradient in the dust-fraction. 
This situation is identical to the pure gas, adiabatic simulations of \cite{nelson13}
where the gas entropy is vertically uniform but has a radial
gradient. The authors indeed find instability. We give a numerical
example in the dusty context in \S\ref{vert_mixed}.   

\subsection{Thermodynamics of dust-drag instabilities}\label{dust_work}
The physical interpretation of \S\ref{grow_osc} is here derived in
detail. Consider constant $c_s$ but $\tstop\neq0$ (so $\delta\mathcal{C}
\neq 0$). Eq. \ref{int_rel} indicates that $\sigma^2$ is generally complex, { unless
the second integral on the right-hand-side is real. Thus } 
we may have growing oscillations, or overstability, due to dust-gas
friction. This is seen by taking the imaginary part of
Eq. \ref{int_rel} 
\begin{align}
%  s = \frac{\imag\int \left(\nabla\cdot\dd\bm{v}^*\right)\dd\mathcal{C}dV}{2\omega\int\rho\left(|\dd
%    v_r|^2 + |\dd v_z|^2\right)dV}, \label{thermal_instability}
  s = \frac{\imag\int \left(\nabla\cdot\dd\bm{v}^*\right)\dd\mathcal{C}dV}{2\omega\mathcal{I}^2}, \label{thermal_instability}
\end{align} 
assuming $\omega\neq0$. %This represents overstability due to dust-gas friction. 
Since drag ($\delta \mathcal{C}$) appears as a source term in our
effective energy equation, the quantity 
$\imag\left[\left(\nabla\cdot\dd\bm{v}^*\right)\dd\mathcal{C}\right]$
represent correlations between compression/expansion and  
heating/cooling.  

It is well-known that such correlations may lead to pulsational
instabilities in stars \citep{cox67}. We thus interpret  
dust-drag overstabilities in a similar way, 
adapting from the treatment of stellar 
pulsations by \cite{cox67} and lecture notes by \cite{samadi15} and 
J. Christensen-Dalsgaard\footnote{\url{http://astro.phys.au.dk/$\sim$jcd/oscilnotes/print-chap-full.pdf}}.        
 
%and disks
%\cite{kato78}. 

\subsubsection{Work done by dusty gas} 

The physical interpretation of Eq. \ref{thermal_instability} is that 
work done by pressure forces in the dusty gas leads to growth { ($s>0$) or decay ($s<0$)} 
%{\bf work done by pressure forces in the dusty gas modifies
in oscillation amplitudes. 
To demonstrate this, we calculate the average
work done assuming periodic oscillations, and show that if the average
work done is positive, then the oscillation amplitude would actually grow. 

Consider oscillations in the dusty gas with period $T_p$. 
The average rate of work done is 
\begin{align}
  \mathcal{W} = \frac{1}{T_p}\int^{t+T_p}_{t}dt^\prime\int_M P
  \frac{D\upsilon}{Dt^\prime} dm \label{work_def} 
\end{align}
\citep[][see their Eq. 4.10 and related discussions]{cox67}, 
where $\upsilon=1/\rho$ is the specific volume of the mixture, $D/Dt$
is the Lagrangian derivative, and the 
spatial integral is taken over the total mass $M$ of the mixture and
$dm$ is a mass element. 

{ Here we consider Lagragian perturbations such that 
  \begin{align}
 P \to P + \real\left(\Delta P e^{-\ii\sigma t}\right)
 \end{align}
is the pressure following a fluid element of the mixture (and similarly for $\rho$). 
 The Lagrangian perturbation $\Delta$ of a variable $X$ is $\Delta X = \delta X + \bm{\xi}\cdot\nabla X$
 and $\bm{\xi}$ is the Lagrangian displacement, so $\xi_{x,z} =  \ii\dd
 v_{x,z}/\sigma$. Inserting the above pressure and density fields
 into Eq. \ref{work_def}, and noting that only products of
 perturbations contribute to $\mathcal{W}$ after time-averaging, we
 find for periodic oscillations (time dependence $e^{\ii\omega t}$ and
 real $\omega$) that:  
}
\begin{align}
  \mathcal{W} = - \frac{\omega}{2} \int \imag\left(\Delta P
  \frac{\Delta \rho^*}{\rho}\right)dV,\label{work_real}
\end{align}
Eq. \ref{work_real} show that a phase difference between gas pressure and
the total density leads to work done
($\mathcal{W}\neq0$).  

Now, Eq. \ref{lin_mass_full}---\ref{lin_energy_full} 
imply the integrand of the numerator in Eq. \ref{thermal_instability} 
is 
\begin{align} 
  \imag \left(\delta \mathcal{C}
  \nabla\cdot\dd \bm{v}^*\right) = 
  -|\sigma|^2\imag \left(\Delta P 
  \frac{\Delta\rho^*}{\rho}\right), \label{pdv}
\end{align}
for constant $c_s$. 
%This states that `PdV' work (the right-hand-side) is 
%provided by dust-drag friction, and can be seen explicitly as 
%follows. 
Then combining Eq. \ref{work_real}, \ref{pdv} and
\ref{thermal_instability} 
gives  
\begin{align}
s = \frac{\mathcal{W}}{\mathcal{I}^2}, \label{growth_work}
\end{align}
where we have set $\sigma= - \omega$ in 
Eq. \ref{pdv}. Eq. \ref{growth_work} states that if the average work
done is positive during an oscillation, $\mathcal{W}>0$, 
then its amplitude would actually grow ($s>0$).  %and not be periodic 

\begin{figure}
  \includegraphics[width=\linewidth,clip=true,trim=0cm 5cm 0cm 0cm]{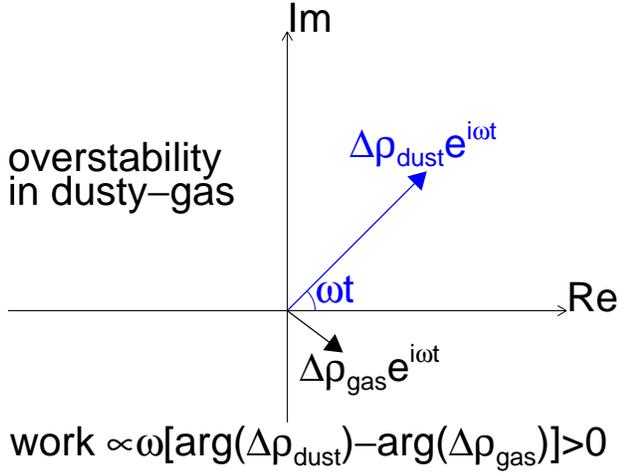}
  \caption{Phase relation for overstable modes in isothermal dusty
    gas. Such modes require (Lagrangian) oscillations in the gas
    pressure, which is directly proportional to the gas density, to
    lag behind that in dust density. The eigenvectors rotate anti-clockwise for $\omega>0$. 
    \label{lag_cartoon}
  }
\end{figure}

Positive work is done by a fluid parcel if $-\omega \imag\left(\Delta  
P\Delta\rho^*\right)>0$. Without loss of generality, take $\omega>0$
and consider a mass element with $\Delta\rho = 1$. Then positive work 
requires $\imag\left(\Delta
P\right)<0$.  This corresponds to Lagrangian pressure perturbations \emph{lagging}
behind that in density, see Fig. \ref{lag_cartoon}
for the case of dusty gas.   

%, since the phase of
%the assumed perturbations increase in time.  
%$\imag\left(\Delta\rho\right)=0$.   

\subsubsection{Physical property of dust-gas drag overstabilities}  

%We now use physical arguments to show that $\mathcal{W}>0$ if pressure
%perturbations (from the gas) lags behind the total density of the gas
%plus dust mixture. 
%The full energy equation is  
%\begin{align*}
%  \frac{DP}{Dt} = \frac{P}{\rho}\frac{D\rho}{D t} + \mathcal{C}. 
%\end{align*} 
%Dust-drag causes a phase difference between the pressure and density 
%evolution of a fluid element. 

%Consider a fluid element at maximum density during an
%oscillation cycle (point $A$), so that $D\rho/ D t = 0$. If dust-gas
%drag provides an effective heating to the fluid element at $A$
%(i.e. gas influx)  then it will experience increasing pressure, $DP/D
%t>0$. It attains pressure maximum (point $B$) \emph{after} reaching
%density maximum. 

%Applying the above argument to isothermal dusty gas, 

The above discussion applies to any single fluid with a pressure and
density. In the case of interest --- dusty gas --- the work done 
is attributed to finite dust-gas drag. 
%For dusty gas, the work done during oscillations 
 %is attributed to finite dust-gas drag. 
The relative 
drift between gas and dust{, which only exists if $\tstop\nabla P \neq 0$ (Eq. \ref{term_vel_approx}),} causes a phase difference between the two
components, and hence between pressure and total density. 

A parcel of the strictly isothermal dusty-gas mixture does 
positive work if  
\begin{align*}
-\sgn\left(\omega\right)\imag\left(\Delta\rhog\Delta\rhod^*\right)>0,
\end{align*}
meaning that gas follows dust (Fig. \ref{lag_cartoon}). Overstabilities 
are thus not possible if the gas does not respond to dust 
(i.e. no back-reaction).% {\bf can we say this means that any effective
%  dust-drag instability require high dust to gas ratios?} 
The pressure-density lag shown in
Fig. \ref{lag_cartoon} is achieved if, just after the total density of a 
parcel maximizes, its gas content
is increasing, which requires a sufficiently large
particle flux \emph{out} of the parcel. See $A$ to $B$ in 
Fig. \ref{pdv_cartoon}. 

%{\bf lagrangian derivative follows the mixture.}
% this highlights one
%  perk of the one-fluid framework. in two-fluid models there are two
%  types of lagragian deriv, not clear which fluid to follow.}

This thermodynamic interpretation does not explain 
why drag forces causes gas pressure to lag behind the dust
density, but shows 
that this \emph{must} be the case for any growing oscillations associated the
dust-gas drag. To rigorously understand how dust-drag causes this 
lag requires an explicit solution to the linearized equations with 
detailed treatment of the function $\mathcal{C}$. However, given the complexity of
$\mathcal{C}$ (see Appendix \ref{lin_dust}), we might generally expect
 dusty disks to  support a range of stable and overstable modes, with the
latter being associated with pressure-density lag. 
%Indeed, \cite{jacquet11} explains the essence of the streaming 
%instability in dusty protoplanetary disks as dust accumulation
%(and hence compression) at a pressure bump, which then drags the 
%gas towards it (and hence heating) to strengthen the pressure
%bump. That is, heating occurs upon compression, as in stellar  
%pulsational instabilities \citep{cox67}.
 In \S\ref{si} we check that the
streaming instability fits into this thermodynamic interpretation in
the strong drag limit.

%A phase difference was also noted in the 
%numerical calculations of the streaming instability by 
%\citet{youdin07b}.    

%Whether the phase
%difference is positive or negative depends on the cooling function
%$\mathcal{C}$, but we may expect a system to generally have both
%modes, and that with a phase lag are unstable. 

\subsection{Locally isothermal gas perfectly coupled to dust}\label{dusty_vsi_int}
If $c_s(r,z)$ is non-uniform but $\tstop=0$, Eq. \ref{int_rel} 
gives  
\begin{align}
%  s = \frac{\imag\int P
%  \left(\nabla\cdot\dd\bm{v}^*\right)\left(\dd\bm{v}\cdot\nabla\ln{c_s^2}\right)dV}{2\omega\int\rho\left(|\dd 
%    v_r|^2 + |\dd v_z|^2\right)dV}. \label{vsi_check} 
  s = \frac{\imag\int P
    \left(\nabla\cdot\dd\bm{v}^*\right)\left(\dd\bm{v}\cdot\nabla\ln{c_s^2}\right)dV}{2\omega\mathcal{I}^2}, \label{vsi_check} 
\end{align} 
again assuming $\omega\neq0$. 
This instability represents VSI caused by vertical shear arising from a radial
temperature gradient \citep{nelson13,barker15,lin15}. We present 
numerical solutions of the VSI with perfectly-coupled dust in \S\ref{results}.

\section{Application to protoplanetary disks}\label{linear_problem}
We now examine instabilities in dusty protoplanetary disks
based on explicit descriptions of the equilibrium state. We first
specify the disk structures to be analyzed. We show  
that sharp radial edges in the dust-to-gas ratio can render the disk unstable. We then 
revisit two well-known instabilities in dusty gas using the
hydrodynamic framework developed thus far, namely secular gravitational (SGI) 
instabilities and the streaming instability (SI).  

\subsection{Disk structure with a prescribed dust distribution}\label{steady_state}  
We assume a Gaussian profile in the
dust-to-gas ratio,    
\begin{align}\label{dust_gauss}
  \epsilon(r,z) = \epsilon_0(r)
  \exp{\left[-\frac{z^2}{2\Htilde^2(r)}\right]}. 
\end{align}
%For simplicity we assume the dust-to-gas ratio at the mid-plane
%$\epsilon_0$, as well as its characteristic scale-height
%$\widetilde{H}$, are both constant.
% We take the mid-plane dust-to-gas ratio
%to be a power-law in radius,  
%\begin{align}
%  \epsilon_0(r) = \epsilon_{00}\left(\frac{r}{r_0}\right)^{-d},  
%\end{align}
%where $\epsilon_{00}$ is the dust-to-gas ratio at the fiducial radius
%$r_0$.  
Inserting Eq. \ref{dust_gauss} into vertical hydrostatic equilibrium,
Eq. \ref{steady_momz} and integrating with the approximate
gravitational potential (Eq. \ref{thin_disk_potential}) we obtain the
gas density as
\begin{align}
  &\rhog(r,z)= \notag\\
&\rho_\mathrm{g0}(r)\exp{\left\{ - \frac{z^2}{2\Hgas^2}
    -\epsilon_0\frac{\Htilde^2}{\Hgas^2}\left[1 -
      \exp{\left(-\frac{z^2}{2\Htilde^2}\right)}\right] \right\}}, 
\end{align}
where
\begin{align}
  \Hgas = \frac{c_s}{\OmK}, \quad \OmK \equiv \sqrt{\frac{GM_*}{r^3}},   
\end{align}
is the gas scale-height in the dust-free limit and $\OmK$ is the
Keplerian frequency, respectively. 

In gas-dominated disks with $\epsilon_0 < 1$ the gas distribution
$\rhog(r,z)$ is close to Gaussian, as in the   
dust-free case, and the dust density is approximately 
\begin{align}
  \rhod \simeq \epsilon_0\rho_\mathrm{g0}(r) \exp
        {\left(-\frac{z^2}{2H_\mathrm{d}^2}\right)}, 
\end{align}
with 
\begin{align}
  \frac{1}{H_\mathrm{d}^2} = \frac{1}{\Htilde^2} + \frac{1}{\Hgas^2}, 
\end{align}
and $H_\mathrm{d}$ is the dust-scale height.% In numerical
%calculations, we  we specify $H_\mathrm{d}< \Hgas$ to obtain 
%$\Htilde$ for input. 

Finally, we define 
\begin{align}
  Z \equiv \epsilon_0\frac{\Hd}{\Hg} \simeq
  \frac{\Sigma_\mathrm{d}}{\Sigma_\mathrm{g}} 
\end{align}
as a measure of the local metalicity, where $\Sigma_\mathrm{d}$ and
$\Sigma_\mathrm{g}$ are the dust and gas surface densities,
respectively. The second equality holds for $\epsilon_0\ll1$.  

%\Sigma_\mathrm{d}$ and the gas
%surface density $\Sigma_\mathrm{g}$.  

\subsubsection{Orbital frequency} 
From Eq. \ref{steady_momr} the disk orbital frequency is 
\begin{align}
  \Omega(r,z) = \OmK(r)\left[1 - \frac{3}{2}\frac{z^2}{r^2} +
    \frac{h_\mathrm{g}^2}{\left(1+\epsilon\right)}\frac{\p}{\p\ln{r}}\ln{\left(c_s^2\rhog\right)}
    \right]^{1/2}, 
\end{align}
where 
\begin{align}
  h_\mathrm{g} \equiv \frac{\Hgas}{r}
\end{align}
is the characteristic disk aspect-ratio, with $h_\mathrm{g}\ll 1$ for protoplanetary disks. 

\subsubsection{Vertical shear}\label{vertshear}
Writing Eq. \ref{vshear} in terms of the gas density and dust-to-gas
ratio with a power-law temperature profile (Eq. \ref{power_temp})
gives the disk's explicit vertical shear profile:
\begin{align}\label{vshear2}
  &r\frac{\p \Omega^2}{\p z}  =
  \frac{c_s^2(r)}{\left(1+\epsilon\right)^2}\left\{
  \frac{\p\epsilon}{\p r}\frac{\p\ln{\rhog}}{\p z}
%  -\frac{\p\epsilon}{\p z}\left[\frac{\p\ln{\rhog}}{\p r} + \frac{q}{r}\right]\right.\notag\\
 -\frac{\p\epsilon}{\p z}\frac{\p\ln{P}}{\p r}\right.\notag\\
  &\phantom{ r\frac{\p \Omega^2}{\p z}  =
    \frac{c_s^2(r)}{\left(1+\epsilon\right)^2}\left\{\right\} }
  \left. -\frac{q}{r} \left(1+\epsilon\right)\frac{\p\ln{\rhog}}{\p z}
  \right\}. 
\end{align}
The first two terms correspond to vertical shear caused by spatial
variations in the dust-to-gas ratio. The third term
proportional to $q$ corresponds to vertical shear due to the 
radial temperature gradient, which survives in the dust-free limit. 

We can compare these sources by 
evaluating them using the equilibrium
solutions in \S\ref{steady_state}. We assume the disk is radially
smooth so that $\p_r\sim 1/r$. % and the dust-to-gas ratio
% $\epsilon\ll1$. 
This gives 

\begin{align}\label{vshear_split}
  \frac{\left|r\p_z\Omega\right|_{\text{
        dust/gas gradient}}}{\left|r\p_z\Omega\right|_{\text{
        temp. gradient}}} \sim
 % \epsilon \frac{\mathrm{max}\left(\delta^2,
  %  1\right)}{\left|q\right|\left(\epsilon + \delta^2\right)},
 \epsilon \frac{\mathrm{max}\left(\delta^2,
    1\right)}{\left|q\right|\left(1+\epsilon\right)\delta^2},
\end{align}
where $\delta\equiv \Htilde/\Hgas$. 
Since $|q|=O(1)$ in PPDs, Eq. \ref{vshear_split} indicates that
vertical shear due to variations in the dust-to-gas ratio dominates 
over that due to the radial temperature gradient for thin dust layers
such that $\delta^2\ll \epsilon$. %Otherwise, vertical shear is
%associated with the radial temperature gradient. %However, axisymmtric
%\emph{instability} always caused by $\p_rc_s^2$.  

\subsubsection{Dusty vertical buoyancy}\label{vbuoyancy}
For the above equilibrium the vertical buoyancy frequency is given
explicitly as 
\begin{align}
  N_z^2 =
  %\frac{c_s^2(r)}{\left(1+\epsilon\right)^2}\frac{\p\ln\rhog}{\p 
%  z}\frac{\p\epsilon}{\p z} \\ &
  %                                =
 \frac{\epsilon}{\left(1+\epsilon\right)^2}\left(\frac{z}{\Htilde}\right)^2\OmK^2\notag,  
\end{align}
where we have used $c_s=c_s(r)$. Then 
\begin{align*}
N_z\lesssim
O\left(\sqrt{\epsilon_0}\OmK\right). 
\end{align*}
However, for well-mixed dust layers such that $H_\epsilon \gg \Hg$, 
$\mathrm{max}\left(N_z\right)$ may occur outside a finite vertical domain.  

\subsection{Instability of dusty edges} 
Here we apply the dusty analog of the Solberg-Hoiland criteria derived
in \S\ref{iso_perfect} assuming strictly isothermal gas. As discussed
there, the first criterion is generally satisfied. Thus we only
consider the second condition, Eq. \ref{dusty_solberg2}.  

We assume the disk is approximately Keplerian so that
$\kappa\simeq\OmK$, and that the vertical gas distribution is   
Gaussian. In terms of the dust-to-gas
ratio $\epsilon$, Eq. \ref{dusty_solberg2}
becomes 
\begin{align}
  % &1 + \frac{\epsilon}{\left(1+
  %   \epsilon\right)^2}\left(h_\mathrm{g}^2\frac{\p\ln{\rhog}}{\p\ln{r}}
  % \frac{\p\ln{\epsilon}}{\p\ln{r}} + 
  % \frac{z^2}{\Htilde^2}\right)>0 \label{stability_est1},   \\ 
% &1 - \frac{\epsilon
%   h_\mathrm{g}^2}{\left(1+\epsilon\right)^2}
%   \frac{\p\ln{\epsilon}}{\p\ln{r}}\left(-\frac{\p\ln{\rhog}}{\p\ln{r}}+\frac{\Htilde^2}{H_\mathrm{g}^2}\frac{\p\ln{\epsilon}}{\p\ln{r}}\right)
%   > 0 \label{stability_est2},
&1 - \frac{\epsilon
  h_\mathrm{g}^2}{\left(1+\epsilon\right)^2}
  \frac{\p\ln{\epsilon}}{\p\ln{r}}\left(-\frac{\p\ln{P}}{\p\ln{r}}+\frac{\Htilde^2}{H_\mathrm{g}^2}\frac{\p\ln{\epsilon}}{\p\ln{r}}\right)
  > 0 \label{stability_est2},
\end{align}
for \emph{stability}. The { first term in  brackets} is usually stabilizing in 
PPDs for reasons given in \S\ref{iso_perfect}. The { second term in
brackets} is 
always destabilizing, but is small in smooth, thin disks with radial
gradients $O(1/r)$, $h_\mathrm{g}\ll 1$, 
provided that $H_\epsilon/\Hg$ is not large (e.g. some dust settling
has occurred). This means that typical dusty PPDs are stable to
axisymmetric perturbations, even for arbitrarily thin
dust layers. 

%%%%%%%%%%%%%%%%%%

To violate Eq. \ref{stability_est2} and obtain instability, notice the 
left hand side is a quadratic in $\p_r\epsilon$. Thus instability is possible for
sufficiently large (in magnitude) radial gradients  in the dust-to-gas
ratio,  
\begin{align}
  \frac{\p\ln{\epsilon}}{\p\ln{r}} > \chi_+ \quad \mathrm{or} \quad 
  \frac{\p\ln{\epsilon}}{\p\ln{r}} < \chi_-,
\end{align}
for \emph{instability}, where
\begin{align}\label{spm}
% \chi_\pm = \frac{1}{2}\frac{H_\mathrm{g}^2}{\Htilde^2} 
%   \left[
%   \frac{\p\ln{\rhog}}{\p\ln{r}} \pm 
%   \sqrt{
%   \left(\frac{\p\ln{\rhog}}{\p\ln{r}}\right)^2 + 
%   4 \frac{\Htilde^2}{H_\mathrm{g}^2}
%   \frac{\left(1+\epsilon\right)^2}{\epsilon h_\mathrm{g}^2}
%   }
%   \,\right]. 
\chi_\pm = \frac{1}{2}\frac{H_\mathrm{g}^2}{\Htilde^2} 
  \left[
  \frac{\p\ln{P}}{\p\ln{r}} \pm 
  \sqrt{
  \left(\frac{\p\ln{P}}{\p\ln{r}}\right)^2 + 
  4 \frac{\Htilde^2}{H_\mathrm{g}^2}
  \frac{\left(1+\epsilon\right)^2}{\epsilon h_\mathrm{g}^2}
  }
  \,\right]. 
\end{align} 
In typical accretion disks where $\p_rP<0$, instability is easier
for increasing dust-to-gas ratios ($\p_r\epsilon > 0$).  

We can neglect pressure gradients in Eq. \ref{spm} 
if $r\p_r\ln{P}\sim O(1)$ and $H_\epsilon/\Hgas\gg
\sqrt{\epsilon}\hgas/2\left(1+\epsilon\right)$. For example, if $\epsilon\simeq 0.01$ and
$\hgas\simeq 0.05$, then we require $H_\epsilon/\Hg\gg
2\times10^{-3}$. This can be met if the dust is not well settled { (e.g. due to a small amount of turbulence).}  
%this turbulence cannot be too big because the analysis itself does not have diffusion/turbulence 
%large diffusion is expected to stabilize 
Then instability requires 
\begin{align}\label{dust_edge}
\left|\frac{\p\ln{\epsilon}}{\p r}\right| \gtrsim
  \frac{1}{\Htilde}\frac{\left(1+\epsilon\right)}{\sqrt{\epsilon}}.
%\simeq
 % \frac{1}{\sqrt{\epsilon}H_\epsilon}, 
\end{align}
%where the second equality applies to $\epsilon\ll 1$. 
That is, if the radial lengthscale of the dust-to-gas ratio 
is much less its vertical lengthscale, $L_\epsilon\ll O(\Htilde)$, 
then the system is  unstable.  
Taking $\epsilon\sim 0.01$,  we find that 
for thin dust layers with $H_\epsilon \simeq \Hd\ll \Hg$,  instability
requires $L_\epsilon\ll O(0.1\Hg)$, i.e. the dust-to-gas ratio must 
vary on an extremely short radial lengthscale. This might be achieved, for
example, at sharp edges associated with gaps opened by giant planets. 

Technically, the above discussion is only applicable when $c_s$ is
constant and $\tstop=0$ (see \S\ref{iso_perfect}). However, since
dusty edges translate to sharp entropy gradients (
\S\ref{dusty_entropy}), we may generally expect sharp features in the
dust distribution of protoplanetary disks to be unstable.

%{\bf mention this is equivalent to sharp entropy gradients?}

\subsection{Radially local problem}
We now specialize further and compute explicit solutions to the linear
problem. We consider radially-localized axisymmetric disturbances of the form  
\begin{align}
  \delta X (r, z) = \delta X_1(r,z)\exp{(\ii k_x r)},
\end{align} 
%and similarly for $\dd P$ and $\dd\bm{v}$. 
where $k_x$ is a real wavenumber such that $|k_xr|\gg 1$, and the 
amplitude $\dd X_1(r,z)$ is 
a slowly-varying function of $r$. Then 
$\p_r\to i k_x$ when acting on the above primitive perturbations, and we may
neglect curvature terms. We take  
$k_x>0$ without loss of generality. Hereafter, we drop the subscript 1
on the amplitudes. 
%The frequency $\sigma = \omega +
%\ii s$ is generally complex, with $\omega$ being the real frequency
%and $s$ is the real growth rate. 

Introducing 
\begin{align}
  W \equiv \frac{\dd\rho}{\rho}, \quad Q \equiv \frac{\dd P}{\rho},
\end{align}
the linearized equations for 
vertically isothermal dusty gas with the pressure
equation in place of the dust-fraction
(Eq. \ref{masseq}, \ref{momeq}, \ref{poisson}, \ref{eff_energy}) are then:    

\begin{align}
  \ii\sigma W &= \ii k_x \dd v_r + \dd v_z^\prime +
  \dd v_r \p_r\ln{\rho} + \dd v_z\p_z\ln{\rho},\label{lin_mass}\\
  -\ii\sigma\dd v_r  &= 2\Omega\dd v_\phi 
% +  \delta\bm{F}\cdot\hat{\bm{r}}
- W F_r - \ii k_x Q - \ii k_x\dd\psi,\label{lin_xmom}\\
  \ii\sigma\dd v_\phi &= \frac{\kappa^2}{2\Omega}\dd v_r + \frac{\p
    v_\phi}{\p z}\dd v_z, \label{lin_ymom}\\
  -\ii\sigma\dd v_z &= - W F_z - \left[Q^\prime + Q
    \left(\ln{\rho}\right)^\prime\right] - \dd\psi^\prime  %\delta\bm{F}\cdot\hat{\bm{z}} 
,\label{lin_zmom}\\
  \ii\sigma Q &= \frac{P}{\rho}\left(\ii k_x \dd v_r + \dd
               v_z^\prime\right) + \frac{1}{\rho}\left(\dd v_r\p_rP + \dd v_z \p_zP\right)\notag\\
                &\phantom{=}-\frac{P}{\rho} \dd v_r\p_r
               \ln{c_s^2} %, \label{lin_energy} 
%+ \dd v_z \p_z\ln{c_s^2}\right),\label{lin_energy} 
               - \frac{\dd\mathcal{C}}{\rho},\label{lin_energy}\\
\dd\psi^{\prime\prime}  &= 4\pi G \rho W + k_x^2\dd\psi, \label{lin_sg}
\end{align}  
where $^\prime \equiv \p_z$ and recall $\bm{F} \equiv -\nabla
P/\rho$. We have temporarily restored self-gravity  to
discuss SGI in the next
section. The linearized dust-diffusion function 
$\dd\mathcal{C}$ is given in  Appendix \ref{lin_dust}. 
% and $\dd\bm{F}$ is the linearized pressure
%force, given in Appendix \ref{lin_press}. 
%Note that we have assumed a temperature profile that only depends on $r$.   

%; and 
%\begin{align}
%  \delta \bm{F} \equiv \frac{\dd\rho}{\rho^2}\nabla P -
%  \frac{1}{\rho}\nabla\dd P, 
%\end{align}
%$\dd\mathcal{C}$ is the linearized dust-duffusion function, given in
%Appendix \ref{lin_dust}. We consider stopping times appropriate for
%small grains in the Epstein regime. Note that for the axisymmetric
%problem, $\dd\bm{F}$ is purely meridional. 

Eq. \ref{lin_mass}---\ref{lin_sg} is a set of ordinary
differential equations in $z$. All coefficients and amplitudes are
evaluated at a fiducial radius $r=r_0$, but their full $z$-dependence
is retained. We next discuss solutions to these equations.  We first
show that the above equations yield the SGI and SI in the strong
drag regime in \S\ref{sgi} and \S\ref{si}, respectively. We then
consider 3D, stratified disks in \S\ref{results} to study how the addition of  dust 
affects the vertical shear instability. 

%for a
%razor-thin, self-gravitating disk in \S\ref{sgi}; and for 3D
%non-self-gravitating, unstratified and stratified disks 
%in \S\ref{si} and \S\ref{results}, respectively.   
 
 % Numerical solutions are generally required for the 
%vertically-stratified problem. 

%\section{Connection to previous results}

\subsubsection{Secular gravitational instability}\label{sgi}
Consider a razor-thin, self-gravitating disk so that $\rho =
\Sigma\delta(z)$, where $\delta$ is the Dirac delta function and $\Sigma$ is
the total surface density. Here $\epsilon = \Sigma_\mathrm{d}/\Sigma_\mathrm{g}$ and 
$\tepsilon=\Sigma_\mathrm{d}/\Sigma$, where $\Sigma_\mathrm{d,g}$ are the dust and gas surface densities. 
The background disk is 
uniform and we neglect the vertical dimension   
in Eq. \ref{lin_mass}---\ref{lin_energy}. The linearized dust-gas drag
term is then $-\delta\mathcal{C}/\rho = \tstop\tepsilon c_s^2k_x^2
Q$. The thin-disk solution to Eq. \ref{lin_sg} is $\dd\psi(z=0) = -2\pi G
\Sigma W/\left|k_x\right|$.%, where $\Sigma$ is the total surface
%density. 

These simplifications yield the dispersion relation
\begin{align*}
  \left(\ii\sigma - \tstop\tepsilon c_s^2k_x^2\right)\left( 2 \pi G
    \Sigma \left|k_x\right|  - \kappa^2 + \sigma^2 \right) = \ii
  \sigma c_s^2 k_x^2\left(1 - \tepsilon\right). 
\end{align*}
Searching for slowly and purely growing modes, $\sigma = \ii s$ with 
$|s|\ll \kappa$, we find 
\begin{align}  
s = \frac{\tstop\tepsilon c_s^2 k_x^2 \left( 2 \pi \Sigma G
    \left|k_x\right| - \kappa^2\right)}{\kappa^2 -2 \pi \Sigma G
    \left|k_x\right| + c_s^2k_x^2\left(1-\tepsilon\right) }. \label{sgi_disp}
\end{align}
This is secular gravitational instability mediated by strong
dust-gas drag with negligible turbulent dust diffusion 
\citep[][ their Eq. 13 becomes our Eq. \ref{sgi_disp} in this limit
after 
a change of variables]{takahashi14}. A similar effect occurs in viscous
self-gravitating gas disks \citep{gammie96,lin16}. In fact, if we
identify $\nu \equiv \tstop \tepsilon c_s^2$ as a kinematic viscosity,
then Eq. \ref{sgi_disp} is identical to \citeauthor{gammie96}'s Eq. 18. 

This exercise shows that the one-fluid framework, further simplified by
the terminal velocity approximation, is sufficient to capture SGI
in the strong drag limit. 

\subsubsection{Streaming instability}\label{si}
We now consider 3D disks without self-gravity. We neglect the vertical     
component of the stellar gravity, appropriate for studying regions
near the disk midplane. This  
allows us to Fourier analyze in $z$ to obtain 
an algebraic dispersion relation of the form  
$\sum_{j=0}^{5}c_j(k_x,k_z)\sigma^j = 0$, where $k_z$ is a real
vertical wavenumber. The coefficients $c_j$ can be read 
off Eq. \ref{streaming_dispersion} in  Appendix \ref{compressible_streaming}. 
There we also show that this dispersion relation reduces to that for
the streaming instability (SI) in the limit of incompressible gas and small
$\tstop$ \citep{youdin05a,jacquet11}.   
 
%\subsection{Numerical examples}

%{\bf note: kappa2 accounts for dust effect assuming smallh=0.05}

In Table \ref{si_compare} we solve the full dispersion relation
(Eq. \ref{streaming_dispersion}) numerically for selected cases where
analytic SI growth rates have been verified with particle-gas
numerical simulations \citep[namely][]{youdin07b,bai10b}.  
Following previous works on SI, we use normalized 
wavenumbers $K_{x,z} = \eta r k_{x,z}$ where
\begin{align} 
  \eta \equiv -\frac{1}{2\rhog r\OmK^2}\frac{\p P}{\p r} = 
  \frac{1}{2\left(1-\tepsilon\right)}\frac{F_r}{r\OmK^2}, 
\end{align} 
measures the pressure offset of Keplerian rotation. We fix 
$\eta=0.05c_s/r\OmK$. In this section section we also quote the
particle stopping time $\tau_\mathrm{s}=\tstop/(1-\tepsilon)$.    

The eigenfrequencies obtained from the one-fluid dispersion relation 
are compared that from a full, two-fluid analysis \citep[similar 
to][]{youdin05a, kowalik13}. As expected eigenfrequencies agree better
with decreasing $\taus$ since in that limit the mixture behaves more 
like a single fluid. Most importantly, we find the work done
$\mathcal{W}>0$ in all cases, and hence find growing oscillations. 

%{\bf note: important to use largrangian pressure pert properly}

% We also checked the growth rates $s$ satisfy 
% \begin{align} 
%   s = \frac{\left|\sigma\right|^2\imag\left(\Delta P
%     \Delta\rho^*/\rho\right)}{2\real\left(\sigma\right)\rho\left(\left|\dd
%   v_x\right|^2+\left|\dd
%   v_z\right|^2\right)}, \label{si_check}
% \end{align}
% as implied by Eq. \ref{thermal_instability} and
% Eq. \ref{pdv}. %, or directly from
% %Eq. \ref{streaming_mass}---\ref{streaming_vz}. 
% Interestingly, we find in dust-rich disks with 
% $\epsilon>1$, Eq. \ref{si_check} can be satisified with $\Delta
% P\simeq \ii\dd v_x \p_rP/\sigma$, i.e. the radial pressure gradient
% is responsible for growth. Conversely, for `linB' with $\epsilon
% < 1$, one can approximate $\Delta P \simeq \dd P$ in Eq. \ref{si_check}.
% This
%suggests that for SI in gas-dominated disks, the distinction between
%Eulerian and Lagragian pressure perturbations is unimportant. 
%{\bf but weak growth in this case}
%However
%in that case the gro 

In Table \ref{si_compare} we also calculate the phase difference
between the Lagrangian pressure perturbation and density perturbations
as    
\begin{align*} 
\varphi&\equiv \sgn{\left[\real{(\sigma)}\right]}\arg\left(\Delta
  P\Delta\rho^*\right).\notag 
%\notag\\
%       &= \sgn{\left[\real{(\sigma)}\right]}\arg\left(\Delta\rhog^*\Delta\rhod\right).\notag
\end{align*}
(Note that it is important to include 
the global radial pressure gradient in $\Delta P = \delta P + \ii \dd v_x \p_rP/\sigma$.) Then 
$\varphi > 0 $ indicates gas pressure lagging behind total density, 
which is true for all the cases. Thus SI is indeed associated with
such a phase lag.

%"generalized stokes number" would include other problem parameters,
%e.g. wavenumber  

%{\bf suppose large scale pressure grad drops outwards. 
%pert the sys by kicking more dust inwards. gas moves out, into region 
%  of lower pressure in the bg. i.e. create pressure bump at larger
%  radius. it attracts the first dust particles back out, plus some more
%  because bump is larger. (move gas from higher density to lower
%  density produce asymmetric bump/trough.}

% \begin{deluxetable*}{llrrrrrr}
%   \tablecolumns{8}
%   \tablecaption{Selected eigenfrequencies for the streaming
%     instability. \label{si_compare}
%   }
%   \tablehead{
%     \colhead{Mode} & 
%     \colhead{$\tau_\mathrm{s}\OmK$} &
%     \colhead{$\epsilon$} &
%      \colhead{$K_{x,z}$} &
%       \colhead{$\sigma/\OmK$ (two-fluid)} &
%     \colhead{$\sigma/\OmK$ (one-fluid)} &
%     \colhead{$\mathcal{W}$ (arbitrary units)} &
%       \colhead{$\Delta P$ lag}
%   }
% \startdata
%  linA, \cite{youdin07b} &  $0.1$       & 3.0 & 30    & $0.3480 +
%  0.4190\ii$ & $0.3640 + 0.4249\ii$ & $0.90$ & $30\degr$\\ 

% linB, \cite{youdin07b} & $0.1$        &  0.2 & 6 & $-0.4999 +
% 0.0155\ii$&   $-0.4981 + 0.0054\ii$  & $1.54$ & $1.2\degr$ \\

% linC,  \cite{bai10b}  & $10^{-2}$   &  2.0 & 1500&   $0.1049 +
% 0.5981\ii$   &  $0.1338 + 0.6650\ii$  & $0.15$& $11\degr$ \\

% linD, \cite{bai10b} &  $10^{-3}$   &  2.0 & 2000 & $0.3225 + 
% 0.3154\ii$& $0.3219 + 0.3154\ii$ &  $1.28$ & $22\degr$ 
% \enddata
% \end{deluxetable*}

\begin{deluxetable*}{lrrrrrr}
  \tablecolumns{8}
  \tablecaption{Selected modes of the streaming
    instability. \label{si_compare}
  }
%  \tablehead{
%    \colhead{Mode} & 
%    \colhead{$\tau_\mathrm{s}\OmK$} &
%    \colhead{$\epsilon$} &
%     \colhead{$K_{x,z}$} &
%    \colhead{Complex frequency} &
%    \colhead{$\sigma/\OmK$} &
%    \colhead{Work done} &
%    \colhead{$\mathcal{W}/\left|\Delta P\Delta\rho^*/\rho\right|$} &
%    \colhead{Pressure-density lag} &
%    \colhead{$\varphi$} 
%  }
\startdata
\hline\hline
\multicolumn{1}{c}{Mode ($\tau_\mathrm{s}\OmK$, $K_{x,z}, \rhog/\rhod$)} &
\multicolumn{2}{c}{Complex frequency, $\sigma/\OmK$} &
\multicolumn{2}{c}{Work done, $\mathcal{W}/\left|\Delta P\Delta\rho^*/\rho\right|$} &
\multicolumn{2}{c}{Pressure-density lag, $\varphi$} \\
\hline
\multicolumn{1}{c}{} &
 \multicolumn{1}{c}{two-fluid} &
 \multicolumn{1}{c}{one-fluid} &
 \multicolumn{1}{c}{two-fluid} &
 \multicolumn{1}{r}{one-fluid} &
 \multicolumn{1}{c}{two-fluid} &
 \multicolumn{1}{r}{one-fluid} \\
 \hline
% \multicolumn{6}{c}{\textcolor{blue}{linA\tablenotemark{$\dagger$}: $\tau_\mathrm{s}\OmK=0.1$, $K_{x,z}=30$, $\rhog/\rhod = 3$}}\\
% \hline
linA\tablenotemark{$\dagger$} ($0.1, 30, 3$) &
$0.3480 + 0.4190\ii$ & $0.3640 + 0.4249\ii$ & $0.078$ & $0.090$ & $27\degr$ & $30\degr$  \\
% \hline
% \multicolumn{6}{c}{\textcolor{blue}{linB\tablenotemark{$\dagger$}: $\tau_\mathrm{s}\OmK=0.1$, $K_{x,z}=6$, $\rhog/\rhod = 0.2$}} \\
% \hline
linB\tablenotemark{$\dagger$} ($0.1, 6,0.2$) &
$-0.4999 + 0.0155\ii$&   $-0.4981 + 0.0054\ii$ & 0.025   & $0.0054$ & $5.8\degr$  &  $1.2\degr$  \\
% \hline
% \multicolumn{6}{c}{\textcolor{blue}{linC\tablenotemark{$\ddagger$}: $\tau_\mathrm{s}\OmK=10^{-2}$, $K_{x,z}=1500$, $\rhog/\rhod = 2$}} \\
% \hline
linC\tablenotemark{$\ddagger$} ($10^{-2}, 1500, 2$) &
$0.1049 + 0.5981\ii$   &  $0.1338 + 0.6650\ii$ & 0.0076 & $0.013$ & $8.3\degr$ &$11\degr$  \\
% \hline
% \multicolumn{6}{c}{\textcolor{blue}{linD\tablenotemark{$\ddagger$}: $\tau_\mathrm{s}\OmK=10^{-3}$, $K_{x,z}=2000$, $\rhog/\rhod = 2$}} \\
% \hline
linD\tablenotemark{$\ddagger$} ($10^{-3}, 2000, 2$) &
$0.3225 +  0.3154\ii$& $0.3219 + 0.3154\ii$ &  $0.061$ & $0.061$  & $22\degr$ & $22\degr$  
\enddata
\tablenotetext{$\dagger$}{\cite{youdin07b}}
\tablenotetext{$\ddagger$}{\cite{bai10b}}
\end{deluxetable*}

\begin{figure}
\includegraphics[width=\linewidth]{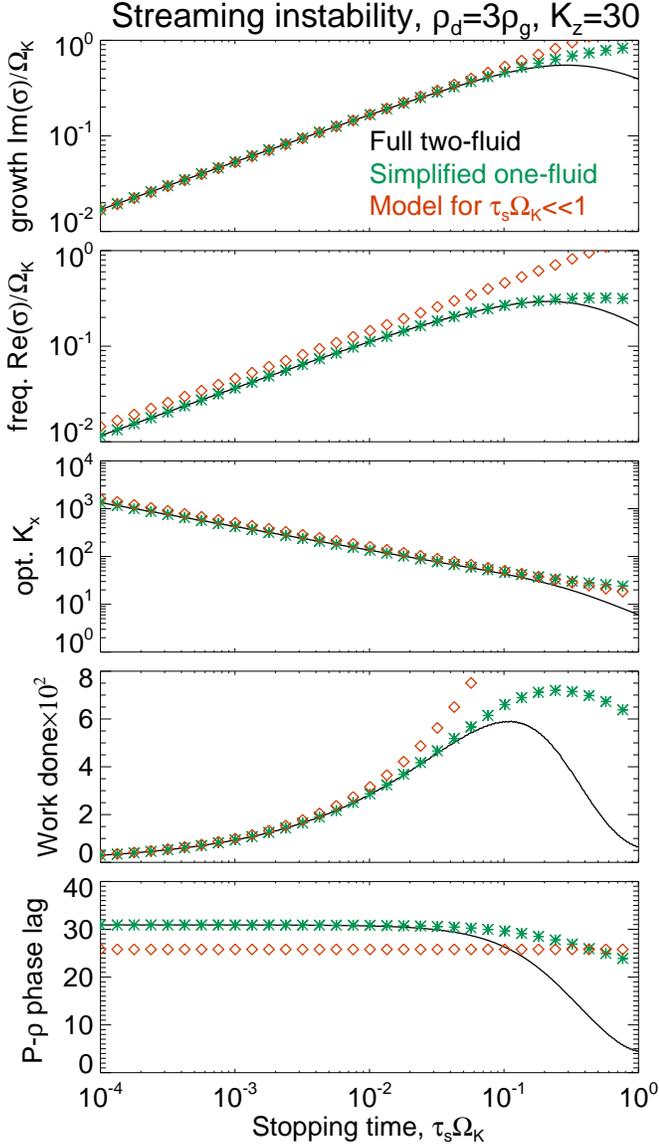}
\caption{Comparison of the linear 
streaming instability between a full two-fluid analysis (solid 
line), the one-fluid framework simplified by the terminal 
velocity approximation (green asterisks) and an analytic solution to
the one-fluid dispersion relation in the dust-rich limit (orange
diamonds, see also Appendix \ref{si_dust_rich}). The vertical 
wavenumber is fixed and growth rates are maximized over $K_x$. 
\label{si_compare_fig}}
\end{figure}

In Fig. \ref{si_compare_fig} we calculate the most unstable SI 
mode as a function of $\taus$ at fixed $K_z=30$ and $\epsilon =3$. Growth rates are maximized over $K_x$. 
We 
compare results between the full, two-fluid linear analysis and the
one-fluid framework. We also include an analytic model, developed in Appendix 
\ref{si_dust_rich}, based on the one-fluid dispersion relation with 
additional approximations (orange diamonds). 

Both results based on the one-fluid framework compares well with the 
full, two-fluid analysis, only breaking down at a relatively large
$\taus\OmK\gtrsim 0.1$. For larger $\taus$ the two-fluid phase
lag drops, along with the growth rate. This suggest that a
non-vanishing phase lag is indeed necessary for instability. However, the
magnitude of the phase lag does not correlate with growth rates. In
fact, $\varphi$ remains finite as $|\sigma|, \taus\to0$ (but
non-zero). This arises because the optimum $K_x\propto \taus^{-1/2}$
diverges (see Appendix \ref{si_dust_rich}).   

In a future work we will perform a more detailed parameter survey to
compare the simplified one-fluid and full two-fluid frameworks in
calculating SI \citep[cf. comparisons using other problems, ][]{laibe14,price15}. 

\section{
  Vertical shear instability with dust}\label{results} 

We now present numerical solutions for vertically stratified,
non-self-gravitating disks, assuming perfectly coupled dust. 
We formally take $\tstop=0$ so the equilibrium defined in \S\ref{eqm} 
are exact steady states. In reality,   
%\subsection{Validity}
%{\bf maybe incorporate into intro to this section?}
%We considered $\tstop=0$ in order to perform stability analysis on an
%exact steady state. 
%When $\tstop\neq0$, a stratified dusty 
%disk initialized as in \S\ref{eqm} would evolve 
%as 
dust settles to the midplane on a timescale $t_\mathrm{settle}\sim 
1/\tstop\OmK^2$ \citep{takeuchi02}. However, we expect the 
perfectly-coupled limit to be valid  
provided timescales of interest $t_\mathrm{grow}\ll
t_\mathrm{settle}$. For the VSI this translates to 
%Since VSI growth rates are $O(h_\mathrm{g}\OmK)$, 
%we require  
%\begin{align}
 $ \tstop\OmK \ll h_\mathrm{g}. $
%\end{align} 
For thin PPDs, this is satisfied for $\tstop\OmK \ll O(10^{-2})$.

We first consider constant
midplane dust-to-gas ratios $\epsilon_0$ and characteristic thickness
$H_\epsilon$. Then the dust-to-gas ratio $\epsilon=\epsilon(z)$. In 
this limit any growing modes must be associated with the imposed
temperature gradient (see \S\ref{limits}). We are then studying the
effect of dust-loading on the VSI previously studied in pure gas
disks \citep[][\citetalias{lin15} in this section]{lin15}. In 
\S\ref{varHd} we allow $\p_r\epsilon\neq 0$, and in 
\S\ref{vert_mixed} we consider VSI driven entirely by radial
gradients in the dust-to-gas ratio (as discussed in
\S\ref{iso_perfect}).  

The parameters for the linear problem includes $\epsilon_0$, the dust
layer thickness $\Hd$, and the perturbation radial wavenumber
$k_x$. We choose a midplane gas density profile 
$\rho_\mathrm{g0}\propto r^{-3/2}$. The fiducial power-law 
index for the temperature profile is $q=-1$; and we set the gas disk
aspect-ratio $h_\mathrm{g}=0.05$. These values were also used by  
\citetalias{lin15}. 

%Two boundary conditions are needed for the $\tstop=0$
%problem \citep[e.g.][]{lubow93}. 
We impose solid vertical boundaries so that $\delta
v_z(\pm\zmax)=0$. We solve the linearized equations as a generalized
eigenvalue problem using a pseudo-spectral code adapted  
from \citetalias{lin15}. Amplitudes are expanded in Chebyshev
polynomials up to order $512$. %We check results using
%Eq. \ref{vsi_check}.      

\subsection{Qualitative expectations}\label{vsi_est}
\citetalias{lin15} found the appropriate way to compare 
vertical shear (destabilizing) and  vertical buoyancy (stabilizing)
is $r\p_z\Omega/\OmK$  against $N_z^2/\OmK^2$. From 
\S\ref{vertshear} and \S\ref{vbuoyancy} we find $|r\p_z\Omega|\sim q
h_\mathrm{g}\OmK$ for a 
thin, gas dominated disk; while $N_z^2\sim \epsilon\OmK^2$. Thus we
expect dust-induced buoyancy forces to stabilize the disk against the
VSI where $\epsilon \gtrsim h_\mathrm{g}$. 

%\begin{align}

\subsection{Effect of dust-loading}
We first vary the midplane dust-to-gas ratio 
$\epsilon_0\in[10^{-3},1]$, fixing the dust thickness to  
$\Hd=0.99\Hg$. Then  $\epsilon$ is roughly constant with height. We
set the vertical domain to $\zmax=5\Hg$.  

\begin{figure}
  \includegraphics[width=\linewidth]{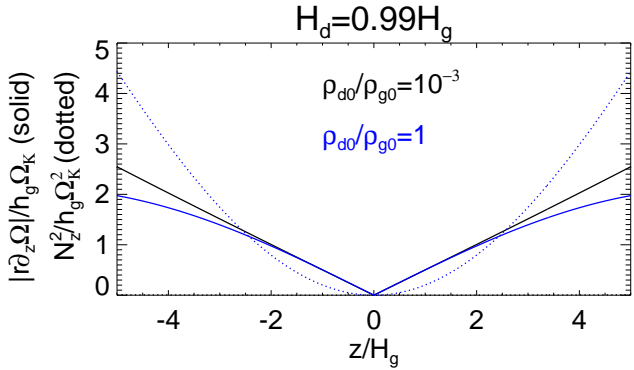} 
  \caption{Vertical shear rate (solid) compared to vertical buoyancy
    (dotted) in a locally isothermal, dusty disk with midplane dust-to-gas ratio
    of $\epsilon_0=10^{-3}$ (black) and $\epsilon_0=1$ (blue). 
    The dust layer thickness is fixed to $\Hd=0.99\Hg$ so the 
    dust-to-gas ratio $\epsilon$ is approximately constant with
    height. Vertical buoyancy here is due to dust-loading. 
    \label{compare_vshear_fixHd}
    }
\end{figure}

Fig. \ref{compare_vshear_fixHd} compares the basic state
vertical shear rate, which is destabilizing, and the vertical buoyancy
frequency, which is stabilizing. For the nearly dust-free disk
$\epsilon_0=10^{-3}$ the vertical shear dominates buoyancy for all
$|z|>0$. However, a heavy dust-load with $\epsilon_0=1$ renders the 
buoyancy to dominate over vertical shear in the disk atmosphere 
$|z|\gtrsim 2.5\Hg$. We thus expect instability all heights for 
$\epsilon_0=10^{-3}$, but to be restricted to the midplane for
$\epsilon_0=1$. 

\begin{figure}
  \includegraphics[width=\linewidth]{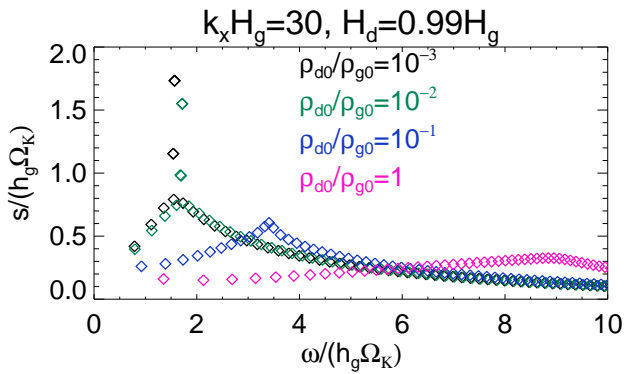} 
  \caption{Unstable modes in a locally isothermal, perfectly coupled
    dusty disk with fiducial parameters
    $(p,q,h_\mathrm{g}, \Hd/\Hg )=(-1.5,-1,0.05, 0.99)$. The real
    frequency $\omega$ and growth rates $s$ are shown for a range of
    midplane dust-to-gas ratios $\epsilon_0=\rho_\mathrm{g0}/\rho_\mathrm{d0}$. 
    \label{vsi_dust_loading}
    }
\end{figure}

Fig. \ref{vsi_dust_loading} show unstable modes for different values
of $\epsilon_0$ with fixed perturbation wavenumber  $k_x\Hg = 30$. The
eigenvalue distributions for $\epsilon_0 \leq 10^{-2}$ are similar to the
dust-free fiducial case considered by \citetalias{lin15}. This is
expected since $\epsilon_0 < \hgas$ (\S\ref{vsi_est}). Eigenvalues 
consists 
of the roughly horizontal `body modes', and the nearly-vertical
`surface modes' \citep[which are associated with the imposed vertical
boundaries, ][]{barker15}.   

We find that increasing the dust-to-gas ratio reduce VSI growth
rates. Notably, surface modes, which are typically fastest growing in
the dust-free case, are suppressed in dusty disks for $\epsilon_0\geq
0.1$ (i.e. $\epsilon_0> \hgas$).   
The body modes' growth rates remain $ O(h_\mathrm{g}\OmK)$ 
but their oscillation frequency increases with
dust-loading, i.e. it increases with the vertical buoyancy. 
The total number of modes do not change 
significantly. This is in contrast with the effect of increasing
cooling times in an adiabatic gas disk, for which \citetalias{lin15}
find fewer unstable modes. 
%until the system is eventually completely stable. 

\begin{figure}
\includegraphics[scale=0.32, clip=true, trim=0.5cm 0cm 3cm 0cm]{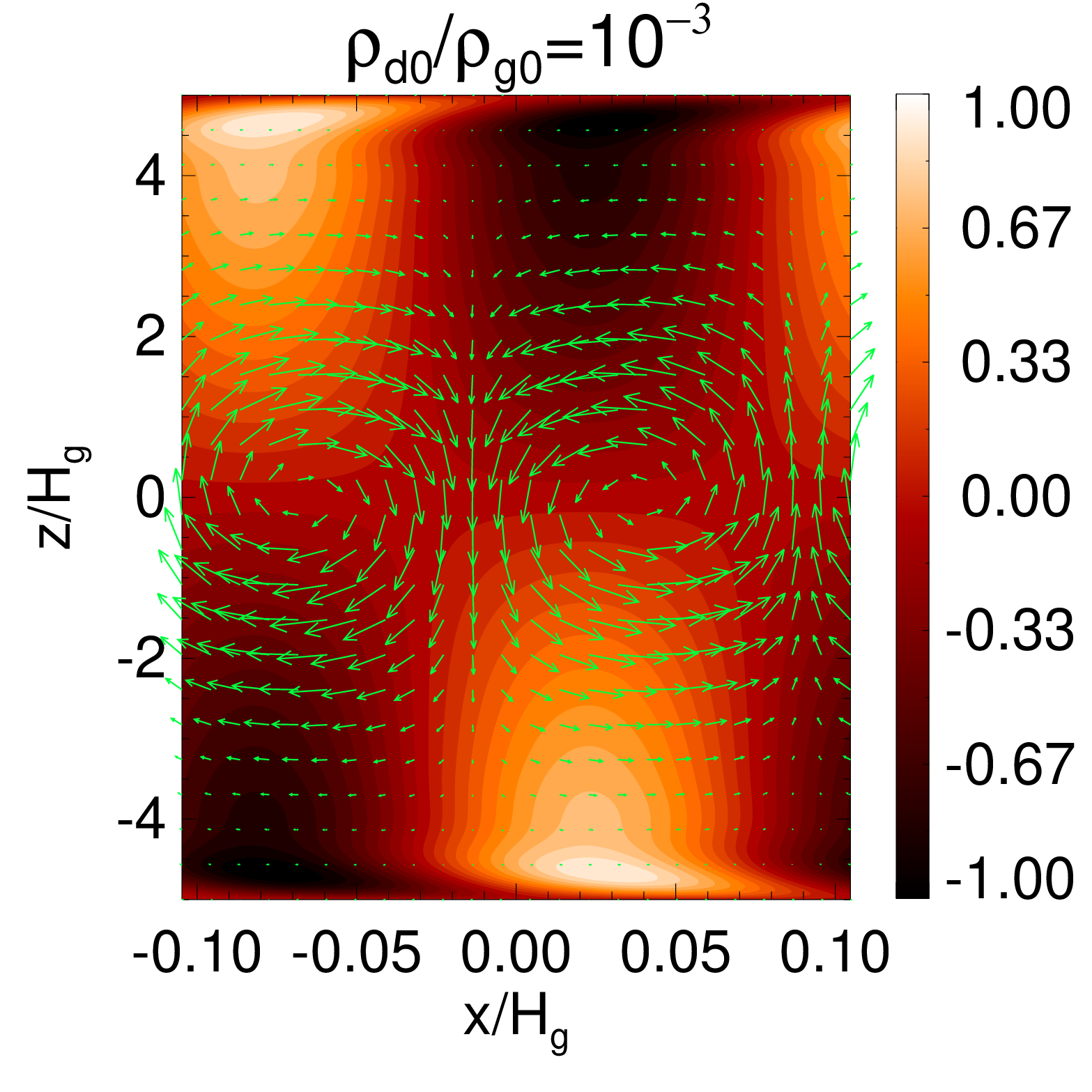}\includegraphics[scale=0.32, clip=true, trim=1.8cm 0cm 0cm 0cm]{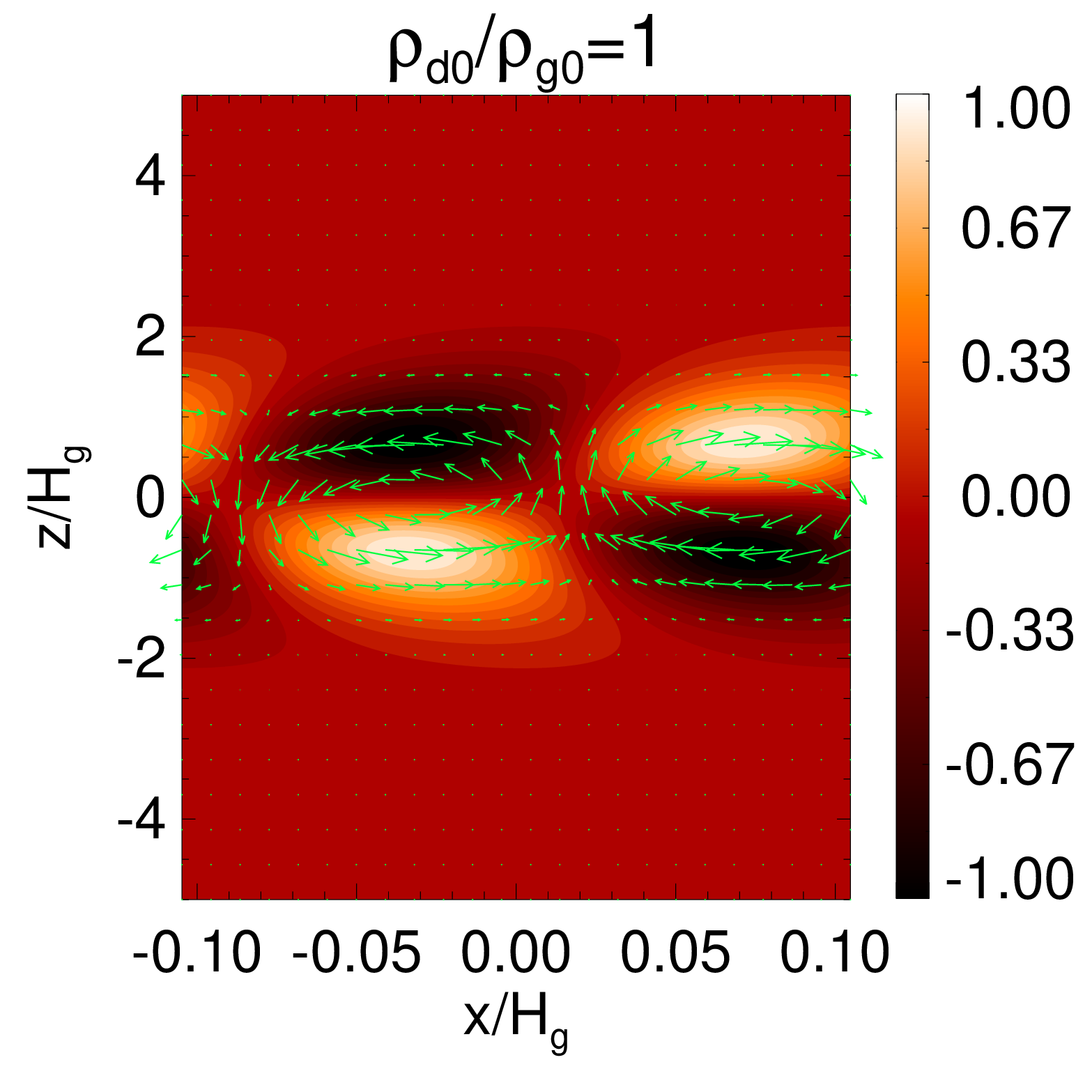}
  \caption{Fundamental dusty VSI mode in real space for midplane dust-to-gas
    ratio $\epsilon_0=10^{-3}$ (left) and $\epsilon_0=1$
    (right). The color scale shows the perturbation to the
    dust-to-gas ratio, $\delta\epsilon$; and the arrows show
    $\sqrt{\rho}\left(\dd v_x, \dd v_z\right)$. 
    \label{vsi_dust_loading2d}
    }
\end{figure}

The lowest frequency `fundamental' body mode is energetically dominant
because the entire disk column is perturbed \citep[cf. surface modes
  which only disturb the disk boundaries,][]{umurhan16c}. In Fig. \ref{vsi_dust_loading2d}
we compare the fundamental mode between the nearly 
dust-free case $\epsilon_0=10^{-3}$ an a dusty disk with
$\epsilon_0=1$. Dust-loading preferentially
stabilizes the disk atmosphere against the VSI, restricting
meridional motions to $|z|\lesssim 2\Hg$. 
This is consistent with Fig. \ref{compare_vshear_fixHd} comparing the basic state vertical
shear and buoyancy. % Notice also maxium perturbtions to the dust-to-gas ratio,
% $\delta\epsilon$, shifts from disk boundaries 
% to the midplane.  

\begin{figure}
  \includegraphics[width=\linewidth]{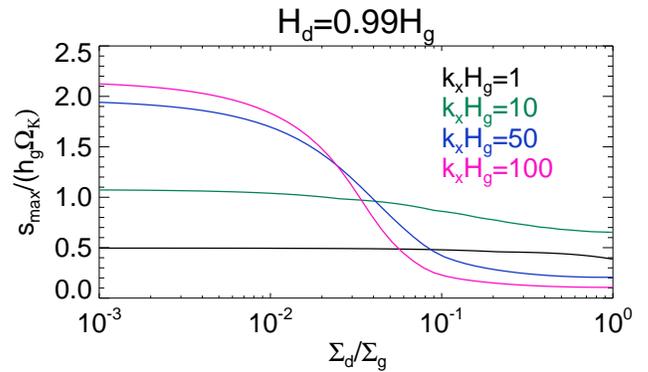} 
  \caption{Maximum growth rate of the dusty VSI as a function of the
    midplane dust-to-gas ratio $\epsilon_0$ for perturbations with
    different radial wavenumbers $k$. The dust layer thickness is
    fixed to $\Hd\simeq \Hg$. 
    \label{vsi_dust_loading_vareps}
    }
\end{figure}

In Fig. \ref{vsi_dust_loading_vareps} we plot the growth rates as a
function of $\epsilon_0$ for different perturbation wavenumbers
$k_x$. Dust-loading stabilizes the VSI more effectively for shorter
wavelength perturbations. This is because for high wavenumbers the
dominant modes are surface modes, which are effectively stabilized by
dust-loading as buoyancy forces are largest near the vertical 
boundaries. The figure suggest that VSI becomes much less efficient
for $\epsilon_0\gtrsim 0.1$ and $k_x\Hg\gtrsim 50$. 

\subsection{Effect of dust layer thickness} 
We now vary $\Hd$ but fix the metalicity 
$Z \equiv \epsilon_0 \Hd/\Hg = 0.03$ to obtain $\epsilon_0$. Since we
will consider thin dust layers, here we use a smaller   
domain with $\zmax=2\Hg$ so that $\epsilon$ does not become
too small. 

\begin{figure}
  \includegraphics[width=\linewidth]{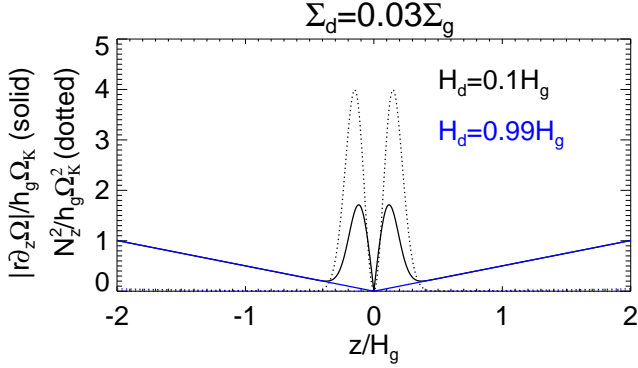} 
  \caption{Vertical shear rate (solid) compared to vertical buoyancy
    (dotted) in a locally isothermal, dusty disk 
    with metalicity $Z=0.03$ and dust thickness $\Hd=0.1\Hg$
    (black) and $\Hd=0.99\Hg$ (blue). 
    \label{compare_vshear_fixZ}
    }
\end{figure}

We analyze two disks with $\Hd=0.1\Hg$ and 
$\Hd=0.99\Hg$. Fig. \ref{compare_vshear_fixZ} compares the vertical
shear  rate and buoyancy frequency. For $|z|\gtrsim 0.4\Hg$ the two
disks have the same profile with vertical shear dominating over
buoyancy. We thus expect perturbations away from the disk midplane in 
both cases. For $|z|\lesssim 0.4\Hg$, a thin dust 
layer with $\Hd=0.1\Hg$ boosts the vertical shear rate, but the
associated buoyancy is larger still, implying the mid-plane should be
stable. 
%Thus we expect the mid-plane of 
%the disk with $\Hd=0.1\Hg$ to have limited vertical motions.  

\begin{figure}
  \includegraphics[scale=0.32, clip=true, trim=0.5cm 0cm 3cm 0cm]{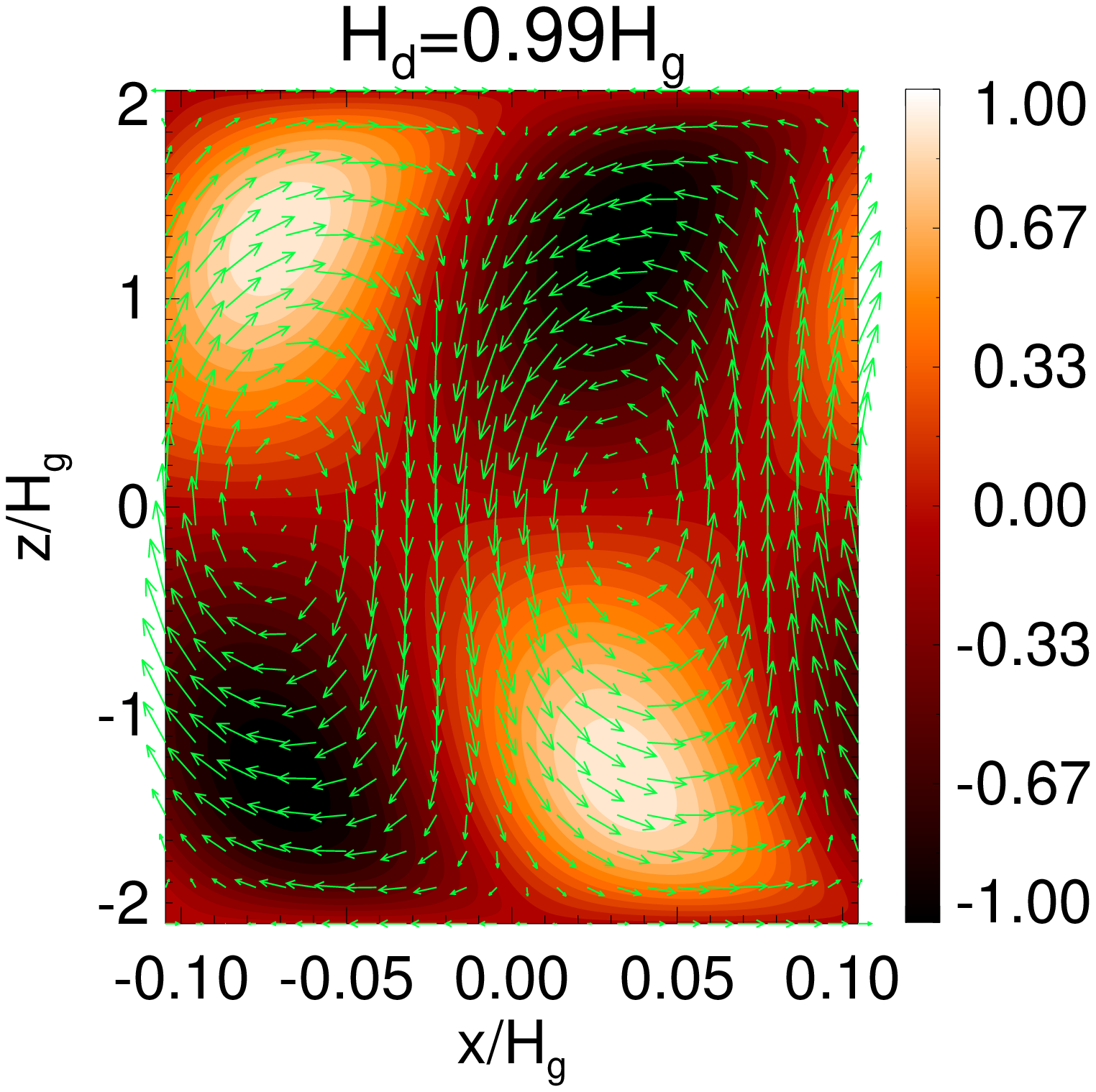}\includegraphics[scale=0.32, clip=true, trim=1.8cm 0cm 0cm 0cm]{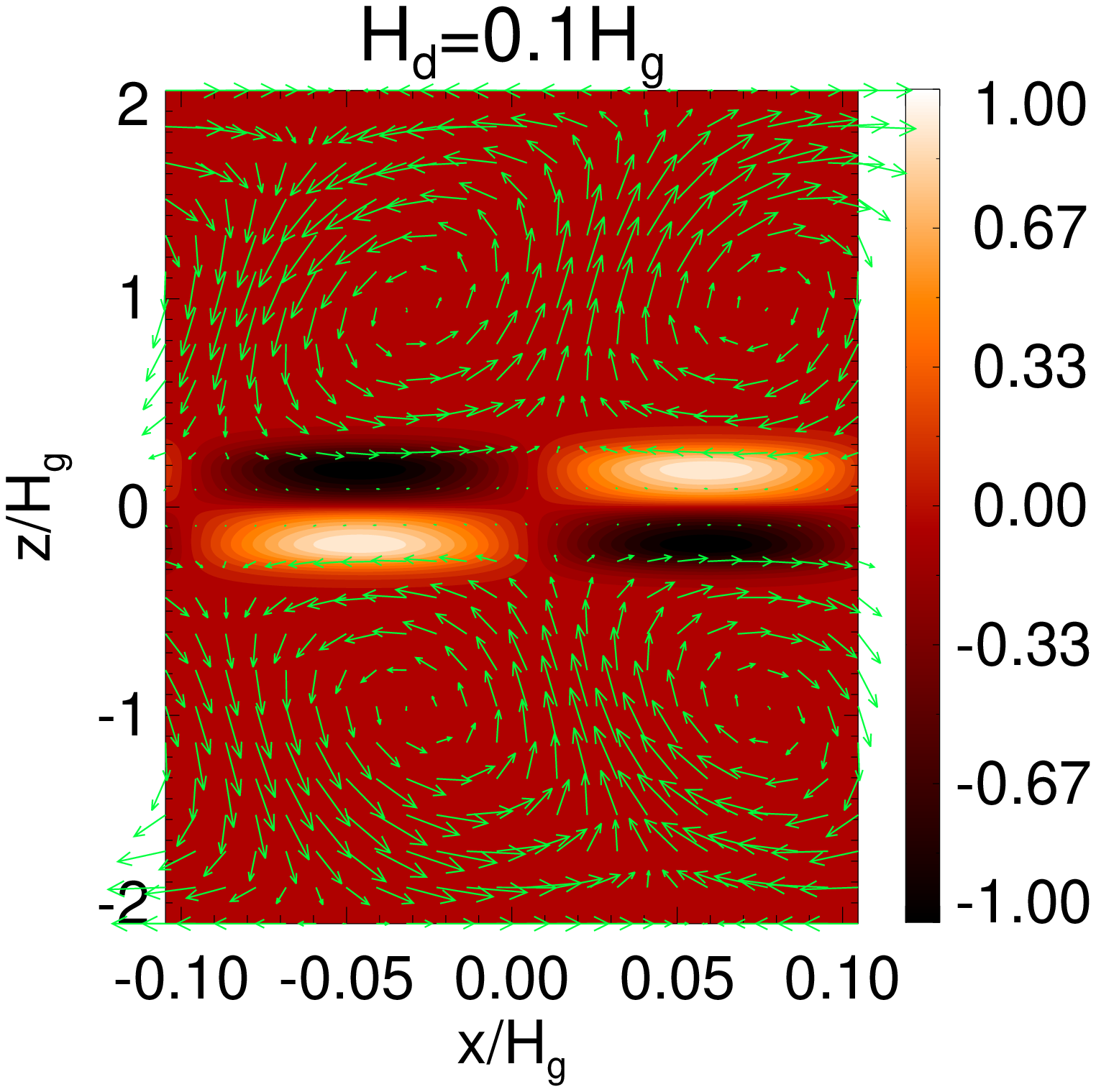} 
  \caption{Fastest-growing dusty VSI mode in real space for midplane
    dust layer thickness $\Hd=0.99\Hg$ (left) and $\Hd=0.1\Hg$
    (right). The dust content is fixed to
    $\Sigma_\mathrm{d}=0.03\Sigma_\mathrm{g}$. 
    The color scale shows the perturbation to the
    dust-to-gas ratio, $\delta\epsilon$; and the arrows show
    $\sqrt{\rho}\left(\dd v_x, \dd v_z\right)$.
    \label{result2d_fixZ}
    }
\end{figure}

Fig. \ref{result2d_fixZ} compares the fastest growing VSI body modes  
with $k_x\Hg=30$ for the two cases above.\citepalias[The thinner domain
  adopted here eliminates surface modes, ][]{lin15}.   
We find very similar mode 
frequencies 
\begin{align*}
  \sigma = \begin{cases}
    \left(0.3053\ii - 0.8142\right)h_\mathrm{g}\OmK & \Hd=0.99\Hg, \\
    \left(0.3178\ii - 1.2237\right)h_\mathrm{g}\OmK & \Hd=0.1\Hg,
  \end{cases}
\end{align*}
since the vertical shear profile is similar throughout most of the
disk. However, meridional motions are suppressed near the midplane of
the $\Hd=0.1\Hg$ disk, as expected from the larger buoyancy frequency
relative to vertical shear there. This leads to a structure
analogous to PPD dead zones: a quiescent midplane between active
surface layers \citep{gammie96}.

\begin{figure}
  \includegraphics[width=\linewidth]{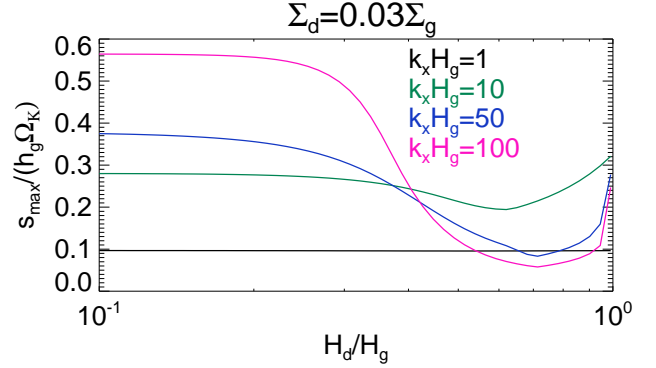} 
  \caption{Maximum VSI growth rate for different perturbation
    wavenumbers $k$ as a function of the dust layer
    thickness $\Hd$ at fixed metalicity $Z=0.03$. 
    \label{compare_eigenvals_fixZ}
    }
\end{figure}

Fig. \ref{compare_eigenvals_fixZ} shows the maximum VSI growth rates
as a function of $\Hd$. As before, we find growth
rates are most affected by the vertical structure of the dust layer
when the perturbation wavenumer is large. Notice VSI growth rates converge as
$\Hd\to 0$. %to dust free values?    
Thus a thin dust layer, however large its associated vertical shear,
does not affect VSI growth rates. The non-monotonic behavior for $\Hd\gtrsim
0.5\Hg$ arises because the vertical buoyancy frequency 
\begin{align*}
N_z^2(H_d;z,Z) \simeq &Z\Hg z^2
\exp{\left(-\frac{z^2}{2\Hg^2}\right)}\OmK^2\notag\\
&\times 
\frac{1}{\Hd}\left(\frac{1}{\Hd^2} -
\frac{1}{\Hg^2}\right)\exp{\left(-\frac{z^2}{2\Hd^2}\right)}  
\end{align*}
is a non-monotonic function of $\Hd$ at fixed $z$. At $z=\Hg$ and $z=2\Hg$
the buoyancy frequency is maximized for $\Hd\simeq0.5\Hg$ and
$\Hd\simeq 0.8\Hg$, respectively. This is consistent with the abscissa
of minima in growth rates in Fig. \ref{compare_eigenvals_fixZ}. 

\subsection{Axisymmetric stability of ultra-thin dust layers}
The discussion in \S\ref{iso_perfect} imply strictly  
isothermal disks with a radially uniform dust-to-gas ratio are
stable against axisymmetric perturbations, no matter how thin the dust
layer is. We now demonstrate this numerically. 

To connect with similar studies, here  we also use 
the Richardson number $\rich \equiv N_z^2/\left(r\p_z\Omega\right)^2$
to label calculations \citep{youdin02}.   
Numerical simulations show    
non-axisymmetric instabilities develops when $\rich\lesssim0.1$, brought about
by very thin dust layers \citep{chiang08, lee10}. We show that \emph{axisymmetric}
instabilities never develop in radially uniform disks, however small $\rich$. 

We shall consider ultra thin dust layers with $\Hd\leq 0.01\Hg$ and thus
restrict the vertical domain to $\zmax = 0.02\Hg$. We fix the 
metalicity $Z=0.01$ so the midpane dust-to-gas ratio $\epsilon_0$ is 
$O(1)$.  We consider radial wavenumbers with $k_x\Hd=1$. 

\begin{figure}
  \includegraphics[width=\linewidth]{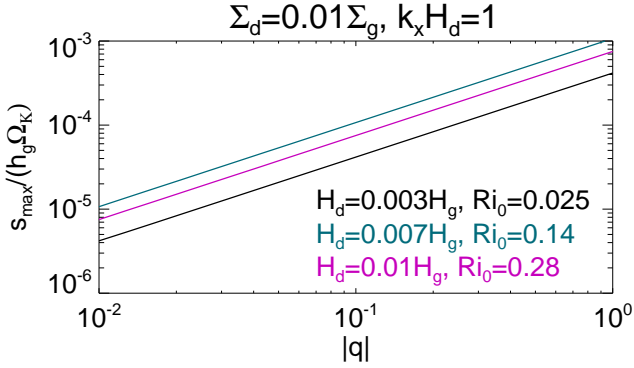} 
  \caption{Maximum VSI growth rate for ultra-thin dust layers $\Hd 
    \leq 0.01\Hg$ (with radially uniform dust-to-gas  ratio). Vertical shear is dominated by that due to 
    vertical dust stratification, but axisymmetric instability is still 
    associated with the radial temperature gradient $q$. 
 Here, $\rich_0$ is the minimum 
    Richardson number in the domain in the limit $q\to0$. The disk is stable to axisymmetric perturbations 
    in the strictly isothermal limit, regardless of the dust layer thickness. 
    \label{ultra_thin}
    }
\end{figure}

Fig. \ref{ultra_thin} shows the maximum growth rate  as a function of
the radial temperature gradient, $q$. For all cases the vertical
shear is dominated by that due to the dust layer (see
\S\ref{vertshear}). However, we see that $s\propto
|q|$, i.e growth rates vanish in the strictly isothermal limit.  
In particular, this holds for $\rich < 0.1$, the critical value 
for non-axisymmetric instabilities. 

Axisymmetric instability here is associated with the thermal
contribution to vertical shear: as $q\to0$,  $\p_z\Omega$ becomes
entirely due to $\p_z\epsilon$ and there is no instability ($s\to
0$). This result is independent of $\rich$, so the Richardson number
is does not characterize the axisymmetric stability of dust
layers. 

\subsection{Effect of a radially-varying dust-to-gas ratio}\label{varHd}
%{\bf bonus stuff, maybe delete in final version}

We now consider a radially-varying dust-to-gas ratio. Specifically we
let $\epsilon_0\propto r^{-1}$ and $H_\epsilon \propto \Hg$
(cf. constant values in the previous calculations). Then
\begin{align*} 
  \frac{\p\epsilon}{\p r} =
  \left(\frac{z^2}{H_\epsilon^2}\frac{d\ln{\Hg}}{dr} -
    \frac{1}{r}\right)\epsilon.   
\end{align*}
Here we fix $Z=0.01$, $\Hd=0.8\Hg$.       

\begin{figure}
  \includegraphics[width=\linewidth]{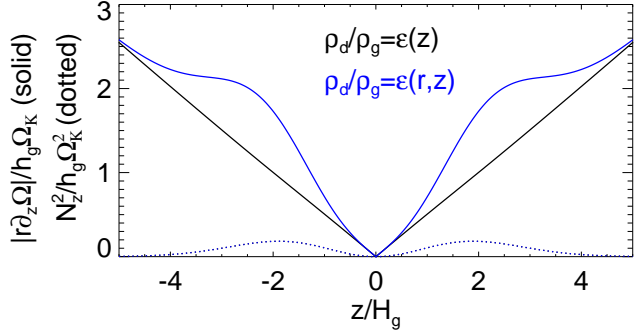} 
  \caption{Vertical shear rate (solid) compared to vertical buoyancy
    (dotted) in a locally isothermal, dusty disk 
    with metalicity $Z=0.01$ and dust thickness $\Hd=0.8\Hg$.
    Black curves assume a dust-to-gas ratio that only depends on
    height; whereas the blue curve also allows a radial dependence in
    $\epsilon$. See text for details.  
    \label{compare_vshear_varHd}
    }
\end{figure}

\begin{figure}
  \includegraphics[width=\linewidth]{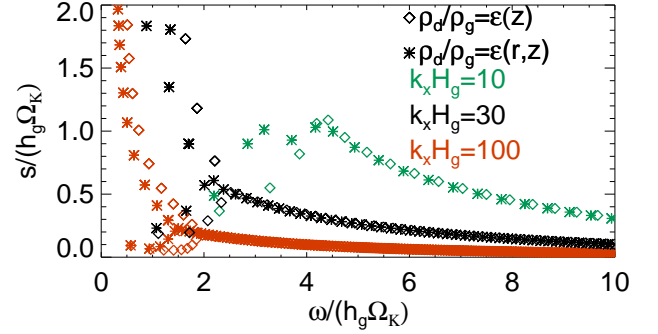} 
  \caption{Unstable VSI modes in the disk models of
    Fig. \protect\ref{compare_vshear_varHd}. { Diamonds (asterisks) 
    are mode frequencies for a disk with radially uniform (varying)
    dust-to-gas ratio. The radial wavenumbers are $k_xH_g = 10$
    (green),$k_xH_g = 30$ (black) and $k_xH_g = 100$ (orange). }
    \label{vsi_dust_varHd}
    }
\end{figure}

Fig. \ref{compare_vshear_varHd} shows that a radially-varying
dust-to-gas ratio increases the magnitude of the vertical shear rate
away from the midplane (see also Eq. \ref{vshear2}). Thus we {
  typically} find higher VSI growth rates, as shown in
Fig. \ref{vsi_dust_varHd}. 
Surfaces modes, { appearing at high $k_x$}, are more effectively
enhanced by the additional vertical shear induced by
$\p_r\epsilon$. { Low frequency body modes with small
  $k_x$ are more affected by radial variations in the 
  dust-to-gas ratio than high-$k_x$, high-frequency body modes. }    

However, all growth rates remain
$O(\hgas\OmK)$. Importantly, the increase in the growth rate of the
fundamental mode is small. We do not expect the radial
dependence in $\epsilon$ to significantly affect the VSI in a dusty
disk. 

\subsection{Pure instability with vertically well-mixed
  dust}\label{vert_mixed} 

In \S\ref{iso_perfect} we found that for strictly isothermal
disks, a vertically-uniform dust-to-gas ratio can be unstable
if $d\epsilon/dr\neq0$.   
{
To demonstrate this numerically we set $q=0$ and $H_\mathrm{d}$ such
that $H_\epsilon=10^3\Hg$. We let $\rhod/\rhog \propto r^{-d}$
with different power indices $d$. The metalicity is fixed to $Z=0.03$.  
In these cases vertical shear $\p_z\Omega$ is 
attributed to the radially-varying dust-to-gas ratio (see
Eq. \ref{vshear2}).    
}

\begin{figure}
  \includegraphics[width=\linewidth]{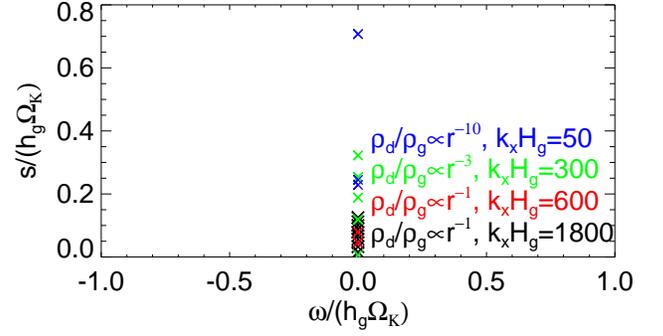} 
  \caption{Purely growing modes in a strictly isothermal disk ($q=0$) with
    vertically uniform dust-to-gas ratio. The instability is due to 
    vertical shear $\p_z\Omega\neq0$ arising from radial variations in
    the dust-to-gas ratio, $d\epsilon/dr\neq 0$. \label{vert_mixed_modes}
    }
\end{figure}

{ We show unstable modes in 
Fig. \ref{vert_mixed_modes}.} 
%In order to obtain appreciable growth   
%rates, we consider super-Solar metalicity $Z=0.03$ and high
%radial wavenumber $k_x\Hg=1800$. 
As expected from the discussion in
\S\ref{iso_perfect}, the disk admits purely growing modes with
$\omega=0$. This is distinct from the growing oscillations associated
with classic VSI discussed above. 

{ We find in smooth disks large 
  $k_x$ is needed for appreciable growth rates, e.g. $k_xH_g=1800$
  with $\rhod/\rhog\propto r^{-1}$. 
  Local shearing box simulations may 
  be required to study the non-linear evolution of such short 
  wavelengths \citep[e.g.][]{bai10b,yang16}. Alternatively, as shown in Fig. \ref{vert_mixed_modes},
  a rapidly varying dust-to-gas ratio permits dynamical instability at
  longer radial wavelengths, which might be resolvable in global disk 
  simulations. 
}

%{\bf
%  \subsection{Implications for protoplanetary disks}
%dust settling, dust-induced buoyancy 
%The above calculations 
%radial d/g -> radial temp profile 
%}

%growth rates not sig changed 
%v shear from thin dust layers are overwhelmed by 
\section{Discussion}\label{discussion}

{

%pros and cons 
%energy analogy not restricted to term vel approx 
%spurious modes

\subsection{Applications and limitations}

The main application we envision for the dusty/adiabatic gas analogy 
is to develop physical interpretations of dust-gas drag instabilities; and to find 
dusty analogs of pure gas instabilities in protoplanetary disks, 
which is discussed in \S\ref{dust_analogs}. One can also exploit the similarity to adapt existing
hydrodynamic codes to simulate dusty protoplanetary disks (see \S\ref{dust_sims}). 

It is important to keep in mind the assumptions 
used to develop our thermodynamic model of dusty gas. 
The terminal velocity approximation, $\bm{v}_\mathrm{d} -
\bm{v}_\mathrm{g} = \tstop \nabla P /\rhog$, was employed from the
outset. 
%It assumes  
%the dust-gas relative drift $\bm{v}_mathrm{d} - \bm{v}_\mathrm{g} =
%\tstop \nabla P /\rhog$ 
This is applicable to small particles with short stopping times and
strongly coupled to the gas. Generally we require $\tstop$ to be the
shortest timescale in the physical problem. For  
example, to study dust settling, we require $\tstop \ll
t_\mathrm{settle}\sim 1/\tstop\OmK^2$, or 
$\tstop\OmK \ll 1$. 

%However, the validity of the simplified one-fluid equations may be
%problem-dependent, and not just on the terminal velocity 
%approximation. 

However, the validity of the terminal velocity approximation may not
only depend on $\tstop$. The value of the dust gas-ratio and the
problem itself may also be important. 
%You can decide if this is a relevant improvement, and whether to
%include. 
For example, Table \ref{si_compare} shows the dust-rich
`linA' mode and dust-poor `linB' mode of the streaming instability
have the same $\tstop$, but the former is accurately captured by the
simplified equations, while the latter is not.  This suggests that for
the streaming instability the simplified equations is better suited
for $\rhod/\rhog>1$. The simplified equations also contain spurious
modes in certain limits (see \S\ref{spurious_epi}). 

%Ultimately,
                                %one must 
                                %c heck against the full equations.  

Note that our thermodynamic interpretation of dust-gas drift does
\emph{not} actually depend on the terminal velocity approximation.   
Once the gas equation of state is fixed and the true energy equation deleted, 
the dust continuity equation can be converted to a new effective energy equation. 
As an example, for strictly isothermal dusty gas we obtain   
\begin{align}
\frac{DP}{Dt} = - P\nabla\cdot\bm{v} +
c_s^2\nabla\cdot\left[\fdust\left(1-\fdust\right)\rho\left(\bm{v}_\mathrm{d}
    - \bm{v}_\mathrm{g}\right)\right]. \label{no_term_vel}  
\end{align}
This equation does not use
the terminal velocity approximation (which gives
Eq. \ref{eff_energy}). Evaluating the right-hand-side generally 
requires solving an evolutionary equation for $\bm{v}_\mathrm{d} - 
\bm{v}_\mathrm{g}$ \citep{laibe14}. Nevertheless,
Eq. \ref{no_term_vel} shows that dust-gas relative drift can be
interpreted as a heat flux within the dust-gas mixture. Thus, even 
without the terminal velocity approximation we can 
regard the dusty-gas as a single ideal fluid subject to cooling.   
}

\subsection{Generalization to locally polytropic disks}\label{gen_poly}

%\subsubsection{Locally polytropic disks}
We can extend the dusty/adiabatic gas correspondence to 
other fixed equations of state. As an example, consider the locally
polytropic disk 
\begin{align}
  P = K(r,z)\rhog^{\Gamma}, 
\end{align}
where $K$ is a prescribed function and $\Gamma$ is the constant
polytropic index. Then eliminating $\tepsilon$ from the dust equation
\ref{dusteq} gives 
\begin{align}\label{poly_energy}
  \frac{D P}{D t} &= - \Gamma P\nabla\cdot\bm{v}  + P \bm{v}\cdot\nabla
  \ln{K} + \frac{\Gamma P}{\rhog}\nabla\cdot\left(\tepsilon\tstop\nabla
  P\right). 
%+ \mathcal{C},\\
%  \mathcal{C}& = \frac{\Gamma P}{\rhog}\nabla\cdot\left(\tepsilon\tstop\nabla
%  P\right).
\end{align}
Thus the dusty gas behaves like a pure gas with adiabatic index
$\Gamma$. The entropy is given by 
\begin{align}
  \seff = \ln{\left[K^{1/\Gamma}(r,z)\left(1 - \tepsilon\right)\right]}.  
\end{align}
The results of \S\ref{iso_perfect}---\ref{dust_work} remain valid for the strictly 
polytropic disk with $K=$constant, while one sets $c_s^2\to K$ in
\S\ref{dusty_vsi_int}. 
%When $K$ is constant, the corresponding Solberg-Hoiland criteria for
%axisymmetric stability may be obtained from that in standard adiabatic 
%hydrodynamics \citep[e.g.][]{tassoul78} by
%using this definition of entropy in
%Eqs. \ref{dusty_solberg1}---\ref{dusty_solberg2}. 

%In particular, we will explore whether this can surpress the suprious
%modes discussed in Appendix \S\ref{spurious_epi}. 

%stratified case require more boundary conditions 

\subsection{Dusty analogs of other gaseous instabilities} \label{dust_analogs} 

%gi
%rwi 

\subsubsection{Classic gravitational instabilities} %vertical structure 
The addition of dust enhances gravitational
instability (GI) because dust particles contribute to the
total disk mass but not thermal pressure, which effectively lowers the
disk temperature \citep[][]{thompson88,shi13}. For typical dust-loading 
$\tepsilon, Z\ll1$, this effect is unimportant. However, if $\epsilon$ is
large (e.g. due to dust settling) then the reduced temperature
$\widetilde{T}~=~T(1 - \tepsilon)$ may be lowered to enable instability.  

As noted in \S\ref{loc_iso_eos}, dust settling causes 
$\widetilde{T}$ to \emph{increase} away from the midplane. This contrasts
to previous studies of GI in vertically stratified disks
 where the temperature decreases
from the midplane \citep[e.g.][]{mamat10, kim12,lin14c}. 
While we expect only the total surface density and
characteristic temperatures are relevant to stability  
\citep{toomre64}, a non-trivial vertical temperature
structure, induced by dust, may modify the vertical structure of 3D
waves and unstable modes. 

Another potential connection to previous results for gas disks is  
gravito-turbulence. Cooling, self-gravitating gaseous disks 
sustain a turbulent state where shock heating due to gravitational 
instabilities are balanced by radiative cooling 
\citep{gammie01}. Since dust-gas drift appears as a diffusion or 
cooling in our framework (Eq. \ref{eff_energy}), it may be conceivable
to have `dusty gravito-turbulence'. As self-gravity increases the
local density, the associated pressure maxima attract dust-particles,
but the back-reaction onto the gas may try to flatten the pressure
bump \citep{taki16}, thus enabling a quasi-steady state. 

%see also
%\S\ref{loc_iso_eos}
%We thus expect the standard Toomre
%parameter to be modified to $Q = c_s\sqrt{1-\tepsilon}\Omega/\pi
%G\Sigma$. 

\subsubsection{Rossby wave instability}
The Rossby wave instability (RWI) is a non-axisymmetric, 2D shear
instability that operates in thin disks when it has radial structure
\citep{lovelace99,li00}. These studies consider adiabatic pure gas and 
show instability is possible if there is an extremum in the generalized
potential vorticity 
%\begin{align}
 $\mathcal{V}_\mathrm{g} =
 \kappa^2\mathcal{S}_\mathrm{g}^{-2/\gamma}/2\Omega\Sigma_\mathrm{g}$, 
%\end{align} 
where $\mathcal{S}_\mathrm{g} = P/\Sigma_\mathrm{g}^\gamma$ is essentially
the gas entropy. Here $P$
should be interpreted as the vertically integrated pressure. The
non-linear result of the RWI is vortex formation \citep{li01}. 

The dusty/adiabatic gas analogy imply the condition for RWI in a
polytropic dusty gas disk is an extremum   
\begin{align}
  \mathcal{V} =
  \frac{\kappa^2}{2\Omega\Sigma}\frac{1}{K^{2/\Gamma}}\left(1+
    \frac{\Sigma_\mathrm{d}}{\Sigma_\mathrm{g}}\right)^2.    
\end{align}
Thus RWI may also be triggered by extrema 
in the dust-to-gas ratio, e.g. narrow dust rings/gaps. This may lead 
to direct formation of dusty vortices, as opposed to dust-trapping by  
a pre-existing gas vortex \citep{barge95,lyra13}.  

%{\bf so may it's not vortex formation then dust-trapping. could just
%  form dusty vortices in one go}

%gap edges. thin dust rings. 

\subsubsection{Convective overstability}
The `convective overstability' (ConO) was discovered in pure gas, non-adiabatic, unstratified disk models    
where the radial buoyancy frequency is such that $N_r^2\equiv
F_r\p_rS/C_P  < 0 $ and cooling times $t_\mathrm{cool}\sim \OmK^{-1}$. This
combination leads to growing epicycles
\citep{klahr14,lyra14,latter16}.     
%with 
%growth rate $\propto  -N_r^2$ 
% Indeed, the effective radial buoyancy frequency in our dusty disk 
% with negligble background density gradient is unstable, 
% \begin{align}
%   N_r^2 \equiv -\frac{1}{\rho}\frac{\p P}{\p r}\frac{\p S_\mathrm{eff}}{\p r} =
%   -\frac{\rho}{P}F_r^2<0.  
% \end{align}

%The disk structure required for ConO might be realized in a locally
%isothermal dusty disk. 

Now consider a strictly isothermal dusty disk where the relevant 
entropy is dust-induced, $\seff =-\ln{(1+\rhod/\rhog)}$. 
%The dusty/adiabatic gas
%analogy implies the replacement $S\to\seff = -\ln{(1+\rhod/\rhog)}$ in
%$N_r$. 
Since $F_r\propto -\p_rP > 0$ in typical disk models, ConO would
require $\p_r\seff<0$, implying the dust-to-gas ratio increases
\emph{outwards}. This might be realized at special radial locations in 
protoplanetary disks, such as planet gaps. Dust-settling itself may
also lead to the mid-plane dust-to-gas ratio to increase outwards
\citep{takeuchi02}. 

%fact, the typical disk models used to study the streaming
%instability have zero background radial density gradient, in which
%case  $N_r^2= -\rho F_r^2/P < 0$.   

Isothermal disks have $\tcool=0$, so the pure gas ConO cannot
exist. However, we have shown that dust-gas drag provides an effective
energy source/sink to the mixture (\S\ref{energy_analogy}). It would be
interesting to explore whether or not dust-gas drag can play the role
of cooling to enable a `dusty convective 
overstability'. 

%Furthermore, 
%the effective radial buoyancy frequency in our dusty disk  
%with negligble background density gradient is unstable, 
%\begin{align}
%   N_r^2 \equiv -\frac{1}{\rho}\frac{\p P}{\p r}\frac{\p S_\mathrm{eff}}{\p r} =
%   -\frac{\rho}{P}F_r^2<0.  
% \end{align}

% One might then think of the streaming instability as a dusty relative
% to the convective  overstability. Explicit solutions in the small
% $\tstop$ limit show that streaming instability growth rates are
% also $\propto |N_r^2|$  \citep{jacquet11}, as is for the convective
% overstability \citep{lyra14,latter16}. Furthermore, both instabilities require
% finite effective cooling times, and both require $k_z\neq0$. 
% %{\bf note: si osc freq not too small compared to kappa} 

% However, the convective overstability is most unstable at $k_x=0$
% \citep{lyra14}  whereas the streaming instability growth rate is maximized at finite $k_x$
% \citep{youdin05a}. 

\subsubsection{Zombie vortices}

The `zombie vortex instability' (ZVI) is a non-axisymmetric,
non-linear instability discovered in pure gas disks
\citep{marcus15,umurhan16d}. It operates by having finite-amplitude perturbations
exciting `critical layers',  where the intrinsic wave frequency of the
perturbation matches the local vertical buoyancy frequency 
\citep{marcus13}. These critical layers roll 
up into vortices, exciting further critical layers. The process
repeats itself and the result is an array of vortices. 

Since the necessary physical ingredient for the ZVI is vertical
buoyancy, it requires an almost purely adiabatic gas with long cooling times ($\tcool\OmK\gg
1$). This limits its applicability in typical protoplanetary disks to 
$\lesssim 1$AU \citep{lesur16, malygin17}. However, this consideration
assumes a dust-free disk as far as the dynamics is concerned. 

On the other hand, we have show that dust-loading induces an effective
buoyancy even in isothermal disks. It is thus natural to ask whether or not
the required adiabatic conditions for ZVI can be realized through
dust-loading (specifically a vertical gradient in the dust-to-gas ratio), 
and thus produce `dusty ZVI'. This may allow the zombie 
vortices to develop $\gtrsim 1$AU in PPDs. 

\subsection{Implications for numerical simulations}\label{dust_sims}
{ The one-fluid model with the terminal velocity approximation, Eq. \ref{masseq}---\ref{momeq}, 
has been applied to simulate dusty protoplanetary disks \citep{dipierro15,ragusa17}. These studies 
explicitly evolve the dust density.  
}

{ However, the dusty/adiabatic gas equivalence identified in this paper
means that one need not implement a separate module for simulating the dust component in
locally isothermal/polytropic gas. The standard energy equation substitutes for the dust continuity 
equation. Thus pure gas dynamic codes can be used to simulate protoplanetary dusty disks. 
} %means that pure 
%gas dynamics codes can be applied to simulate locally 
%isothermal/polytropic dusty gas. One need not implement a separate 
%equation for the dust component: 
%for strictly isothermal or polytropic gas perfectly-coupled to dust:  
% the standard energy equation substitutes for the dust continuity
%equation. 
When { the sound-speed} $c_s$ (or $K$) is not constant and/or the dust-gas coupling is imperfect, $\tstop\neq0$, one
should add corresponding source terms in the energy equation
(e.g. Eq. \ref{poly_energy}). 
The source term associated with dust-gas drag, $\mathcal{C}$, is 
analogous to radiative diffusion \citep{price15} or thermal 
conduction, which is also common in modern hydrodynamics codes.       

We have taken advantage of the dusty/adiabatic gas equivalence to 
convert the popular
\textsc{Pluto}\footnote{\url{http://plutocode.ph.unito.it/}}      
hydrodynamics code \citep{mignone07} into a dusty gas dynamics code appropriate for
simulating protoplanetary disks coupled to small dust. { In fact,
  our code is unaware of the fact that it is modeling dusty gas.} 
We will apply 
this modified code ---d\textsc{Pluto}--- to study dusty disk-planet interaction (Lin et al., 
in preparation). 
Preliminary results { show our approach} %indicate the simplified one-fluid 
reproduce features such as dusty rings associated with 
planet-induced gaps similar to that obtained from explicit two-fluid
simulations \citep[e.g.][]{dong17}.

\section{Summary}\label{summary}
{
In this paper, we examine the conditions under which the presence of
dust can trigger instabilities in gaseous protoplanetary disks.  To this end, 
we develop an analogy between isothermal  dusty-gas and pure ideal
gas. 
} 
The correspondence arises
because drag forces reduce the relative velocity between gas and
dust. In the limit of perfect dust-gas coupling, with stopping time $\tstop \to 0$,  
 dust is entrained in 
the gas. Then the dust-to-gas ratio $\rhod/\rhog$ is conserved
following the flow. This property is analogous to entropy conservation 
following an adiabatic, pure gas. 

For finite drag, $\tstop\neq 0$, the dust content of a 
parcel of the dusty-gas mixture is no longer conserved. The parcel 
can exchange dust particles with neighboring parcels. % owing 
%to dust-gas relative drift. 
This is analogous to heat exchange between a parcel of pure gas and
its surroundings.    

We explicitly show that for a fixed gas equation of state, the  
evolutionary equation for $\rhod/\rhog$ may be replaced by an 
effective energy equation. This leads to a 
natural definition of the effective entropy of isothermal dusty gas as  
\begin{align*}
  \seff  = \ln{\left(\frac{c_s^2\rhog}{\rhog+\rhod}\right)},  
\end{align*}
%This allows us to define the buoyancy frequency of an isothermal 
%dusty gas. 
which implies that a non-uniform dust-to-gas ratio induces buoyancy forces.  
The effect of finite dust-gas friction appears as an energy
source term.  This analogy with standard  
hydrodynamics with cooling/heating allow us to find dusty analogs of gaseous
instabilities, and provide thermodynamical interpretations of  
dust-drag instabilities.

We obtain the equivalent Solberg-Hoiland criteria for the 
axisymmetric stability of strictly isothermal, perfectly-coupled dusty gas.  
Applying this to typical protoplanetary disks, we find that 
the vertical shear associated with dust 
 layers \emph{cannot} lead to axisymmetric  
  instabilities, however thin the dust layer is.    
Instead, sharp radial  edges in the dust-to-gas ratio could destabilized the
disk, as these imply sharp gradients in the disk's effective entropy
profile. Alternatively, if the dust is vertically well-mixed, then any
radial gradient in $\rhod/\rhog$ can destabilize the disk. 

 % {\bf presumably this limits how sharp dust rings can be} 

We  apply our thermodynamic framework to interpret
the streaming instability \citep[SI, ][]{youdin05a,jacquet11} and to generalize the gaseous vertical shear
instability \citep[VSI, ][]{nelson13,lin15} to dusty disks. We 
explicitly show in SI the evolution of gas 
pressure lags behind dust density. In fact, this is a general property of 
overstabilities driven by dust-gas drag. 
It takes a finite time for the gas to respond to
the dust motion. A lag implies there exists a time interval where 
the gas pressure of a parcel of the dusty-gas mixture is increasing whilst dust is already being expelled.  
The dusty gas then does positive work that amplifies 
oscillations. This interpretation is analogous to stellar pulsational
instabilities \citep{cox67}. 

For the VSI we find dust-loading is generally stabilizing. In our disk models 
dust-loading does not affect VSI growth rates significantly, but
meridional motions may be suppressed where the dust-induced
vertical buoyancy dominates over vertical shear, consistent with our
previous study \citep{lin15}. 
Since the dust-induced buoyancy forces
increase away from the midplane, we find dust-loading can stabilize
`surface modes'  of the VSI, that would otherwise have the largest
growth rates. { We also show that radial variations in
  $\rhod/\rhog$ can trigger a type of VSI, even when the 
  usual sources of vertical shear -- vertical dust gradients and radial 
  temperature gradients -- are negligible.  
  %effectively translates to a non-uniform radial temperature
  %profile, and the resulting vertical shear can lead to instability,
  %even if the true sound-speed is constant. 
}  

%implications for realistic disks
{ In a realistic disk, dust particles settle on a timescale $
  t_\mathrm{settle} \sim 1/\tstop\OmK^2$ \citep{takeuchi02}, compared 
  with typical VSI growth timescales, $t_\mathrm{grow}\sim 
  1/h_\mathrm{g}\OmK$. This suggest that particles with $\tstop\OmK\lesssim 
  h_\mathrm{g}$ cannot settle against the VSI. On the 
  other hand, larger particles with $\tstop\OmK\gtrsim h_\mathrm{g}$ should
  settle to form a dusty midplane. In fact, our calculations suggest that
  settling would stabilize the disk against VSI, and allow
  further settling. The result may be a quiet, dusty midplane \citep[unless non-axisymmetric
  instabilities develop, ][]{chiang08} with VSI-turbulent gaseous
  atmospheres. 
}

We also discuss future applications of our thermodynamic framework to 
study dusty protoplanetary disks. Because isothermal dusty gas has an
effective entropy, we suggest that purely hydrodynamic processes, such
as the Rossby Wave Instability \citep{li00} or the `Zombie Vortex
Instability' \citep{marcus15}, where entropy plays a role, could have dusty counter-parts.   
Furthermore, hydrodynamic instabilities driven by thermal cooling, such as VSI in
stably-stratified disks \citep{lin15} or the `Convective
Overstability' \citep{klahr14, lyra14},  may also find dusty analogs 
because finite dust-gas drag is equivalent to a heat sink/source. 

The dusty/adiabatic gas equivalence also offers a simple way to simulate dusty protoplanetary disks 
using purely hydrodynamic codes. All that is required is a re-interpretation of the fluid variables and additional 
source terms in the usual energy equation. The latter is already available in many public codes. In a follow-up work
we will apply this approach to study dusty disk-planet interaction.

%In particular, we emphasize the 
%physical origin of any instability driven by strong dust-gas
%drag is a phase lag between the pressure and total density evolution
%of the dusty gas mixture. This leads to growing modes because 
%positive work is done during an oscillation cycle, provided by
%dust-gas drag, which is equivalent to a `heat' 
%source. 

%We demonstrate a potentially new dust-drag instability in
%stratified disks that may explain 
 
%We also applied
%our framework to find unstable modes in dusty, stratified disks
%similar to that reported recently by \cite{loren15}. 

\acknowledgements
We thank P. Loren-Aguilar for initial discussions that motivated this study. We also thank 
C. Baruteau, R. Dong, J. Fung, S. Inutsuka, G. Laibe, W. Lyra, S.-J. Paardekooper, 
M. Pessah, and O. Umurhan for comments during the course of this project. 
This work is supported by the Theoretical Institute for Advanced Research in Astrophysics   
(TIARA) based in Academia Sinica Institute of Astronomy and
Astrophysics (ASIAA), NASA Astrophysics Theory Program grant
NNX17AK59G, and  
the Steward Theory Fellowship at the University of Arizona.

\appendix
\section{Variational principle}\label{var_prin}

Here we consider the more general equation of state for the gas with
$P = K \rhog^{\Gamma}$ where $K$ is a prescribed function of position
and $\Gamma$ is a constant. The effective energy equation is then
Eq. \ref{poly_energy}, which generalizes
Eq. \ref{eff_energy}. Assuming axisymmetry throughout, we linearize this equation along with Eq. \ref{masseq},  
\ref{momeq},  and \ref{poisson} to give:   

% We also include the self-gravitational
% potential $\psi$ of the gas
% plus dust mixture, which satisfies 
% %\begin{align}
% $  \nabla^2 \psi  = 4 \pi G \rho. $
% %\end{align} 

\begin{align}
%  \ii\sigma\frac{\delta\rho}{\rho} &= \nabla\cdot\dd\bm{v} +
%  \dd\bm{v}\cdot\nabla\ln{\rho},\label{lin_mass_full}\\
 \ii\sigma\frac{\Delta\rho}{\rho} &= \ii\sigma\frac{\delta\rho}{\rho} - 
  \dd\bm{v}\cdot\nabla\ln{\rho} =  \nabla\cdot\dd\bm{v},\label{lin_mass_full}\\
 %  \ii\sigma\frac{\delta P}{P} &= \Gamma \nabla\cdot\dd\bm{v} +
 % \dd\bm{v}\cdot\nabla\ln{P} - \dd\bm{v}\cdot\nabla\ln{K} - \frac{\dd\mathcal{C}}{P}.\label{lin_energy_full}\\
  \ii\sigma\frac{\Delta P}{P} &= \ii\sigma\frac{\delta P}{P} -
  \dd\bm{v}\cdot\nabla\ln{P} = 
\Gamma \nabla\cdot\dd\bm{v}  - \dd\bm{v}\cdot\nabla\ln{K} - \frac{\dd\mathcal{C}}{P},\label{lin_energy_full}\\
  -\ii\sigma\dd v_r  &= 2\Omega\dd v_\phi + 
  \hat{\bm{r}} \cdot \delta\bm{F} -  \hat{\bm{r}} \cdot \nabla\dd\psi ,\\
  \ii\sigma\dd v_\phi &= \frac{\kappa^2}{2\Omega}\dd v_r + \frac{\p
    v_\phi}{\p z}\dd v_z,\\
  -\ii\sigma\dd v_z &=  \hat{\bm{z}} \cdot \delta\bm{F}  -  \hat{\bm{z}} \cdot \nabla\dd\psi, \\ 
\nabla^2 \delta\psi & = 4\pi G \dd\rho. 
\end{align}  
where the linearized pressure force $\delta \bm{F}$ is given in
Appendix \ref{lin_press}. Recall $\Delta = \delta +
\bm{\xi}\cdot\nabla$ is the Lagrangian perturbation and $\bm{\xi}$ is the
Lagrangian displacement. In addition, 
$\ii\sigma\bm{\xi}\cdot\nabla = -\dd\bm{v}\cdot\nabla$ for axisymmetric flow.  These equations do not assume the
radially-local approximation used in numerical computations. Note that
Eq. \ref{lin_mass_full}---\ref{lin_energy_full} imply 
\begin{align}
\left|\sigma\right|^2\frac{\Delta P}{P}\frac{\Delta \rho^*}{\rho} =
\Gamma\left|\nabla\cdot\dd\bm{v}\right|^2 -
\nabla\cdot\dd\bm{v}^*\left(\dd\bm{v}\cdot\nabla\ln{K} + \frac{\dd\mathcal{C}}{P}\right),
\end{align}
so the first (real) term on the right hand side does not contribute to
$\imag{\left(\Delta P\Delta\rho^*\right)}$. Setting $K$ to constant
and taking the imaginary part gives Eq. \ref{pdv}. 

From the
linearized meridional momentum equations, we find  
\begin{align}
  \sigma^2\int\rho\left(|\dd v_r|^2 + |\dd v_z|^2\right)dV = \int\left( \rho
  \kappa^2 |\dd v_r|^2 + \rho r\frac{\p \Omega^2}{\p z} \dd v_z \dd
  v_r^*  + \ii\sigma \rho \dd \bm{F}\cdot\dd\bm{v}^*
  - \ii \sigma \rho \nabla\dd\psi\cdot  \dd\bm{v}^* 
  \right)dV,\label{meridional}
\end{align}
where the integral is taken over the volume of the fluid. Integrating
by parts and ignoring surface integrals, the last term is
\begin{align}
  \int \ii\sigma\rho \dd \bm{F}\cdot\dd\bm{v}^*dV &=\int\left( \ii\sigma \frac{\dd
    \rho}{\rho}\dd\bm{v}^*\cdot \nabla P - \ii\sigma \dd
  \bm{v}^*\cdot\nabla \dd P  \right)dV\notag\\
&= \int\left( \ii\sigma \frac{\dd
    \rho}{\rho}\dd\bm{v}^*\cdot\nabla P + \ii\sigma \dd
 P  \nabla\cdot \dd \bm{v}^*  \right)dV\notag\\
 & = \int\left[
%   \left(
%   \ii\sigma \frac{\dd P}{P} - \dd\bm{v}\cdot\nabla s + \dd
%   \bm{v} \cdot \nabla \ln{c_s^2} + \frac{\dd\mathcal{C}}{P}
%   \right)\dd\bm{v}^*\cdot\nabla P  
   \left(
   \ii\sigma \frac{\dd P}{\Gamma P} - \dd\bm{v}\cdot\nabla \seff + \frac{\dd
   \bm{v}}{\Gamma} \cdot \nabla \ln{K} +
   \frac{\dd\mathcal{C}}{\Gamma P} \right)\dd\bm{v}^*\cdot\nabla P
   + \ii\sigma \dd P  \nabla\cdot \dd \bm{v}^*
   \right]dV\notag\\
 & = \int\left[
%   \ii\sigma\frac{\dd P}{P}\nabla\cdot\left(P\dd\bm{v}^*\right) +
%   \left(\dd\bm{v}^*\cdot\nabla P\right) \left(\dd\bm{v} \cdot \nabla
%   \ln{c_s^2} + \frac{\dd\mathcal{C}}{P}   -  \dd\bm{v}\cdot\nabla s\right) 
%   \right]dV\notag\\
   \ii\sigma\frac{\dd P}{\Gamma P}\left(\dd\bm{v}^*\cdot\nabla P +
   \Gamma P\nabla\cdot\dd\bm{v}^*   \right)
   +\left(\dd\bm{v}^*\cdot\nabla P\right) \left( \frac{\dd\bm{v}}{\Gamma} \cdot \nabla
   \ln{K} + \frac{\dd\mathcal{C}}{\Gamma P}   -  \dd\bm{v}\cdot\nabla \seff\right) 
   \right]dV\notag\\
 & = \int\left\{
%   \left[\frac{1}{P}\nabla\cdot\left(P\dd\bm{v}\right) -
%   \dd\bm{v}\cdot\nabla\ln{c_s^2} - \frac{\dd\mathcal{C}}{P}  \right]\nabla\cdot\left(P\dd\bm{v}^*\right)
%   +\left(\dd\bm{v}^*\cdot\nabla P\right)\left(\dd\bm{v} \cdot \nabla
%   \ln{c_s^2} + \frac{\dd\mathcal{C}}{P}   -  \dd\bm{v}\cdot\nabla s \right)
%   \right\}dV \notag\\
 \left[
   \nabla\cdot\dd\bm{v} +
  \frac{\dd\bm{v}}{\Gamma}\cdot\nabla\ln{P} -
  \frac{\dd\bm{v}}{\Gamma}\cdot\nabla\ln{K} -
  \frac{\dd\mathcal{C}}{\Gamma P}. 
  \right]
 \left(\dd \bm{v}^*\cdot\nabla P +
 \Gamma P\nabla\cdot\dd\bm{v}^*   \right)\right. \notag\\
 &\phantom{=\int\left\{\right\}}\left. 
 + \left(\dd\bm{v}^*\cdot\nabla P\right)\left(\frac{\dd\bm{v}}{\Gamma} \cdot \nabla
 \ln{K}
 + \frac{\dd\mathcal{C}}{\Gamma P}   -  \dd\bm{v}\cdot\nabla \seff \right)
 \right\}dV \notag\\
 &=\int\left[
%     \frac{1}{P}\left|\nabla\cdot\left(P\dd\bm{v}\right)\right|^2 -
%     \left(\dd\bm{v}^*\cdot\nabla P\right)\left(\dd\bm{v}\cdot\nabla s
%     \right) -
%     P\left(\nabla\cdot\dd\bm{v}^*\right)\left(\dd\bm{v}\cdot\nabla\ln{c_s^2} + \frac{\dd\mathcal{C}}{P} \right)  
    \frac{1}{\Gamma P} \Bigl\lvert \dd \bm{v}\cdot\nabla P +
 \Gamma P\nabla\cdot\dd \bm{v}    \Bigr\rvert ^2 - 
     \left(\dd\bm{v}^*\cdot\nabla P\right)\left(\dd\bm{v}\cdot\nabla \seff
     \right) -
      P\left(\nabla\cdot\dd\bm{v}^*\right)\left(\dd\bm{v}\cdot\nabla\ln{K} + \frac{\dd\mathcal{C}}{P} \right) 
     \right]dV,
\end{align}
where  $\seff\equiv\ln{(P^{1/\Gamma}/\rho)}$. 
The self-gravitational part of Eq. \ref{meridional} is, again
neglecting surface terms when integrating by parts, 
\begin{align}
-\int \ii\sigma \rho \nabla\dd\psi \cdot\dd  \bm{v}^* &= \int \ii\sigma
  \dd\psi \nabla\cdot\left(\rho\dd\bm{v}^*\right) dV
 = \int \left|\sigma\right|^2 \dd\psi \dd \rho^*dV 
  = -\frac{1}{4\pi G}\int   \left|\sigma\right|^2
   \left|\nabla\psi\right|^2 dV,  
\end{align}
where the linearized continuity and Poisson equations have been used. 

Hence,
\begin{align}\label{integral_ex}
  &\sigma^2\int\rho\left(|\dd v_r|^2 + |\dd v_z|^2\right)dV \notag\\
&=  \int\left\{
  \rho |\dd v_r|^2 \left(\kappa^2 - \frac{1}{\rho}\frac{\p P}{\p
    r}\frac{\p \seff}{\p r}\right)
  + \rho |\dd v_z|^2\left(-\frac{1}{\rho}\frac{\p P}{\p
    z}\frac{\p \seff}{\p z}\right)
   + \rho \dd v_z \dd v_r^*\left(r\frac{\p\Omega^2}{\p z} -
  \frac{1}{\rho}\frac{\p P}{\p
    r}\frac{\p \seff}{\p z}\right) 
  + \rho \dd v_z^*\dd v_r \left(-\frac{1}{\rho}\frac{\p P}{\p
    z}\frac{\p \seff}{\p r}\right)\right. \notag\\
&\phantom{==\int}\left.+
%  \frac{1}{P}\left|\nabla\cdot\left(P\dd\bm{v}\right)\right|^2
 \frac{1}{\Gamma P}\Bigl\lvert \dd\bm{v}\cdot\nabla P +
 \Gamma P\nabla\cdot\dd\bm{v}     \Bigr\rvert^2
 - \frac{1}{4\pi G}\left|\sigma\nabla\dd\psi\right|^2 
  \right\}dV 
%  P\left(\nabla\cdot\dd\bm{v}^*\right)\left(\dd\bm{v}\cdot\nabla\ln{c_s^2}\right)dV
-\int    P\left(\nabla\cdot\dd\bm{v}^*\right)\left(\dd\bm{v}\cdot\nabla\ln{K}\right)dV
  -\int \left(\nabla\cdot\dd\bm{v}^*\right)\dd\mathcal{C}dV.  
\end{align}
Note that the coefficient of $\delta v_z\delta v_r^*$ and $\delta
v_z^*\delta v_r$ are equal owing to the equilibrium state
(Eq. \ref{vshear}). The non-self-gravitating case with $\Gamma=1,\, K = c_s^2$ gives
Eq. \ref{int_rel}. Similar integral relations are given by 
\cite{kato78, kley93,latter06}.  

%If $c_s$ is constant then the second integral vanishes, $\sigma^2$
%is real, and the first integral leads to the usual Solberg-Hoiland
%criteria. However, for a stationary but non-uniform temperature
%profile, the second  

\section{Linearized pressure forces and its divergence}\label{lin_press}
%Define
%\begin{align}
 % \bm{F} \equiv - \frac{\nabla P}{\rho}. 
%\end{align}
The linearized form of the pressure force $\bm{F} = -\nabla P/\rho$ is  
\begin{align}
  \delta \bm{F} = - \frac{\dd\rho}{\rho}\bm{F} -
  \frac{1}{\rho}\nabla\dd P,
\end{align}
with divergence
\begin{align}
  \nabla\cdot\dd\bm{F} = -
  \nabla\left(\frac{\dd\rho}{\rho}\right)\cdot\bm{F} -
  \frac{\dd\rho}{\rho}\nabla\cdot\bm{F} +
  \nabla\ln{\rho}\cdot\frac{\nabla\dd P}{\rho} -
  \frac{1}{\rho}\nabla^2\dd P. 
\end{align}

%\subsection{Explicit expressions}

The explicit forms for $\dd\bm{F}$ and $\nabla\cdot\dd\bm{F}$, in the
radially-local approximation, are
\begin{align}
  &\dd F_r = - W F_r - \ii k_x Q,\\
  &\dd F_z = - W F_z - \left[Q^\prime + Q
    \left(\ln{\rho}\right)^\prime\right]  = - \ii\sigma \dd v_z + \dd\psi^\prime, 
\end{align} 
where the last equality is the linearized
vertical momentum equation; and 
\begin{align}
  \nabla\cdot\dd\bm{F} &= F_r\left(\p_r\ln{\rho} - \ii k_x \right)W - F_z
  W^\prime - W \nabla\cdot\bm{F} + \ii k_x Q \p_r\ln{\rho} +
  \left(\ln{\rho}\right)^\prime\left[Q^\prime + Q
    \left(\ln{\rho}\right)^\prime\right] \notag\\
  & \phantom{=} + k_x^2 Q - \left\{Q^{\prime\prime} + 2
    Q^\prime\left(\ln{\rho}\right)^\prime
    + Q\left[\left(\ln{\rho}\right)^{\prime\prime}+\left(\ln{\rho}\right)^{\prime
      2}\right]\right\}\notag\\
    &= \left[\left(\p_r\ln{\rho}-\ii k_x\right)F_r - \nabla\cdot\bm{F} +
      F_z^\prime\right]W + \left(\ii k_x \p_r\ln{\rho} + k_x^2\right)Q  -
      \ii\sigma \dd v_z^\prime + \dd\psi^{\prime\prime}.
\end{align}

\section{Linearized dust diffusion}\label{lin_dust}
We consider small grains in the Epstein regime, with fixed internal
density and size, so that
\begin{align}
  \tstop   =   \frac{\mathcal{K}}{\rho c_s}\label{epstein}
\end{align}
\citep{price15}, where $\mathcal{K}$ is a constant. Then the dust diffusion
function becomes
\begin{align}
  \mathcal{C} \equiv c_s^2\nabla\cdot\left(\tepsilon\tstop\nabla
  P\right) = - \mathcal{K} c_s^2 \nabla\cdot
  \left(\frac{\tepsilon}{c_s}\bm{F}\right) =
  -\mathcal{K}c_s \left(\bm{F}\cdot\nabla\tepsilon + \tepsilon \nabla\cdot\bm{F}
  - \frac{1}{2}\tepsilon\bm{F}\cdot\nabla\ln{c_s^2}\right).  
\end{align}
Linearizing,
\begin{align}
  - \frac{\dd\mathcal{C}}{\mathcal{K}c_s} = \bm{F}\cdot\nabla\dd\tepsilon +
  \dd\bm{F}\cdot\nabla\tepsilon + \dd\tepsilon\nabla\cdot\bm{F} +
  \tepsilon\nabla\cdot\dd\bm{F} -
  \frac{1}{2}\nabla\ln{c_s^2}\cdot\left(\bm{F}\dd\tepsilon + \tepsilon
  \dd\bm{F}\right). 
\end{align}
The linearized dust-fraction $\dd\tepsilon$ and its derivatives in the
radially-local approximation are given by
\begin{align}
  \dd\tepsilon      &= (1-\tepsilon)W - \frac{Q}{c_s^2}, \\
  \p_r\dd\tepsilon &= \left[(1-\tepsilon)\left(\ii k_x -
    \p_r\ln{\rho}\right) - \p_r\tepsilon\right]W + \left(\p_r\ln{c_s^2}
  + \p_r\ln{\rho} - \ii k_x \right)\frac{Q}{c_s^2},\\
  \dd\tepsilon^\prime &= (1-\tepsilon)W^\prime - \tepsilon^\prime W -
  \left(\frac{Q}{c_s^2}\right)^\prime. 
\end{align}

%\section{Conversion formulae}
%The gas density $\rhog$ and dust-to-gas ratio $\epsilon$ is related
%to the total density $\rho$ and dust-fraction $\tepsilon$ by
%\begin{align}
%  \ln{\rho} &= \ln{\rhog} + \ln{\left(1 + \epsilon\right)},\\
%  \nabla\ln{\rho} &= \nabla\ln{\rhog} + \frac{\nabla\epsilon}{1+
%    \epsilon},\\
%  \nabla^2\ln{\rho} &= \nabla^2\ln{\rhog} +
%  \frac{\nabla^2\epsilon}{1+\epsilon} -
%  \frac{\left|\nabla\epsilon\right|^2}{\left(1+\epsilon\right)^2}, 
%\end{align}
%and
%\begin{align}
%  \tepsilon &= \frac{\epsilon}{1+ \epsilon},\\
% \nabla\tepsilon & =
% \frac{\nabla\epsilon}{\left(1+\epsilon\right)^2},\\
% \nabla^2\tepsilon &=
% \frac{\nabla^2\epsilon}{\left(1+\epsilon\right)^2} -
% \frac{2\left|\nabla\epsilon\right|^2}{\left(1+\epsilon\right)^3}. 
%\end{align}

\section{One-fluid dispersion relation for the streaming instability}\label{compressible_streaming}
We consider an unstratified disk with $\Phi = \Phi (r)$ (by setting $z=0$ in Eq. \ref{thin_disk_potential})  
so that $\p_zP = \p_z\rho = 
0$. The background $\p_rP/\rho$, $\tepsilon$ are constant 
input parameters. %Then for the Epstein drag law, Eq. \ref{epstein}, we
%have $\mathcal{C}=0$ in the background state. 
We Fourier analyze in $r$ and $z$ so that $\p_z\to \ii k_z$ and
$\p_r\to \ii k_x$ when acting on perturbations, and denote $|\bm{k}|^2
\equiv k_x^2 + k_z^2$. We consider large $k_x$ and thus neglect background
gradients when compared to that of perturbations.  
%neglect the radial density gradient compared to that of pertuburbations 
%(formally setting $\p_r\rho=0$). 
 This is also done in most local studies of
dusty disks \citep[e.g.][]{youdin07b}.  
The linearized
equations are, after eliminating the azimuthal velocity: 

\begin{align}
  \ii \sigma W &=\nabla\cdot \dd \bm{ v} = \ii k_x \dd v_x + \ii k_z \dd v_z,\label{streaming_mass}\\
    \sigma^2 \dd v_x &= \kappa^2 \dd v_x - \ii \sigma F_r W +
    k_x\sigma Q,\label{streaming_vx}\\
  -\ii\sigma\dd v_z &= -\ii k_zQ,\label{streaming_vz}\\
\ii \zeta \sigma  Q & = \frac{P}{\rho} \nabla \cdot \dd\bm{v}   -
  \frac{\dd \mathcal{C}}{\rho}. 
\end{align}
%{\bf bg pressure gradient neglected in energy eq (is this ok? could
%  fix, but get more complicated dispersion}
The artificial factor $\zeta = 1$  is inserted to keep track of the
left-hand-side of the energy equation.  Setting $\zeta$ to zero is
equivalent to assuming incompressible gas \citep{jacquet11}. 
The linearized dust diffusion
function (Appendix \ref{lin_dust}), 
under the above approximations, is 
\begin{align}
-\frac{\dd \mathcal{C}}{\rho}  = \tstop c_s^2 \left[\ii k_x F_r\left( 1-2\tepsilon\right)
  W + \tepsilon |\bm{k}|^2 Q\right]. \label{lin_drag_si}
\end{align}
%{\bf only the second term contributes to work done? dont think so}

We eliminate the velocity perturbations to obtain
\begin{align}
  \left(\ii \zeta \sigma - \tstop c_s^2 \tepsilon |\bm{k}|^2\right)Q &=
  \ii \left[
  \frac{P}{\rho}\sigma + k_x\tstop c_s^2 F_r\left(1 -
  2\tepsilon\right)\right]W, \label{si_QW1}\\
   \sigma^2 \left( \kappa^2 - \sigma^2 - \ii k_x F_r\right)W &=
    \left(k_z^2\kappa^2 - \sigma^2|\bm{k}|^2\right)Q, 
\end{align}
which yields the dispersion relation
\begin{align}
  &\frac{\zeta}{c_s^2|\bm{k}|^2}\sigma^5 + \ii \tepsilon \tstop
  \sigma^4 - \left[ \frac{\zeta}{c_s^2|\bm{k}|^2} \left(\kappa^2 - \ii
  k_x F_r\right) + \left(1 - \tepsilon\right)\right]\sigma^3 - \ii
  \tstop \left[\tepsilon \kappa^2 - \ii k_x F_r\left( 1 -
  \tepsilon\right)\right]\sigma^2 \notag\\ 
  &+ \left(1 -
  \tepsilon\right)\left(\frac{k_z\kappa}{|\bm{k}|}\right)^2\sigma +
  k_x \tstop F_r\left(1 - 2\tepsilon\right)
  \left(\frac{k_z\kappa}{|\bm{k}|}\right)^2  = 0.\label{streaming_dispersion}
\end{align}
The equation of state $P = c_s^2\left( 1 - \tepsilon\right)\rho$
was used.

\subsection{Incompressible { gas} limit}

 We can set $\zeta = 0$ or consider $c_s^2\to \infty$ to
obtain the incompressible { gas} limit. Using $\tau_\mathrm{s} \equiv
\tstop/\left(1 - \tepsilon\right)$, we obtain: 
\begin{align}
\ii \tepsilon \tau_\mathrm{s}
  \sigma^4 - \sigma^3 - 
  \tau_\mathrm{s} \left[ \ii \tepsilon \kappa^2 + k_x F_r\left( 1 -
  \tepsilon\right)\right]\sigma^2 
  + \left(\frac{k_z\kappa}{|\bm{k}|}\right)^2\sigma + 
  k_x \tau_\mathrm{s} F_r\left(1 - 2\tepsilon\right)
  \left(\frac{k_z\kappa}{|\bm{k}|}\right)^2  = 0,\label{streaming_incompressible}
\end{align}
as derived by \cite{jacquet11} and \cite{laibe14} for the streaming
instability. 
If $|\sigma|/\OmK = O(\tau_\mathrm{s}\OmK)$ and
$\tau_\mathrm{s}\OmK\ll 1$, then the quartic term
is small and may be neglected. In that case we obtain the cubic
dispersion relation of \citet{youdin05a}.

\subsection{Approximate solutions in the dust-rich limit}\label{si_dust_rich}
Here we seek analytic solutions for the streaming instability by 
examining limiting cases and with additional approximations. We will
fix $k_z, \fdust$ and maximize growth rates over $k_x$. It turns out
simple solutions exist in the dust-rich case with 
$\fdust$ near unity. We consider small stopping times, 
$\tau_\mathrm{s}\to0$, and assume that the corresponding optimum
$k_x\to\infty$.        

We begin with the incompressible dispersion
relation, Eq. \ref{streaming_incompressible}. We assume a Keplerian
disk ($\Omega=\kappa=\OmK$), and low frequency  modes 
($|\sigma|\ll \OmK$), which allows us to neglect the quartic  
term. (This is also necessary to avoid spurious modes, see
\S\ref{spurious_epi}.)  In dimensionless form, the dispersion relation is 
\begin{align}
  \nu^3 + \st\left[\ii\tepsilon + 2 K_x \fg \left(1-\fdust\right)\right]\nu^2
  - \left(\frac{K_z}{K_x}\right)^2\nu - 2\st K_x \fg
  \left(1-2\fdust\right)\left(\frac{K_z}{K_x}\right)^2 = 0,\label{dimless_cubic}
\end{align}
where $\nu = \sigma/\OmK$, $\st = \taus\OmK$, $\fg = 1-\fdust$ (the gas
fraction), and recall $K_{x,z}=\eta r k_{x,z}$. We have used $F_r =
2\eta \fg r \Omega^2$ and considered large $K_x$.  

Analytic expressions for cubic roots are unwieldy. To keep the problem
tractable, let us \emph{assume} at this stage that the quadratic term
can be neglected compare to the last term. That is, 
%This amounts 
%to assuming a  sufficiently \emph{large} $K_z$ such that :
\begin{align}
\left|\frac{\ii\fdust}{2K_x\fg} + \left(1 -
      \fdust\right)\right| \left|K_x\nu\right|^2 \ll
  K_z^2\left(2\fdust - 1\right). \label{min_kz}
\end{align}
(The imaginary term can be
neglected for fixed $\fdust$ but allowing $K_x\to\infty$.) We show in \S\ref{si_dust_rich_approx} that this
simplification is valid for dust-rich disks.  

Eq. \ref{dimless_cubic} now becomes the depressed cubic 
\begin{align}
  \nu^3 - \calP \nu = \calQ, \label{depressed_cubic}
\end{align}
where $\calP$ and $\calQ$ can be read off Eq. \ref{dimless_cubic}. This can be
solved with Vieta's substitution 
\begin{align}
\nu = \mu + \frac{\calP}{3\mu},\label{si_vieta}
\end{align}
then Eq. \ref{depressed_cubic} becomes a quadratic for $\mu^3$,
\begin{align}
\mu^6 - \calQ\mu^3 + \frac{\calP^3}{27} = 0. 
\end{align}
The solution is
\begin{align}
  \mu^3 = \frac{\calQ}{2}\left(1\pm \sqrt{1 -
      \frac{4\calP^3}{27\calQ^2}}\right) 
   = \st \fg (1- 2\fdust)\frac{K_z^2}{K_x}\left\{1 \pm \sqrt{1 -
      \frac{K_z^2}{27K_x^2\left[\st K_x \fg
          \left(1-2\fdust\right)\right]^2}}\right\}.  
\end{align}
At this point we assume the term $\propto K_z^2$ inside the square
root may be neglected. That is, 
\begin{align}
  K_z^2 \ll 27 \left(K_x^2\st\right)^2 \left[\fg
          \left(1-2\fdust\right)\right]^2. \label{max_kz}
%27 K_x^2 \left[\st K_x \fg
%          \left(1-2\fdust\right)\right]^2.
\end{align}
We show in \S\ref{si_dust_rich_approx} that this is readily satisfied. Then
$\mu \simeq \mathcal{Q}^{1/3}$.  

Now, remembering that we are considering the dust-rich limit with 
$1-2\fdust<0$, the explicit solution for $\mu$ is    
\begin{align}
\mu = \left[2\st\fg \left(2\fdust
    -1\right)\frac{K_z^2}{K_x}\right]^{1/3}e^{\ii\pi/3}. \label{mu_explicit} 
\end{align}
We have chosen the complex root for instability. Inserting this into
Eq. \ref{si_vieta} gives the eigenfrequency $\nu$:
\begin{align}
\real{(\nu)} %&= \frac{1}{2}\left(|\mu| + \frac{K_z^2}{3
%    K_x^2|\mu|}\right)\notag\\
&= \frac{1}{2}\left(\calA \st^{1/3}K_x^{-1/3} + \calB
    \st^{-1/3}K_x^{-5/3}\right),\\
\imag{(\nu)} %&= \frac{\sqrt{3}}{2}\left(|\mu| - \frac{K_z^2}{3
%    K_x^2|\mu|}\right),\notag\\
&= \frac{\sqrt{3}}{2}\left(\calA \st^{1/3}K_x^{-1/3} - \calB
    \st^{-1/3}K_x^{-5/3}\right),
\end{align}
%Inserting the expression for $|\mu|$ into $\imag{\nu}$, the
%explicit expression for the growth rate may be written as 
%\begin{align}
%  s = \frac{\sqrt{3}}{2}\left(\calA \st^{1/3}K_x^{-1/3} - \calB
%    \st^{-1/3}K_x^{-5/3}\right)\OmK, 
%\end{align}
where
\begin{align}
  &\mathcal{A} = \left[2\fg \left(2\fdust
      -1\right)K_z^2\right]^{1/3} ,\\
  &\mathcal{B} = \frac{K_z^2}{3\calA}. 
\end{align}

Maximizing growth rates over $K_x$ by setting $\p \imag{(\nu)}/\p K_x
= 0$, we find the optimum wavenumber is given by
\begin{align}
  K_x = \left(\frac{5\calB}{\calA}\right)^{3/4}\st^{-1/2} =
  \left(\frac{5}{3}\right)^{3/4}\sqrt{\frac{K_z}{2\fg\left(2\fdust-1\right)\st}}. \label{si_dust_rich_optkx}
\end{align}
The maximum growth rate is then 
\begin{align}
  \mathrm{max}\left[\imag{(\nu)}\right] =
  \frac{2\sqrt{3}\calA^{5/4}}{5^{5/4}\calB^{1/4}}\sqrt{\st} =
  \frac{2\sqrt{2}\times3^{3/4}}{5^{5/4}}\sqrt{\fg\left(2\fdust-1\right)K_z\st}.\label{si_dust_rich_grow}
\end{align}
and the real frequency $\real{(\nu)} =
(\sqrt{3}/2)\imag{(\nu)}$. We see that as $\taus\to0$, the most unstable
radial wavenumber diverges, $K_x\propto \taus^{-1/2} \to \infty$ with 
a vanishing growth rate, $\mathrm{max}(s)\propto \sqrt{\taus}\to
0$. This corresponds to instability at arbitrarily small radial
length scales.    

\subsubsection{Finite phase lag as $\taus\to 0$}  
An interesting property of the special solutions described above is
that there is always a phase lag between the Lagrangian pressure and
density perturbations. Since $\real{(\nu)}>0$, the phase lag may be
defined as $\varphi = \arg{\left(\Delta P \Delta \rho^*\right)}$. %For
%SI we neglect the background density gradients, so $\Delta\rho
%\simeq  \delta\rho$ but retain the background pressure gradient. 
(Note that we may set $\Delta\rho = 1$ without loss of generality.) 
It may be shown that in the limit of large $K_x$  that $\Delta P
\simeq \ii \delta v_x \p_rP/\sigma$. To obtain $\dd v_x$ we use
Eq. \ref{streaming_vx} in the low frequency limit. The expression for $\dd v_x$ involves
the Eulerian pressure perturbation, $\delta P$, for which we use 
Eq. \ref{si_QW1} and assume incompressibility ($\zeta\to0$).  

With these additional simplifications the phase lag $\varphi$ is given
via  
\begin{align}
  \tan{\varphi} = \frac{\imag{\left(\Delta
        P\Delta\rho^*\right)}}{\real{\left(\Delta
        P\Delta\rho^*\right)}}\simeq\frac{K_x\imag{(\nu)}}{2\fg^2\st
      K_x^2 + K_x \real{(\nu)}}. 
\end{align}
From here it is clear for the above solutions, where
$K_x\propto \st^{-1/2}$ and $\nu \propto \st^{1/2}$, that $\varphi$ is
a constant. Explicitly inserting the solutions give 
\begin{align}
\tan{\varphi} = \frac{2\times 3^{3/2}\left(2\fdust-1\right)}{16 -
  7\fdust}. 
\end{align}
For the case shown in Fig. \ref{si_compare_fig} with $\fdust=0.75$, we
have $\varphi\simeq 26\degr$, comparable to the $\sim 30\degr$
obtained from numerical solutions. 

\subsubsection{Consistency check}\label{si_dust_rich_approx}
{ Here we check if the above explicit solutions are
  consistent with the assumptions used to obtain them.  
} 
Inserting the solutions for the most unstable mode 
%(Eq. \ref{si_dust_rich_optkx}---\ref{si_dust_rich_grow} with ) 
{ we find Eq. \ref{min_kz} becomes}
\begin{align}
  \fdust \gg \frac{12}{17}\simeq 0.7.
\end{align}
Since $\fdust \leq 1$, this requirement can only be marginally
satisfied. However, we find that this mostly give errors in the real
frequency (see Fig. \ref{si_compare_fig}), while growth rates are
still captured correctly. 
On the other hand, the assumption of Eq. \ref{max_kz} becomes the trivial
inequality \begin{align} 
  1\ll \frac{27}{4}\left(\frac{5}{3}\right)^3\simeq 30,
\end{align}
so Eq. \ref{max_kz} is satisfied, which justifies the approximate solution for 
$\mu$ in Eq. \ref{mu_explicit}. 

{ Ultimately, the validity of these assumptions are justified
   a posteriori by comparison with the solution to the full equations
   in Fig. \ref{si_compare_fig}. 
}

\subsection{Spuriously growing epicycles}\label{spurious_epi}

A caveat of the dispersion relations 
Eq. \ref{streaming_dispersion} and \ref{streaming_incompressible} is 
that they admit spurious unstable modes with $|\sigma|\gtrsim \Omega$. 
This violates the one-fluid approximation to model dusty gas. 
We demonstrate this below by considering the 
incompressible dispersion relation. (We checked numerically that
compressibilty has negligible effects on the modes examined.) 

Consider the limit $k_z=0$. Then Eq. \ref{streaming_incompressible}
becomes 
\begin{align}      %streaming_incompressible
%\ii \sigma^2 - \frac{\sigma}{\epsilon \tstop}  - \left(\ii \kappa^2 +
%  \frac{k_xF_r}{\epsilon}\right) = 0. \label{kz_zero_si}
\ii\nu^2 - \frac{\nu}{\fdust\st} - \left[\ii + 
  2K_x\frac{\fg}{\fdust}\left(1-\fdust\right)\right]=0\label{kz_zero_si} 
\end{align} 
in dimensionless form. Neglecting the quadratic term leads to
stability, $\imag{(\nu)}<0$. This is consistent with a full two-fluid analysis 
\citep{youdin05a}.  
 
% However, the real and imaginary parts of Eq. \ref{kz_zero_si}  are
% \begin{align}
% \omega\left(2 s + \frac{1}{\epsilon\tstop}\right) - \frac{k_xF_r}{\epsilon} & =
%                                                                      0, \label{epi_overs1} \\
% \omega^2 - s^2 - \frac{s}{\epsilon\tstop}  - \kappa^2 &= 0, \label{epi_overs2}
% \end{align}
% so the growth rate satisfies 
% \begin{align*}
% \left( \frac{k_xF_r\tstop}{1 + 2 s\epsilon\tstop}\right)^2 - s^2
%   - \frac{s}{\epsilon \tstop} = \kappa^2. %\omega = \frac{k_xF_r\tstop}{1 + 2 s\epsilon\tstop}
% \end{align*} 
However, solving Eq. \ref{kz_zero_si} explicitly assuming $\left|\fdust\taus
\imag{(\nu)}\right| \ll 1$ would yield  
\begin{align}
%  |\omega| &\simeq \left|k_xF_r\right| \tstop \sim \kappa, \\
%  s &\simeq \epsilon \tstop \left[\left(k_xF_r\tstop\right)^2 - \kappa^2\right],\label{epi_grow}
\real{(\nu)}&\simeq 2K_x\left(1-\fdust\right)^2\st \sim O(1)\\
\imag{(\nu)}&\simeq \fdust \st \left[4K_x^2\left(1-\fdust\right)^4\st^2 -
1\right].\label{epi_grow}
\end{align}
Accordingly, growth is possible if 
\begin{align}
%\tstop > \frac{\kappa}{k_x\left|F_r\right|}\equiv t_{\mathrm{s}0}.\\%\label{epi_grow_criterion}
\st > \frac{1}{2K_x\left(1-\fdust\right)^2}\label{epi_grow_criterion}
\end{align}
%The mode is marginally stable when $\tstop =
%t_{\mathrm{s}0}$. 
Alternatively, for fixed $\taus$ growth is enabled
by a sufficiently large $K_x$. Unlike the streaming instability,
which is strongly suppressed when $\fdust=1/2$, these modes can still
grow at equal dust-to-gas ratio.  

These growing epicycles with $|\omega|\gtrsim \kappa$ are absent in
full two-fluid models \citep{youdin05a}. The discrepancy lies in the
fact that the one-fluid equations, to first order in $\taus$, are
only valid for low-frequency waves with $|\sigma|\lesssim 
\Omega$. Thus only low-frequency modes should be retained from
analyses based on Eqs. \ref{masseq}---\ref{tempeq}. 

%When this is violated,
%terms of $O(\tstop^2)$ should be retained in deriving the one-fluid
%equations (Eq. \ref{masseq}---\ref{tempeq}).  

\begin{figure}
  \includegraphics[width=\linewidth]{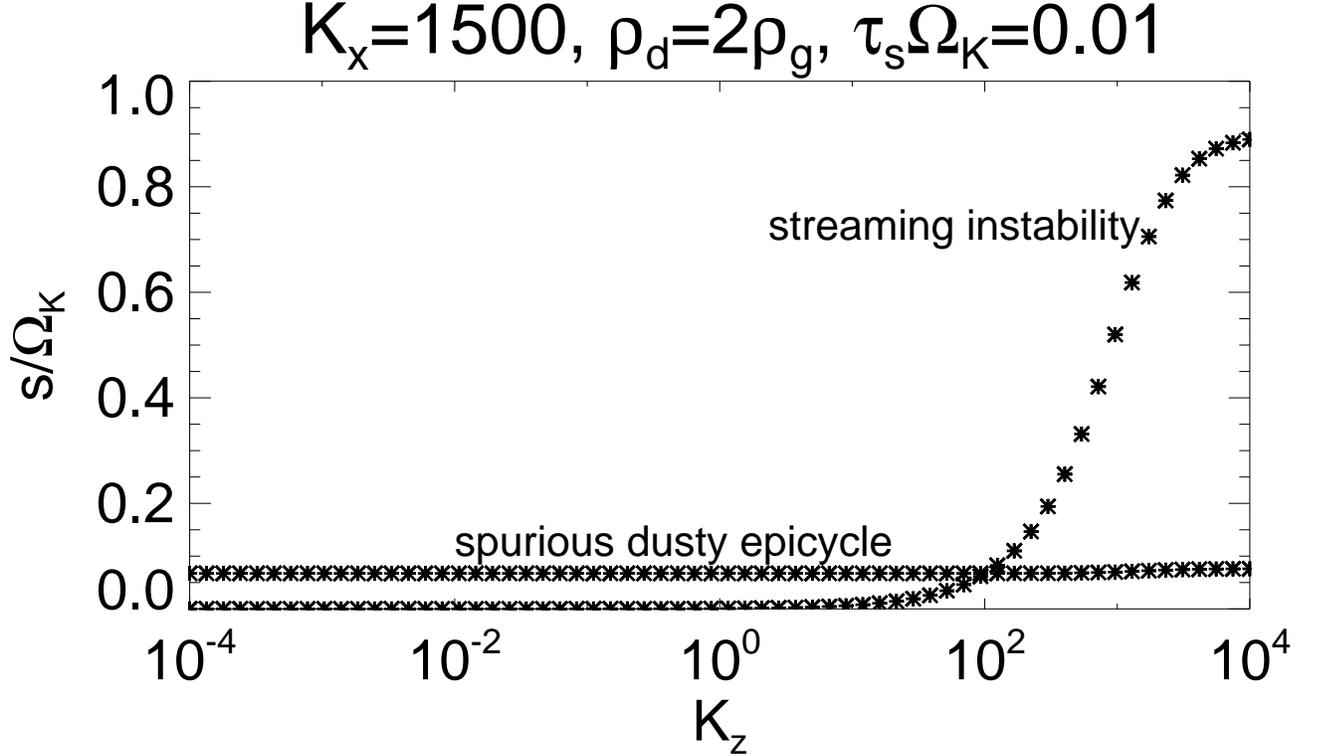}
  \caption{Growth rates of dust-drag instabilities with fixed radial
    wavenumber and dust-to-gas ratio, as a function of
    the vertical wavenumber at fix stopping time. \label{dusty_growth1}
  }
\end{figure}

Fig. \ref{dusty_growth1} show unstable modes found from
Eq. \ref{streaming_dispersion} as a function of $K_z$ at fixed $K_x 
= 1500$, $\epsilon=2$ and $\taus\OmK=0.01$.  
The (spurious) overstable dusty epicycles' growth
rates are weakly dependent on $K_z$ and are well approximated by
that in the $K_z=0$ limit, Eq. \ref{epi_grow}. 
%, which becomes 
%\begin{align*} 
%  s = \epsilon \left(1 - \tepsilon
%  \right)\taus\OmK^2\left[4K_x^2\left(1-\tepsilon\right)^4\left(\taus\OmK\right)^2
%  -1 \right]
%\end{align*}
%in the above notation. 
For the case considered in Fig. \ref{dusty_growth1} we find
$s\simeq0.067\OmK$, as observed. By contrast, the streaming
instability requires $K_z>0$, and dominates when
$K_z\gtrsim 100$. 

Care must be taken if the first-order one-fluid equations are used
to simulate dusty gas. 
%One should ensure that the physical dust-gas instabilty
%(which should have low-frequency in order for the first-order one-fluid
%approximation) has a larger growth rate than that of the spurious
%modes, which is given by Eq. \ref{epi_growth}. 
%Fig. \ref{dusty_growth1} suggests that the
%streaming instability would dominate, provided there is sufficient
%resolution to resolve high $K_z$ modes.  
For example, 2D, razor-thin disks would allow the spurious epicycles
to dominate, since in that case the streaming instability cannot
operate. However, Fig. \ref{dusty_growth1} show that in realistic 3D
disks where a range of $K_z$ is allowed, the streaming instability
should dominate. 

Thus simulations based on the first-order one-fluid equations should
be set up to suppress these spurious epicycles. This might be
achieved, for example, through physical or numerical viscosity to
eliminate high-$k_x$ modes, since for fixed disk/dust parameters these
spurious epicycles only operate at sufficiently small radial
wavelengths. Alternatively, one needs to ensure the physical
instabilities of interest have larger growth rates than the spurious
epicycles.

\end{document}